\documentclass[11pt,letterpaper]{article}

\usepackage[T1]{fontenc}
\usepackage[utf8]{inputenc}
\usepackage{lmodern}
\usepackage{microtype}

\usepackage[margin=1in]{geometry}
\usepackage{setspace}

\usepackage{amsmath, amsthm, amssymb, amsfonts}

\usepackage{algorithm}

\usepackage{graphicx}
\graphicspath{{Figure/}}
\usepackage{caption}
\usepackage{float}

\usepackage{booktabs}
\usepackage{multirow}

\usepackage{natbib}
\usepackage{xurl}

\usepackage{algpseudocode}

\usepackage{xcolor}
\usepackage{enumitem}

\definecolor{linkblue}{RGB}{24,64,112}
\usepackage[colorlinks=true,allcolors=linkblue,breaklinks=true,pdfencoding=auto,psdextra]{hyperref}
\hypersetup{pdftitle={Associating High-Dimensional Longitudinal Datasets through an Efficient Cross-Covariance Decomposition},pdfauthor={Jianbin Tan and Pixu Shi}}
\floatstyle{ruled}
\restylefloat{algorithm}
\newtheorem{Ex}{Example}

\newtheorem{Le}{Lemma}
\newtheorem{theorem}{Theorem}

\newtheorem{assumption}{Assumption}
\newtheorem{remark}{Remark}

\newcommand{\argmin}{\mathop{\rm arg\min}}
\newcommand{\argmax}{\mathop{\rm arg\max}}
\newcommand{\wtilde}{\widetilde}
\def\bm{\boldsymbol}

\newtheorem{theoremS}{Theorem}

\newcommand{\emailx}[1]{\href{mailto:#1}{#1}}

\title{Associating High-Dimensional Longitudinal Datasets through an Efficient Cross-Covariance Decomposition}

\author{Jianbin Tan\thanks{Email: \emailx{jianbin.tan@duke.edu}}\\
{\small Department of Biostatistics and Bioinformatics, Duke University, Durham, NC, U.S.A.}\\
and\\
Pixu Shi\thanks{Email: \emailx{pixu.shi@duke.edu}}\\
{\small Department of Biostatistics and Bioinformatics, Duke University, Durham, NC, U.S.A.}
}

\date{}

\begin{document}
\maketitle

\begin{abstract}
Understanding associations between paired high-dimensional longitudinal datasets is a fundamental yet challenging problem arising across scientific domains, including longitudinal multi-omic studies. The difficulty stems from the complex, time-varying cross-covariance structure coupled with high dimensionality, which complicates both model formulation and statistical estimation for association discovery. To address these challenges, we propose a new framework called Functional-Aggregated Cross-covariance Decomposition (FACD), tailored to analyzing associations in paired high-dimensional longitudinal datasets through a statistically efficient procedure. Unlike existing methods that are limited to low-dimensional data or rely on parametric modeling of temporal dynamics, FACD is capable of learning temporal structure by aggregating high-dimensional signals and accommodating variable selection to identify the most temporally associated features across datasets. We establish statistical guarantees for FACD and demonstrate its advantages over existing approaches through extensive simulation studies. Finally, we apply FACD to a longitudinal multi-omic human study, revealing blood molecules with time-varying associations across omic layers during acute exercise.
\end{abstract}

{\small \textsc{Keywords:} {\em Canonical correlation analysis, high-dimensional functional data, longitudinal multi-omic data, sparse singular value decomposition}}

\setstretch{1.08} 
\section{Introduction}\label{sec:intro}

Understanding associations between paired high-dimensional longitudinal datasets is increasingly important in biology and medicine. The growing availability of longitudinal multi-omic studies, which collect multiple types of high-dimensional molecular data over time, enables the investigation of dynamic processes across molecular layers such as the metabolome, microbiome, lipidome, proteome, and transcriptome. Analyses of such data have provided valuable insights into temporal adaptations and coordinated molecular activities during disease progression and in response to external stimuli. Notable examples include the monitoring of inflammatory bowel disease \citep{ng2023has}, the progression of COVID-19 infection \citep{diray2023multi}, and responses to environmental stressors such as high altitude \citep{han2024longitudinal} or short-term spaceflight \citep{tierney2024longitudinal}. However, these datasets are often characterized by high dimensionality, irregular and sparse temporal sampling, and complex temporal patterns, making it challenging to extract interpretable, time-varying associations between paired data modalities.

In cross-sectional settings, canonical correlation analysis (CCA) is a classical tool for identifying associations between paired datasets by seeking linear projections that maximize their correlation \citep{hardle2015canonical}. CCA has been extended to various settings, including high-dimensional data \citep{witten2009extensions, parkhomenko2009sparse, lin2013group} and longitudinal data \citep{shin2015canonical,gorecki2017correlation, hao2017identification, lee2023longitudinal}. However, extending CCA to data that are both high-dimensional and longitudinal is far from straightforward. Existing CCA methods for longitudinal data \citep{shin2015canonical, gorecki2017correlation, lee2023longitudinal} are mainly developed for low-dimensional regimes and may not scale well to modern high-dimensional studies. Moreover, many of these approaches require prespecified models to characterize temporal structure, limiting their flexibility in real-world applications where the underlying dynamics may not be well captured by fixed model forms.
 
To account for general temporal patterns, functional data approaches \citep{he2003functional,eubank2008canonical,yang2011functional,hsing2015theoretical} provide a flexible framework for uncovering associations between longitudinal datasets. In general, these approaches first estimate the cross-covariance kernel from the observed data and then perform a kernel decomposition to extract the dominant modes of dependence. While these procedures can, in principle, be extended to high-dimensional longitudinal settings, these extensions require factorizing cross-covariance kernels defined over high-dimensional functional spaces. Such decompositions are computationally intractable, statistically inefficient, and often require large sample sizes.

\begin{figure}
\centering
\includegraphics[width=0.8\textwidth]{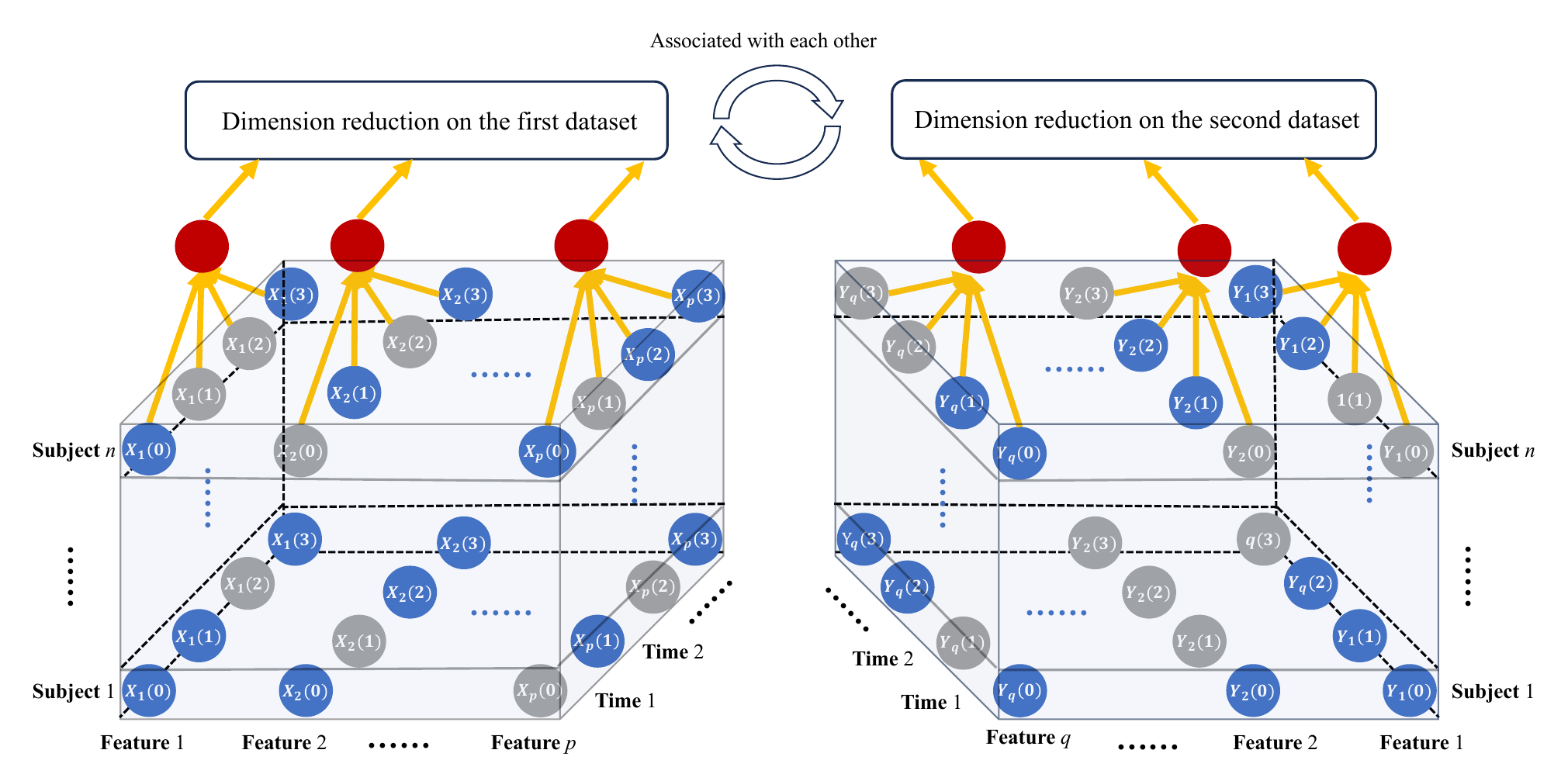}
\caption[Illustration of longitudinal associations]{
Diagram illustrating associations between two high-dimensional longitudinal datasets.
Each dataset is represented by a three-dimensional array indexed by feature, time, and subject.
Yellow arrows represent time-varying functional loadings that aggregate features into lower-dimensional representations, denoted by red circles, whose cross-covariance across the two datasets is maximized.
Gray circles indicate missing observations at certain time points.
}
\label{illu_task}
\end{figure}

To fill this methodological gap, we propose Functional-Aggregated Cross-covariance Decomposition (FACD), a new framework for characterizing interpretable time-varying associations between paired high-dimensional longitudinal datasets. Typically, FACD captures the dominant association structure through a joint, time-dependent dimension reduction of the two datasets, as illustrated in Figure~\ref{illu_task}. We first show that the extraction of associated components can be formulated as a spectral decomposition of a high-dimensional cross-covariance kernel, which in turn admits a representation through functional basis expansion. To characterize the cross-covariance structure, we introduce a functional aggregation technique that integrates signals across high-dimensional features, enabling a data-adaptive basis representation for capturing complex temporal associations in real-world data.

By leveraging the relationship between data-adaptive bases and cross-covariance decomposition, FACD avoids direct spectral decomposition of high-dimensional kernels by transforming the problem into a matrix singular value decomposition, yielding a computationally tractable and statistically efficient implementation. To improve interpretability in high dimensions, we further incorporate sparsity into the FACD procedure, enabling the selection of key features across datasets. These procedures accommodate irregular and sparse temporal observations, making FACD applicable to a wide range of longitudinal settings.

We provide theoretical support and statistical guarantees for the FACD framework. Through extensive simulation studies, we demonstrate the superior performance of FACD over related methods in recovering canonical components. Finally, we apply FACD to a longitudinal multi-omic study profiling blood molecules in healthy human subjects undergoing acute physical exercise \citep{contrepois2020molecular}. The analysis reveals meaningful temporal associations across omic layers at both the global and molecular levels, underscoring the potential of FACD to advance our scientific understanding of dynamic biological interactions.

The rest of this article is organized as follows. In Section~\ref{sec:method}, we introduce the theoretical framework of FACD for high-dimensional functional data. In Section~\ref{sec:imp}, we describe the implementation of FACD for observed longitudinal data. We conduct simulation studies to evaluate the effectiveness of FACD in Section~\ref{sec:sim} and apply FACD to real-world data in Section~\ref{sec: data}. Finally, we provide statistical analysis in Section~\ref{Sec: sta_the} and conclude with a discussion in Section~\ref{sec:dis}. Additional supporting results are provided in the Supplementary Materials.
The codes and datasets are publicly available at
\url{https://github.com/Tan-jianbin/Functional-Aggregated-Cross-covariance-Decomposition-FACD}.
An R package named \texttt{FACD} can be found at
\url{https://github.com/Tan-jianbin/FACD}.

\section{Functional-Aggregated Cross-covariance Decomposition (FACD)}\label{sec:method}

For \( \bm{f}(t), \bm{g}(t) \in \mathbb{R}^p \) with \( t \in \mathcal{T} \), define the inner product
\(
\langle \bm{f}, \bm{g} \rangle_p := \int_{\mathcal{T}} \bm{f}^\top(t) \bm{g}(t)\, \mathrm{d}t
\). Without loss of generality, we set \( \mathcal{T} = [0,1] \).  
Let \( L^2(\mathbb{R}^p) \) denote the Hilbert space of square-integrable functions on \(\mathcal{T}\) valued in \( \mathbb{R}^p \), equipped with inner product \( \langle \cdot, \cdot \rangle_p \) and norm \( \|\cdot\|_p := \sqrt{\langle \cdot, \cdot \rangle_p} \). 
When \( p = 1 \), we simplify \( \|\cdot\|_p \) and $\langle \cdot, \cdot \rangle_p$ to \( \|\cdot\| \) and $\langle \cdot, \cdot \rangle$, respectively.
{We use $\|\cdot\|_{E}$ to denote the Euclidean norm for real vectors} and let \( \|\cdot\|_F \) denote the Frobenius norm of a matrix, and write \( \operatorname{span}\{f_1, f_2, \dots\} \) for the closure of the functional space spanned by \( \{f_1, f_2, \dots\} \). Define \( \mathbb{I}(\cdot) \) as the indicator function. 
Let \( \mathbb{S}_p \) and \( \mathbb{S}_q \) denote the unit spheres in \( L^2(\mathbb{R}^p) \) and \( L^2(\mathbb{R}^q) \).
For a positive-definite kernel $\bm R(t,u)=\sum_{k\ge1}\lambda_k\,\bm\phi_k(t)\bm\phi_k^\top(u)$ with $\lambda_k>0$, we define its square-root inverse
$\bm R^{-1/2}(t,u)$ as $\sum_{k\ge1}\lambda_k^{-1/2}\,\bm\phi_k(t)\bm\phi_k^\top(u)$.

\subsection{Quantifying High-Dimensional Longitudinal Association}

Let \( \bm{X}(t) = (X_{1}(t), \dots, X_{p}(t))^\top \) denote a \( p \)-dimensional random function representing the values of \( p \) features at time \( t \), where \( t \in \mathcal{T} \). Similarly, let \( \bm{Y}(s) \in \mathbb{R}^{q} \) for \( s \in \mathcal{S} \) denote a second data source. For simplicity, we assume \( \mathcal{T} = \mathcal{S} = [0,1] \), though our method also applies when \( \mathcal{T} \) and \( \mathcal{S} \) are different intervals. 

Define the positive-definite kernels $\bm{R}_X(t_1, t_2) := \operatorname{Cov}\big\{\bm{X}(t_1), \bm{X}(t_2)\big\} \in \mathbb{R}^{p \times p}$ for $t_1, t_2 \in \mathcal{T}$ and
$\bm{R}_Y(s_1, s_2) := \operatorname{Cov}\big\{\bm{Y}(s_1), \bm{Y}(s_2)\big\} \in \mathbb{R}^{q \times q}$ for $s_1, s_2 \in \mathcal{S}$, and the cross-covariance kernel
$\bm{R}_{XY}(t, s) :=\operatorname{Cov}\big\{\bm{X}(t), \bm{Y}(s)\big\} \in \mathbb{R}^{p \times q}$ for $t\in \mathcal{T}$ and $s\in \mathcal{S}$. For any kernel $\bm{R}(t,s)$, denote its singular value decomposition (SVD) as
\begin{equation}\label{SVD}
\bm{R}(t,s) = \sum_{r=1}^{\infty}\lambda_r\, \wtilde{\bm A}^{\text{left}}_r(t)\big\{\wtilde{\bm A}^{\text{right}}_r(s)\big\}^\top,
\end{equation}
where \( \lambda_1\ge\lambda_2\ge\cdots\ge 0 \) are singular values and
\( \wtilde{\bm A}_r^{\text{left}}\in L^2([0,1];\mathbb R^p)\), \( \wtilde{\bm A}_r^{\text{right}}\in L^2([0,1];\mathbb R^q)\)
are the left and right singular functions, normalized in \(L^2\)-norm.
The decomposition \eqref{SVD} exists as long as \( \bm{R}(t,s) \) is continuous with respect to \( t \) and \( s \) (see Theorem~4.6.2 in \citet{hsing2015theoretical}).

In this article, our goal is to quantify the dependence between \( \bm X \) and \( \bm Y \) over time. A natural strategy is to construct low-dimensional summaries of the two longitudinal data sources by projecting them onto time-varying loading functions and seeking projected scores that are maximally associated. Different choices of association measures lead to different notions of longitudinal association. Below, we present two representative examples.

\begin{Ex}[Canonical Correlation Analysis]\label{ex:cca}
Canonical correlation analysis maximizes the correlation between two projected scores. The first pair of canonical loading functions is defined as
\begin{eqnarray}\label{FCCA}
(\wtilde{\bm A}^X_1,\wtilde{\bm A}^Y_1)=\argmax_{\bm{A}_X \in \mathbb{S}_p,\ \bm{A}_Y \in \mathbb{S}_q}
\operatorname{Corr}^2\!\bigg(\int_0^1 \bm{A}^\top_X(t)\bm{X}(t)\,\mathrm{d}t,\ 
\int_0^1 \bm{A}^\top_Y(s)\bm{Y}(s)\,\mathrm{d}s\bigg).
\end{eqnarray}
It can be shown that $\wtilde{\bm A}^X_1$ and
$\wtilde{\bm A}^Y_1$ can be obtained by transforming
and normalizing the leading singular functions obtained
by applying the SVD in \eqref{SVD} to the correlation
kernel $\bm R_{\mathrm{corr}}$:
\[
\bm R_{\mathrm{corr}}(t,s):= 
\iint_{\mathcal T\times \mathcal T}
\bm R_X^{-1/2}(t,u)\,\bm R_{XY}(u,v)\,\bm R_Y^{-1/2}(s,v)^\top\,\mathrm{d}v\,\mathrm{d}u
\in\mathbb R^{p\times q};
\]
see Part~\ref{FCCA_irre} of the Supplementary Materials for the proof.
CCA is a conventional method for associating functional/longitudinal data \citep{he2003functional,eubank2008canonical,hsing2015theoretical}. However, it may suffer from degeneracy when the variance becomes singular or nearly singular. 
For example, in the functional setting \eqref{FCCA}, the variance 
\(
\operatorname{Var}\!\left(\int_0^1 \bm{A}_X^\top(t)\bm{X}(t)\,dt\right)
=\iint_0^1 \bm{A}_X^\top(t)\,\bm{R}_X(t,s)\,\bm{A}_X(s)\,dt\,ds
\)
can be made arbitrarily small by choosing $\bm A_X$ as an eigenfunction of $\bm R_X$ associated with a sufficiently small eigenvalue. This singularity makes CCA favor variance-shrinking loadings and can render $\bm R_{\mathrm{corr}}$ ill-defined, so a CCA solution may be unreasonable or may even fail to exist; see Part~\ref{FCCA_irre} of the Supplementary Materials for further discussion.
\end{Ex}

\begin{Ex}[Canonical Cross-Covariance Analysis]\label{ex:ccca}
Canonical cross-covariance analysis, also called singular component analysis, is a common substitute for CCA in high-dimensional or functional settings \citep{witten2009extensions,parkhomenko2009sparse,lin2013group,yang2011functional}. Canonical cross-covariance analysis directly seeks to maximize the covariance between paired data modalities, thereby avoiding the evaluation of variances. The first pair of canonical loading functions is defined as
\begin{eqnarray}\label{FSCA}
({\wtilde{\bm A}^X_1,\wtilde{\bm A}^Y_1}) 
= \argmax_{\bm{A}_X \in \mathbb{S}_p,\ \bm{A}_Y \in \mathbb{S}_q} 
\operatorname{Cov}^2\!\left(
\int_0^1 \bm{A}^\top_X(t)\bm{X}(t)\,\mathrm{d}t,\ 
\int_0^1 \bm{A}^\top_Y({s})\bm{Y}(s)\,\mathrm{d}{s}
\right).
\end{eqnarray}
It can be shown that \((\wtilde{\bm A}^X_1,\wtilde{\bm A}^Y_1)\) are also the leading singular functions obtained by applying SVD \eqref{SVD} to the cross-covariance kernel 
\(
\bm{R}_{XY}(t, s)
\)
(see Part~\ref{sec: prd} of the Supplementary Materials for the proof).  
To remove the dependence on feature scales, it is common to normalize the components prior to analysis, e.g., replacing $X_j$ by $X_j/\mathbb{E}\|X_j\|$, and similarly for $\bm Y$. This normalization plays a role analogous to inverse-variance normalization in CCA, while avoiding the degeneracy issue.
\end{Ex}

Examples~\ref{ex:cca} and \ref{ex:ccca} indicate that \eqref{SVD} includes common methods for associating functional data.
Given the {loadings} extracted from \eqref{SVD}, we define the $r$th canonical scores for $\bm{X}$ and $\bm{Y}$ by
\begin{equation}\label{XY_score}
    \begin{aligned}
        z_{r}^X&:=\int_0^1 \{\wtilde{\bm{A}}^X_{r}(t)\}^\top\big\{\bm{X}(t)-\bm{\mu}_X(t)\big\}\, \mathrm{d}t,\quad 
        z_{r}^Y&:=\int_0^1 \{\wtilde{\bm{A}}^Y_{r}(t)\}^\top\big\{\bm{Y}(t)-\bm{\mu}_Y(t)\big\}\, \mathrm{d}t,
    \end{aligned}
\end{equation}
where $\bm{\mu}_X$ and $\bm{\mu}_Y$ are the mean functions of $\bm{X}$ and $\bm{Y}$, respectively. 
The pair $(z_r^X,z_r^Y)$ summarizes the overall dependence between $\bm{X}$ and $\bm{Y}$ by aggregating information across time and features, with contributions weighted by the canonical loadings $\wtilde{\bm{A}}^X_{r}$ and $\wtilde{\bm{A}}^Y_{r}$.

To obtain $\wtilde{\bm{A}}^X_{r}$ and $\wtilde{\bm{A}}^Y_{r}$, the SVD of the kernels $\bm R_{\mathrm{corr}}$ or $\bm R_{XY}$ in Examples~\ref{ex:cca} and \ref{ex:ccca} is typically achieved through spectral decompositions.
Taking $\bm R_{XY}$ as an example, we define
\begin{eqnarray*}
    \bm{L}_X(t_1,t_2)&:=&\int_{0}^1\bm{R}_{XY}(t_1,t)\bm{R}^\top_{XY}(t_2,t)\ \allowbreak\mathrm{d}t
=\sum_{r=1}^\infty \lambda_r^2 \wtilde{\bm{A}}^X_{r}(t_1)\big\{\wtilde{\bm{A}}^X_{r}(t_2)\big\}^\top\in \mathbb{R}^{p\times p},\\
\bm{L}_Y(t_1,t_2)&:=& \int_{0}^1\bm{R}^\top_{XY}(t,t_1)\bm{R}_{XY}(t,t_2)\ \mathrm{d}t
=\sum_{r=1}^\infty \lambda_r^2 \wtilde{\bm{A}}^Y_{r}(t_1)\big\{\wtilde{\bm{A}}^Y_{r}(t_2)\big\}^\top\in \mathbb{R}^{q\times q},
\end{eqnarray*}
where $\lambda_r^2$ is the $r$th eigenvalue of $\bm{L}_X$ and $\bm{L}_Y$, and $\wtilde{\bm{A}}^X_{r}$ and $\wtilde{\bm{A}}^Y_{r}$ are the corresponding eigenfunctions.
However, the spectral decompositions of $\bm{L}_X(t_1,t_2)$ and $\bm{L}_Y(t_1,t_2)$ are impractical when $p$ and $q$ are large, as they involve high-dimensional eigenfunction computations for complicated kernels.
This makes existing methods {inapplicable} to high-dimensional functional settings.
Our goal is therefore to develop an efficient procedure for decomposing such high-dimensional kernels to associate $\bm{X}$ and $\bm{Y}$.

\subsection{Functional-Aggregated Decomposition of High-Dimensional Kernels}\label{sec:FACDCC}

In this subsection, we propose an efficient approximation of the singular functions
\( \wtilde{\bm{A}}^X_{r} \) and \( \wtilde{\bm{A}}^Y_{r} \) that avoids high-dimensional
kernel spectral decomposition. As CCA may not be well defined or stable in high-dimensional functional settings, we illustrate our method using the SVD of the cross-covariance kernel $\bm{R}_{XY}$. 
Nonetheless, our method applies more broadly to the decomposition of high-dimensional kernels in general settings.

We begin by presenting the following theorem.

\begin{theorem}\label{theo: svd_mar}
 Define $\mathcal H_1 :=
\operatorname{span}
\{\wtilde A_{rj}^X:
j=1,\ldots,p,\ r\ge1\}\subset L^2({\mathbb{R}})$ and $\mathcal H_2 :=
\operatorname{span}
\{\wtilde A_{rm}^Y:
m=1,\ldots,q,\ r\ge1\}\subset L^2({\mathbb{R}})$, where $(\wtilde{A}_{r1}^X,\ldots,\wtilde{A}_{rp}^X)=\wtilde{\bm{A}}_{r}^X$ and $(\wtilde{A}_{r1}^Y,\ldots,\wtilde{A}_{rq}^Y)=\wtilde{\bm{A}}_{r}^Y$. Then, for any orthonormal bases $\{\psi_{k_1}^X; k_1 \geq 1\}$ and $\{\psi_{k_2}^Y; k_2 \geq 1\}$ in $L^2(\mathbb{R})$ such that $ \mathcal{H}_1\subset \operatorname{span}\{\psi_{k_1}^X, k_1 \geq 1\} \ \text{and}\ \allowbreak \mathcal{H}_2\subset\operatorname{span}\{\psi_{k_2}^Y, k_2 \geq 1\}$,
\begin{eqnarray*}
    \bm{R}_{XY}(t_1,t_2)&=&\sum_{k_1=1}^{\infty}\sum_{k_2=1}^{\infty}\bm{\Lambda}^{XY}_{k_1,k_2}\psi_{k_1}^X(t_1) \psi_{k_2}^Y(t_2),
\end{eqnarray*}
where $\bm{\Lambda}^{XY}_{k_1,k_2}=\int_0^1\int_0^1 \bm{R}_{XY}(t_1,t_2)\psi_{k_1}^X(t_1)\psi_{k_2}^Y(t_2)\ \mathrm{d}t_1\mathrm{d}t_2\in \mathbb{R}^{p\times q}$.
\end{theorem}

Theorem~\ref{theo: svd_mar} shows that \( \bm{R}_{XY}(t_1, t_2) \) can be represented using the basis functions \( \{\psi^X_k\}_{k \geq 1} \) and \( \{\psi^Y_k\}_{k \geq 1} \).
These bases capture the functional variations in the singular functions \( \wtilde{\bm{A}}_{r}^X \) and \( \wtilde{\bm{A}}_{r}^Y \), allowing \( \bm{R}_{XY}(t_1, t_2) \) to be expanded in the tensor product basis \( \{\psi_{k_1}^X(t_1)\psi_{k_2}^Y(t_2);\, k_1, k_2 \geq 1\} \).
While similar approaches are used in functional and longitudinal analyses \citep{gorecki2017correlation, gorecki2020independence, lee2023longitudinal,tan2024graphical}, our method differs in that we do not pre-specify the basis functions but instead estimate them from the kernel $\bm{R}_{XY}$.

To obtain the bases of \( \mathcal{H}_1 \) and \( \mathcal{H}_2 \), we define the positive-definite kernels
\begin{eqnarray}
H_X(t_1,t_2) &:=& \operatorname{tr}\{\bm{L}_X(t_1,t_2)\}/(pq)
= \frac{1}{pq}\sum_{j=1}^p\sum_{m=1}^q \int_0^1 R_{jm}^{XY}(t_1,t)R_{jm}^{XY}(t_2,t)\,\mathrm{d}t,\label{H_X}\\
H_Y(t_1,t_2) &:=& \operatorname{tr}\{\bm{L}_Y(t_1,t_2)\}/(pq)
= \frac{1}{pq}\sum_{j=1}^p\sum_{m=1}^q \int_0^1 R_{jm}^{XY}(t,t_1)R_{jm}^{XY}(t,t_2)\,\mathrm{d}t,\label{H_Y}
\end{eqnarray}
where \( R_{jm}^{XY}(t_1,t_2) \) is the \((j,m)\)th element of \( \bm{R}_{XY}(t_1,t_2) \).  
Here, \( H_X \) can be viewed as a marginal kernel of \( R_{jm}^{XY}(t_1,t)R_{jm}^{XY}(t_2,t) \), aggregating over the feature indices \( j,m \) and the time index \( t \) from \( \bm{Y} \), and capturing the variation in \( \bm{X} \).  
Because \( H_X(t_1,t_2) \) can also be written as  
\(
H_X(t_1,t_2) 
= \frac{1}{pq}\sum_{r=1}^{\infty} \lambda_r^2 
\sum_{j=1}^{p} \wtilde{A}_{rj}^X(t_1)\wtilde{A}_{rj}^X(t_2)
\) (since $\bm{L}_X(t_1,t_2)=\sum_{r=1}^\infty \lambda_r^2 \wtilde{\bm{A}}^X_{r}(t_1)\big\{\wtilde{\bm{A}}^X_{r}(t_2)\big\}^\top$),
it aggregates all functional variation in the functions \( \wtilde{A}_{rj}^X\) spanning \( \mathcal{H}_1 \). Therefore, we can use \( H_X(t_1,t_2) \) to extract functional bases for \( \mathcal{H}_1 \), and similarly \( H_Y(t_1,t_2) \) for \( \mathcal{H}_2 \), as stated in the following theorem.

\begin{theorem}\label{The_opt_bas}
Assume the conditions in Theorem~\ref{theo: svd_mar} hold.  
Let the spectral decompositions of \( H_X(t_1,t_2) \) and \( H_Y(t_1,t_2) \) be  
\[
H_X(t_1,t_2) = \sum_{k\geq 1}\zeta_k^X \wtilde{\psi}_{k}^X(t_1)\wtilde{\psi}_{k}^X(t_2), 
\qquad
H_Y(t_1,t_2) = \sum_{k\geq 1}\zeta_k^Y \wtilde{\psi}_{k}^Y(t_1)\wtilde{\psi}_{k}^Y(t_2),
\]
where \( \wtilde{\psi}_{k}^X \) and \( \wtilde{\psi}_{k}^Y \) are the \( k \)th eigenfunctions with positive eigenvalues \( \zeta_{k}^X \) and \( \zeta_{k}^Y\), respectively.  
Then, the functional spaces $\mathcal{H}_1$ and $\mathcal{H}_2$ satisfy: 
\(
\mathcal{H}_1 \subset \operatorname{span}\{\wtilde{\psi}_{k_1}^X : k_1 \geq 1 \}
\)
and
\( 
\mathcal{H}_2 \subset \operatorname{span}\{\wtilde{\psi}_{k_2}^Y : k_2 \geq 1\}.
\)
Define  
\(
\wtilde{\bm{\Lambda}}^{XY}_{k_1,k_2}
= \int_0^1 \!\!\int_0^1 \bm{R}_{XY}(t_1,t_2)\wtilde{\psi}_{k_1}^X(t_1)\allowbreak\wtilde{\psi}_{k_2}^Y(t_2)\,\mathrm{d}t_1\mathrm{d}t_2.
\)
Then, we have
\begin{equation}\label{eqa_mar_svd}
\bm{R}_{XY}(t_1,t_2)
= \sum_{k_1=1}^{\infty}\sum_{k_2=1}^{\infty}
\wtilde{\bm{\Lambda}}^{XY}_{k_1,k_2}\wtilde{\psi}_{k_1}^X(t_1)\wtilde{\psi}_{k_2}^Y(t_2).
\end{equation}
\end{theorem}

Theorem~\ref{The_opt_bas} shows that the bases of \( \mathcal{H}_1 \) and \( \mathcal{H}_2 \) can be obtained via the spectral decompositions of \( H_X(t_1, t_2) \) and \( H_Y(t_1, t_2) \), respectively. 
Note that \( H_X(t_1, t_2) \) and \( H_Y(t_1, t_2) \) are standard kernels that do not depend on the ambient dimensions \(p\) and \(q\). 
Therefore, the basis extraction in Theorem~\ref{The_opt_bas} is directly applicable to high-dimensional settings. 

Next, we show that, with these bases, the spectral decompositions of \( \bm{L}_X(t_1,t_2) \) and \( \bm{L}_Y(t_1,t_2) \) can be transformed into the SVD of a matrix.

\noindent\textbf{Truncated Approximation of $\bm{R}_{XY}$.} 
Assuming $\{\wtilde{\psi}^X_{k}; k \geq 1\}$ and $\{\wtilde{\psi}^Y_{k}; k \geq 1\}$ have eigenvalues {that decay} with $k$, we can approximate $\wtilde{\bm{A}}_{r}^X$ and $\wtilde{\bm{A}}_{r}^Y$ by working with a truncated version of \eqref{eqa_mar_svd} to approximate $\bm{R}_{XY}(t_1,t_2)$: 
\(
\bm{R}_{XY}(t_1,t_2)\approx\sum_{k_1=1}^{\kappa_X}\sum_{k_2=1}^{\kappa_Y}\wtilde{\bm{\Lambda}}^{XY}_{k_1,k_2}\wtilde{\psi}_{k_1}^X(t_1) \wtilde{\psi}_{k_2}^Y(t_2),
\)
where $\kappa_X$ and $\kappa_Y$ are finite numbers.
Let $\bm{\Gamma}$ be a block matrix with the $(k_1,k_2)$th block given by 
$\wtilde{\bm{\Lambda}}^{XY}_{k_1,k_2}$
for $k_1=1,\dots,\kappa_X$ and $k_2=1,\dots,\kappa_Y$. 
Assume the SVD of $\bm{\Gamma}\in \mathbb{R}^{p\kappa_X\times q\kappa_Y}$ is 
\begin{gather}\label{SVD_Gamma}
   \bm{\Gamma} =  \sum_{r=1}^{R}
   \eta_{r}\cdot \bm{a}_{r}\bm{b}_{r}^{\top},
\end{gather}
where $R$ is the rank of $\bm{\Gamma}$, and $\eta_{r}$, $\bm{a}_{r}$, and 
$\bm{b}_{r}$ are the $r$th singular {value and corresponding singular vectors} of $\bm{\Gamma}$, respectively.
Denote 
$\bm{a}_{r}=((\bm{a}_{r}^{(1)})^\top,\dots,(\bm{a}_{r}^{(\kappa_X)})^\top)^\top\in \mathbb{R}^{p\cdot\kappa_X}$ and 
$\bm{b}_{r}=((\bm{b}_{r}^{(1)})^\top,\dots,(\bm{b}_{r}^{(\kappa_Y)})^\top)^\top\in \mathbb{R}^{q\cdot\kappa_Y}$.
The truncated approximation ${\bm{R}}_{XY}^{(\kappa_X,\kappa_Y)}(t_1,t_2):= \sum_{k_1=1}^{\kappa_X}\sum_{k_2=1}^{\kappa_Y}\wtilde{\bm{\Lambda}}^{XY}_{k_1,k_2}\wtilde{\psi}_{k_1}^X(t_1) \wtilde{\psi}_{k_2}^Y(t_2)$ can be rewritten as
\begin{eqnarray}\label{SVD_appro}
   {\bm{R}}_{XY}^{(\kappa_X,\kappa_Y)}(t_1,t_2)
  & =
&   \sum_{r=1}^{R}
   \eta_{r}\cdot 
   \bigg\{\sum_{k_1=1}^{\kappa_X}\bm{a}_{r}^{(k_1)}\cdot\wtilde{\psi}_{k_1}^X(t_1)\bigg\}
\bigg\{\sum_{k_2=1}^{\kappa_Y}\bm{b}_{r}^{(k_2)}\cdot\wtilde{\psi}_{k_2}^Y(t_2)\bigg\}^{\top}\nonumber \\
&:=& \sum_{r=1}^{R}
   \eta_{r}\cdot \bm{\wtilde{A}}_{r,\text{appr}}^{X}(t_1)
   \big\{\bm{\wtilde{A}}_{r,\text{appr}}^{Y}(t_2)\big\}^{\top},\ t_1,t_2\in [0,1]. 
\end{eqnarray}
This representation takes the same form as \eqref{SVD}.
Consequently, the decomposition of {the cross-covariance kernel} $\bm{R}_{XY}(t_1,t_2)$ is simplified to the SVD of the matrix $\bm{\Gamma}$.

Note that the signs of \( \wtilde{\bm{A}}_r^X \) and \( \wtilde{\bm{A}}_r^Y \) are not identifiable. 
Without loss of generality, we align them with the signs of \( \bm{\wtilde{A}}_{r,\text{appr}}^{X} \) and \( \bm{\wtilde{A}}_{r,\text{appr}}^{Y} \), respectively, such that 
\( \langle \bm{\wtilde{A}}_{r}^{X}, \bm{\wtilde{A}}_{r,\text{appr}}^{X} \rangle_p \geq 0 \) and 
\( \langle \bm{\wtilde{A}}_{r}^{Y}, \bm{\wtilde{A}}_{r,\text{appr}}^{Y} \rangle_{q} \geq 0 \). 
The accuracy of \( \bm{\wtilde{A}}_{r,\text{appr}}^{X} \) and \( \bm{\wtilde{A}}_{r,\text{appr}}^{Y} \) then depends on the decay rates of the eigenvalues 
\( \{\zeta_k^X, k > \kappa_X\} \) and 
\( \{\zeta_k^Y, k > \kappa_Y\} \), as shown in the following theorem.

\begin{theorem}\label{Theorem_appr}
Assume the conditions in Theorem~\ref{theo: svd_mar} hold. Suppose 
\[
\max_{1\le j\le p,\,1\le m\le q}\int_0^1\int_0^1\{R^{XY}_{jm}(t,s)\}^2\ \mathrm{d}t\mathrm{d}s
\]
is bounded by a constant, and that the singular values $\lambda_r$ of $\bm{R}_{XY}(t_1,t_2)$ are distinct.
Then,
\begin{eqnarray*}
  \max\big\{\|\wtilde{\bm{A}}^X_r- \bm{\wtilde{A}}_{r,\text{appr}}^{X}\|_p,\ \|\wtilde{\bm{A}}^Y_r- \bm{\wtilde{A}}_{r,\text{appr}}^{Y}\|_q\big\}
  \leq 
  \frac{C\cdot pq}{\inf_{r^{'}\neq r}|\lambda_r^2-\lambda_{r^{'}}^2|}
  \sqrt{\sum_{k_1>\kappa_X} \zeta_{k_1}^X +  \sum_{k_2>\kappa_Y} \zeta_{k_2}^Y},
\end{eqnarray*}
where $C$ is a constant independent of $p$, $q$, $\kappa_X$, and $\kappa_Y$.
\end{theorem}

The rate in Theorem~\ref{Theorem_appr} indicates that the truncation levels $\kappa_X$ and $\kappa_Y$ should be chosen based on the decay behavior of the eigenvalues of $H_X(t_1,t_2)$ and $H_Y(t_1,t_2)$ to ensure the accuracy of the approximated singular functions. Here, the upper bound acts like a perturbation bound: 
$\sum_{k_1>\kappa_X}\zeta^X_{k_1}+\sum_{k_2>\kappa_Y}\zeta^Y_{k_2}$ quantifies the truncation perturbation, 
the $pq$ factor reflects dimension-averaged accumulation across the $p\times q$ coordinates, 
and the eigen-gap $\inf_{r'\neq r}\lvert \lambda_r^2-\lambda_{r'}^2\rvert$ governs stability by amplifying the error when it is small.
Consequently, our approach leverages cross-coordinate commonality to efficiently implement the SVD of high-dimensional kernels.

\begin{remark}[Data-Adaptive Bases vs.\ Pre-Specified Models]
In much of the {CCA} literature, association discovery for functional/longitudinal data typically begins with an explicit representation model for the observed trajectories. A common approach is a truncated basis expansion:
\begin{equation*}
\bm{X}(t) \;=\; \sum_{k=1}^{K} \bm{c}_{k}\,\phi_k(t),\quad 
\bm{Y}(t) \;=\; \sum_{k=1}^{K} \bm{d}_{k}\,\phi_k(t),
\qquad t\in\mathcal{T},
\end{equation*}
where $\{\phi_k\}_{k=1}^K$ are basis functions on $\mathcal{T}$, and $\bm{c}_{k}\in\mathbb{R}^p$ and $\bm{d}_{k}\in\mathbb{R}^q$ are coefficient vectors.
Existing methods typically rely on pre-specified bases (e.g., linear functions as used in random intercept and slope models \citep{lee2023longitudinal} and other bases \citep{gorecki2017correlation}), often for modeling convenience.
These methods can be restrictive and may even distort the main signals for association discovery when the true dynamics of the data are complex.
In contrast, our FACD framework allows the bases used for association discovery to be extracted from \(H_X\) and \(H_Y\), as shown in Theorem~\ref{The_opt_bas}.
This procedure adapts to the cross-covariance between datasets and {allows} the bases for \(\bm{X}\) and \(\bm{Y}\) to differ, offering greater flexibility for capturing associations in more general longitudinal settings.
\end{remark}

\section{FACD on Observed Data}\label{sec:imp}

In this section, we consider FACD between two longitudinally observed high-dimensional datasets. We assume that
\begin{eqnarray}
\bm{x}_{ig}&=&\bm\mu_X(T_{ig})+\bm{\varepsilon}^X_i(T_{ig})+\bm{\tau}^X_{ig},\ \quad i=1,\dots,n,\ \quad g=1,\dots,N^X_i,\label{model_X_data}\\
\bm{y}_{ih}&=&\bm\mu_Y({T_{ih}})+\bm{\varepsilon}^Y_i({T_{ih}})+\bm{\tau}^Y_{ih},\ \quad i=1,\dots,n,\ \quad h=1,\dots,N^Y_i,\label{model_Y_data}
\end{eqnarray}
where $\{\bm{\varepsilon}_i^X, \bm{\varepsilon}_i^Y\}$ are correlated components for subject $i$ but are independent across different $i$'s, for $i = 1, \dots, n$; $N_i^X$ and $N_i^Y$ are the numbers of observed time points for the $i$th subject's functions; $\{T_{ig};g=1,\dots,N_{i}^X\}$ and $\{T_{ih};h=1,\dots,N_{i}^Y\}$ are the observed time points for the two datasets, which may vary across different subjects; $\bm{\tau}^X_{ig}:=({\tau}^X_{i1g},\dots,{\tau}^X_{ipg})^\top\in \mathbb{R}^p$ and $\bm{\tau}^Y_{ih}:=({\tau}^Y_{i1h},\dots,{{\tau}^Y_{iqh}})^\top\in \mathbb{R}^q$ {are} vectors of mean-zero noise terms that are uncorrelated with $\{\bm{\varepsilon}_i^Y, \bm{\varepsilon}_i^X\}$. These noise terms are independent across different subjects and across the two datasets, but may be dependent across different features and times within the same dataset. 

The goal is to estimate $\{\wtilde{A}_{rj}^X;j=1,\dots, p\}$ and $\{\wtilde {A}_{rm}^Y;m=1,\dots, q\}$ and their associated scores based on the data $\bm{x}_{ig}$s and $\bm{y}_{ih}$s.
To this end, we adopt the cross-covariance framework to explore the association.
To avoid the influence of differing feature scales, we standardize each feature as a preprocessing step, in parallel with the normalization step for canonical cross-covariance analysis.

\subsection{Adaptive Basis Estimation}\label{sec: Opt_basis}
To perform FACD, we first estimate the functional bases 
\( \{\wtilde{\psi}_{k_1}^X : k_1 = 1, \dots, \kappa_X\} \) and 
\( \{\wtilde{\psi}_{k_2}^Y : k_2 = 1, \dots, \kappa_Y\} \). 
We begin by introducing spline function spaces. 
Let \( \mathcal{B}_{L,Z} \) denote the \( L \)-dimensional B-spline space of order \( Z \) with equally spaced knots over \( \mathcal{T} \), and let 
\( \mathcal{B}_{L,Z} \otimes \mathcal{B}_{L,Z} \) be its tensor product space, consisting of bivariate spline functions on \( \mathcal{T} \times \mathcal{T} \).

We first estimate the mean functions \( \bm{\mu}_X \) and \( \bm{\mu}_Y \) of \( \bm{X}_i \) and \( \bm{Y}_i \) from the observed data \( \bm{x}_{ig} \) and \( \bm{y}_{ih} \). 
Let \( \mu_j^X \) and \( x_{ijg} \) denote the \( j \)th component of \( \bm{\mu}_X \) and the \( j \)th element of \( \bm{x}_{ig} \), respectively, and define \( \mu_m^Y \) and \( y_{imh} \) analogously for \( \bm{\mu}_Y \) and \( \bm{y}_{ih} \). 
Following \citet{hsing2015theoretical} and \citet{xiao2020asymptotic}, we estimate \( \mu_j^X \) and \( \mu_m^Y \) via spline regression on the observations \( x_{ijg} \) and \( y_{imh} \) with respect to their observation times. 
For each \( j \), we compute
\[
\hat{\mu}_j^X := \argmin_{f \in \mathcal{B}_{L_{\mu},4}} 
\frac{1}{n} \sum_{i=1}^n \frac{1}{N_i^X} \sum_{g=1}^{N_i^X} \{x_{ijg} - f(T_{ig})\}^2 + \nu_{\mu,j}^X \mathcal{L}(f),
\]
where \({\mathcal{B}_{L_{\mu},4}} \) is the cubic spline space, \( \mathcal{L}(\cdot) \) is a smoothness penalty, and \( \nu_{\mu,j}^X \) is its tuning parameter. 
We define \( \hat{\mu}_m^Y \) similarly and set \( \hat{\bm{\mu}}_X = (\hat{\mu}_1^X, \ldots, \hat{\mu}_p^X)^\top \) and \( \hat{\bm{\mu}}_Y = (\hat{\mu}_1^Y, \ldots, \hat{\mu}_q^Y)^\top \).

Next, we estimate the basis functions
$\{\wtilde{\psi}_{k_1}^X{:\ }k_1 = 1, \dots, \kappa_X\}$ and
$\{\wtilde{\psi}_{k_2}^Y{:\ }k_2 = 1, \dots, \kappa_Y\}$
given the estimated mean functions.
Recall that the $(j,m)$th element of $\bm{R}_{XY}(t_1, t_2)$ is denoted by $R^{XY}_{jm}(t_1, t_2)$. 
Define
\begin{eqnarray}\label{G_kernel}
    G^{XY}_{i,jm}(T_{ig}, T_{ih})
:= \big\{x_{ijg} - \hat{\mu}^X_j(T_{ig})\big\}
   \cdot 
   \big\{y_{imh} - \hat{\mu}^Y_m(T_{ih})\big\},
\end{eqnarray}
which is an empirical realization of $R^{XY}_{jm}(T_{ig}, T_{ih})$ for the $i$th subject.
Define
$U_i^X(T_{ig_1},T_{ig_2}):=\frac{1}{pq} \sum_{j=1}^p\sum_{m=1}^q\frac{1}{N_i^Y}\sum_{h=1}^{N_i^Y}\allowbreak G_{i,jm}^{XY}(T_{ig_1},T_{ih})\cdot G_{i,jm}^{XY}(T_{ig_2},T_{ih}),$
which is an empirical realization of the aggregated kernel
$H_X(T_{ig_1},T_{ig_2})=\frac{1}{pq}\sum_{j=1}^p\allowbreak\sum_{m=1}^q \int_0^1 R_{jm}^{XY}(t_1,t)R_{jm}^{XY}(t_2,t)\,\mathrm{d}t$
for the $i$th subject. 
Using this quantity, we estimate $H_X(t_1,t_2)$ via the bivariate spline regression
\begin{eqnarray*}
 \hat{H}_X(\cdot,\cdot)&{:=}&\argmin_{f\in \mathcal{B}_{L_H,4} \otimes \mathcal{B}_{L_H,4}}  \frac{1}{n}\sum_{i=1}^n\frac{1}{(N_i^X-1)N_i^X}\sum_{1\leq g_1<g_2\leq N_i^X} \big\{U_i^X(T_{ig_1},T_{ig_2})-f(T_{ig_1},T_{ig_2})\big\}^2\\
 &+&\nu_H^X \mathcal{L}(f),
\end{eqnarray*}
where $\nu_H^X$ is a tuning parameter. 

For \( \hat{\mu}_j^X \) and \( \hat{\mu}_m^Y \), we adopt the roughness penalty
\(
\mathcal{L}(f) = \int_0^1 \{ f^{(2)}(t) \}^2 \, \mathrm{d}t
\).
For the kernels \( \hat{H}_X \) and \( \hat{H}_Y \), we use the bivariate roughness penalty
\(
\mathcal{L}(f) = \int_0^1 \int_0^1
\left[
\left( \frac{\partial^2 f}{\partial t_1^2} \right)^2 +
\left( \frac{\partial^2 f}{\partial t_2^2} \right)^2
\right] \mathrm{d}t_1 \, \mathrm{d}t_2,
\)
which separately penalizes roughness in the \(t_1\) and \(t_2\) directions \citep{wood2006low}.
Such spline smoothing approaches are widely used in functional data analysis \citep{hsing2015theoretical,xiao2020asymptotic,tan2024functional}.
Here, we select the tuning parameters for mean and kernel estimation using generalized cross-validation (GCV) \citep{gu2013smoothing}.

After estimating \( H_X(t_1, t_2) \), we perform spectral decomposition of \( \hat{H}_X(t_1, t_2) \) to obtain \( \hat{\zeta}_k^X \) and \(\hat{\psi}_k^X\). 
The number of components \( \kappa_X \) can be selected based on the cumulative proportion of eigenvalues, 
\( \sum_{k=1}^{\kappa_X} \hat{\zeta}_k^X / \int_0^1 \hat{H}_X(t,t)\,\mathrm{d}t \), 
for example, by requiring it to exceed a threshold such as 0.95. 
The same procedure applies to \( \hat{H}_Y \) to obtain \(\hat{\psi}_k^Y\) and \( \kappa_Y \) for the second dataset.

\subsection{Estimation of Functional Loadings and Scores}\label{sec:est_loading_score}

In this subsection, we demonstrate the estimation procedure for the canonical functional loadings $\{\bm{A}^X_{r}\}$ and $\{\bm{A}^Y_{r}\}$, as well as their scores, given the bases $\hat{\psi}_k^X$, $k\leq \kappa_X$, and $\hat{\psi}_k^Y$, $k\leq \kappa_Y$.
Recall that 
$\bm{\Lambda}^{XY}_{k_1,k_2}=\iint_0^1 \bm{R}_{XY}(t_1,t_2)\psi_{k_1}^X(t_1)\psi_{k_2}^Y(t_2)\ \mathrm{d}t_1\mathrm{d}t_2$ according to Theorem~\ref{theo: svd_mar}. We denote $\bm{G}_{i}^{XY}(T_{ig},T_{ih}):=({G}_{i,jm}^{XY}(T_{ig},T_{ih}))_{{j=1,\ldots,p,\ m=1,\ldots,q}}\in \mathbb{R}^{p\times q}$ and define 
\begin{eqnarray*}
    \hat{\bm{\Lambda}}^{XY}_{i,k_1,k_2}:=\frac{1}{N_i^XN_i^Y}\sum_{g=1}^{N_i^X}\sum_{h=1}^{N_i^Y}\bm{G}_{i}^{XY}(T_{ig},T_{ih}) \hat{\psi}_{k_1}^X(T_{ig})\hat{\psi}_{k_2}^Y(T_{ih}),
\end{eqnarray*}
which is an empirical realization of $\bm{\Lambda}^{XY}_{k_1,k_2}$ for the $i$th subject.
Subsequently, we estimate $\bm{\Lambda}^{XY}_{k_1,k_2}$ by 
$
\hat{\bm{\Lambda}}^{XY}_{k_1,k_2}= \sum_{i=1}^n\hat{\bm{\Lambda}}^{XY}_{i,k_1,k_2}/n.
$

In parallel with \eqref{SVD_Gamma}, define \( \hat{\bm{\Gamma}} \in \mathbb{R}^{p\kappa_X \times q\kappa_Y} \) as a block matrix whose \((k_1,k_2)\)th block is \( \hat{\bm{\Lambda}}^{XY}_{k_1,k_2} \) for \( k_1=1,\dots,\kappa_X \) and \( k_2=1,\dots,\kappa_Y \). 
We perform SVD on \( \hat{\bm{\Gamma}} \):
\(
\hat{\bm{\Gamma}} = \sum_{r=1}^{R} \hat{\eta}_{r}\, \hat{\bm{a}}_{r}\hat{\bm{b}}_{r}^{\top},
\)
where \( \hat{\eta}_r \) is the \( r \)th singular value, 
\( \hat{\bm{a}}_{r} = ((\hat{\bm{a}}_{r}^{(1)})^\top, \dots, (\hat{\bm{a}}_{r}^{(\kappa_X)})^\top)^\top \in \mathbb{R}^{p\kappa_X} \), and 
\( \hat{\bm{b}}_{r} = ((\hat{\bm{b}}_{r}^{(1)})^\top, \dots,\allowbreak (\hat{\bm{b}}_{r}^{(\kappa_Y)})^\top)^\top \allowbreak\in \mathbb{R}^{q\kappa_Y} \) are the corresponding \( r \)th singular vectors. 
Based on \eqref{SVD_appro}, the estimated loadings are
\begin{eqnarray}\label{est_functional_loading}
\bm{\hat{A}}^X_{r}=\sum_{k=1}^{\kappa_X}\hat{\bm{a}}^{(k)}_{r}\cdot\hat{\psi}_{k}^X\quad \text{and}\quad
\bm{\hat{A}}^Y_{r}=\sum_{k=1}^{\kappa_Y}\hat{\bm{b}}_{r}^{(k)}\cdot\hat{\psi}_{k}^Y.
\end{eqnarray}

In high-dimensional settings where \( p \) and \( q \) are large, we assume that only a small subset of features in \( \bm{X} \) and \( \bm{Y} \) contribute to their covariance. 
Specifically, we define
\begin{eqnarray}
\mathcal{S}_X &:=& \{ j \in \{1, \dots, p\} :
\|\wtilde{A}_{rj}^X\| \neq 0
\text{ for some } r \geq 1 \text{ with } \lambda_r > 0 \},
\label{sparse_condition}\\
\mathcal{S}_Y &:=& \{ m \in \{1, \dots, q\} :
\|\wtilde{A}_{rm}^Y\| \neq 0
\text{ for some } r \geq 1 \text{ with } \lambda_r > 0 \},
\label{sparse_condition_Y}
\end{eqnarray}
where \( \mathcal{S}_X \) and \( \mathcal{S}_Y \) are small subsets of \( \{1,\dots,p\} \) and \( \{1,\dots,q\} \), respectively. 
This sparsity assumption alleviates the curse of dimensionality and enhances interpretability by identifying key features linking the high-dimensional longitudinal datasets.  
Under this assumption, instead of directly applying SVD to \( \hat{\bm{\Gamma}} \), we solve
\begin{eqnarray}\label{opt_SVD}
\{\hat{\bm{a}}_1,\hat{\bm{b}}_1\}
= \argmin_{\bm{a}\in \mathbb{R}^{p\kappa_X},\,\bm{b}\in \mathbb{R}^{q\kappa_Y}}
\big\|\hat{\bm{\Gamma}} - \bm{a}\bm{b}^\top\big\|_F^2
+ \varrho_X \sum_{j=1}^p \|A_{1j}^X\|
+ \varrho_Y \sum_{m=1}^q \|A_{1m}^Y\|
\nonumber\\
\text{subject to } \bm{a}^\top\bm{a} = 1,\ \bm{b}^\top\bm{b} = 1,
\end{eqnarray}
where \( (A_{11}^X,\dots,A_{1p}^X)^\top = \bm{A}_1^X \) and \( (A_{11}^Y,\dots,A_{1q}^Y)^\top = \bm{A}_1^Y \) are constructed from \eqref{est_functional_loading}, using \( \bm{a} \) and \( \bm{b} \) in place of \( \hat{\bm{a}}_1 \) and \( \hat{\bm{b}}_1 \), and \( \varrho_X, \varrho_Y \) are tuning parameters that induce sparsity in the functional loadings through Lasso-type penalties \citep{hastie2009elements}. 
As \( \varrho_X \) and \( \varrho_Y \) increase, the norms of the functional loadings \( \hat{A}_j^X \) and \( \hat{A}_m^Y \) shrink toward zero, thereby identifying the important features.

{
Optimization~\eqref{opt_SVD} can be solved using a group-Lasso algorithm \citep{hastie2009elements}, as detailed in Part~\ref{algo_sec} of the Supplementary Materials. Specifically, we initialize \(\widehat{\bm a}_1\) and \(\widehat{\bm b}_1\) using the leading left and right singular vectors of \(\widehat{\bm\Gamma}\), respectively, and then apply an alternating minimization algorithm to obtain the final estimates. A numerical analysis of computational efficiency and initialization stability is provided in Figure~\ref{FACD_com} in Part~\ref{algo_sec} of the Supplementary Materials. The results show that the algorithm has a relatively short overall runtime and converges rapidly under the SVD initialization, with the median number of iterations below 20 across all considered settings. Moreover, we compare the solution obtained using the SVD initialization with the best converged solution among 20 perturbed SVD initializations. The third panel of Figure~\ref{FACD_com} shows consistently small rank-one differences between the resulting solutions, indicating that convergence is stable under perturbations of the SVD initialization. Detailed simulation settings and implementation procedures are provided in Part~\ref{algo_sec} of the Supplementary Materials.
}

After obtaining $\hat{\bm{a}}_{1}=((\hat{\bm{a}}_{1}^{(1)})^\top,\dots,(\hat{\bm{a}}_{1}^{(\kappa_X)})^\top)^\top$ and $\hat{\bm{b}}_{1}=((\hat{\bm{b}}_{1}^{(1)})^\top,\dots,(\hat{\bm{b}}_{1}^{(\kappa_Y)})^\top)^\top$ from \eqref{opt_SVD} and $\bm{\hat{A}}^X_{1}$ and $\bm{\hat{A}}^Y_{1}$ from \eqref{est_functional_loading},
we can iteratively proceed to $r>1$. Specifically, once we obtain $\{\hat{\bm{a}}_r, \hat{\bm{b}}_r\}$ for $r \leq R$, we estimate the $(R+1)$th singular vectors of $\hat{\bm{\Gamma}}$ by replacing $\hat{\bm{\Gamma}}$ in \eqref{opt_SVD} with $\hat{\bm{\Gamma}} - \sum_{r=1}^R \hat{\eta}_r \hat{\bm{a}}_r \hat{\bm{b}}_r^\top$, where $\hat{\eta}_r = \hat{\bm{a}}_r^\top \hat{\bm{\Gamma}} \hat{\bm{b}}_r$.
The $r$th loadings are then estimated using \eqref{est_functional_loading}.
In parallel with \eqref{XY_score}, the $r$th scores for the $i$th subject are estimated by
\begin{eqnarray*}
    \hat{z}_{ir}^X:=\sum_{j=1}^p\frac{1}{N_i^X}\sum_{g=1}^{N_i^X}\hat{A}^X_{rj}(T_{ig})\{x_{ijg}-\hat{\mu}_j^X(T_{ig})\},\quad
    \hat{z}_{ir}^Y:=\sum_{m=1}^q\frac{1}{N_i^Y}\sum_{h=1}^{N_i^Y}\hat{A}^Y_{rm}({T_{ih}})\{y_{imh}-\hat{\mu}_m^Y(T_{ih})\}.
\end{eqnarray*}
The entire procedure for estimating the FACD components is summarized in Algorithm~\ref{algo:total_algorithm}. 
{Here, we select $R$ by maximizing $\frac{\hat{\eta}_{r}}{\hat{\eta}_{r+1}}$ over $r \leq R_{\text{max}}$, where $R_{\text{max}}$ is a pre-specified upper bound for $R$. }

\begin{algorithm}[ht!]
\caption{FACD of high-dimensional longitudinal data}
\label{algo:total_algorithm}
\footnotesize
\raggedright

\noindent \textbf{Input:} Observed data $\{\bm{x}_{ig}; i=1,\dots,n,\ g=1,\dots,N_i^X\}$ and 
$\{\bm{y}_{ih}; i=1,\dots,n,\ h=1,\dots,N_i^Y\}$; rank $R$; tuning parameters 
$\nu_H^X$, $\nu_H^Y$, and 
$\varrho_r^X$, $\varrho_r^Y$. \par

\noindent \textbf{Step 1: Estimation of aggregation kernels.} Compute $G_{i,jm}^{XY}(T_{ig_1},T_{ih})$ based on \eqref{G_kernel}. Solve 
\begin{align}
 \hat{H}_X &= \argmin_{f\in \mathcal{B}_{L_H,4} \otimes \mathcal{B}_{L_H,4}} 
 \frac{1}{n}\sum_{i=1}^n\frac{1}{(N_i^X-1)N_i^X}
 \sum_{1\le g_1<g_2\le N_i^X}
 \{U_i^X(T_{ig_1},T_{ig_2})-f(T_{ig_1},T_{ig_2})\}^2
 + \nu_H^X \mathcal{L}(f), \label{min_opt_basis_X}\\
 \hat{H}_Y &= \argmin_{f\in \mathcal{B}_{L_H,4} \otimes \mathcal{B}_{L_H,4}} 
 \frac{1}{n}\sum_{i=1}^n\frac{1}{(N_i^Y-1)N_i^Y}
 \sum_{1\le h_1<h_2\le N_i^Y}
 \{U_i^Y(T_{ih_1},T_{ih_2})-f(T_{ih_1},T_{ih_2})\}^2
 + \nu_H^Y \mathcal{L}(f), \label{min_opt_basis_Y}
\end{align}
where
\(
U_i^X(T_{ig_1},T_{ig_2}) :=\frac{1}{pq} \sum_{j=1}^p\sum_{m=1}^q\frac{1}{N_i^Y}\sum_{h=1}^{N_i^Y}G_{i,jm}^{XY}(T_{ig_1},T_{ih})\cdot G_{i,jm}^{XY}(T_{ig_2},T_{ih}),
\)
and \( {\mathcal{L}(f) = \int_0^1 \int_0^1 \left[ \left( \frac{\partial^2 f}{\partial t_1^2} \right)^2 + \left( \frac{\partial^2 f}{\partial t_2^2} \right)^2 \right] \, \mathrm{d}t_1 \, \mathrm{d}t_2.} \)

\medskip
\noindent \textbf{{Step 2}: Estimation of adaptive bases.}
Perform spectral decomposition of $\hat{H}_X(t_1,t_2)$ and $\hat{H}_Y(t_1,t_2)$ 
to obtain eigenvalues $\hat{\zeta}_k^X$, $\hat{\zeta}_k^Y$ and eigenfunctions 
$\hat{\psi}_k^X$, $\hat{\psi}_k^Y$.  
Select $\kappa_X$ and $\kappa_Y$ such that 
$\sum_{k=1}^{\kappa_X}\hat{\zeta}_k^X / \int_0^1 \hat{H}_X(t,t)\,\mathrm{d}t > 0.95$ 
and similarly for $\kappa_Y$.

\medskip
\noindent \textbf{Step 3: Estimation of projection matrix.}
Define $\hat{\bm{\Gamma}}$ as a block matrix whose $(k_1,k_2)$th block is 
$\hat{\bm{\Lambda}}^{XY}_{k_1,k_2}$, where
\(
\hat{\bm{\Lambda}}^{XY}_{k_1,k_2}
= \frac{1}{n}\sum_{i=1}^n
\left\{
\frac{1}{N_i^XN_i^Y}\sum_{g=1}^{N_i^X}\sum_{h=1}^{N_i^Y}
\bm{G}_{i}^{XY}(T_{ig},T_{ih})
\hat{\psi}_{k_1}^X(T_{ig})\hat{\psi}_{k_2}^Y(T_{ih})
\right\},
\)
with $\bm{G}_{i}^{XY}(T_{ig},T_{ih})
= (G_{i,jm}^{XY}(T_{ig},T_{ih}))_{1\le j\le p,\,1\le m\le q}$.

\medskip
\noindent \textbf{{Step 4}: Iterative estimation of canonical components.} For $r = 1,\ldots,R$:
\begin{align*}
\{\hat{\bm{a}}_r,\hat{\bm{b}}_r\} =
\argmin_{\bm{a}\in \mathbb{R}^{p\kappa_X},\,\bm{b}\in \mathbb{R}^{q\kappa_Y}}
\|\hat{\bm{\Gamma}}-\bm{a}\bm{b}^\top\|_F^2
+\varrho^X_r\sum_{j=1}^p\|A^X_{rj}\|
+\varrho^Y_r\sum_{m=1}^q\|A^Y_{rm}\|,\ 
\text{subject to } \bm{a}^\top\bm{a}= 1, \quad \bm{b}^\top\bm{b}= 1,
\end{align*}
where $(A^X_{r1},\dots,A^X_{rp})^\top=\sum_{k=1}^{\kappa_X}{\bm{a}}^{(k)}\cdot\hat{\psi}_{k}^X$ and $(A^Y_{r1},\dots,A^Y_{rq})^\top=\sum_{k=1}^{\kappa_Y}{\bm{b}}^{(k)}\cdot\hat{\psi}_{k}^Y$, with ${\bm{a}}=(({\bm{a}}^{(1)})^\top,\dots,({\bm{a}}^{(\kappa_X)})^\top)^\top\in \mathbb{R}^{p\cdot\kappa_X}$ and ${\bm{b}}=(({\bm{b}}^{(1)})^\top,\dots,({\bm{b}}^{(\kappa_Y)})^\top)^\top\in \mathbb{R}^{q\cdot\kappa_Y}$.

\medskip
\qquad
Update $\hat{\bm{\Gamma}} \leftarrow 
\hat{\bm{\Gamma}} - \hat{\eta}_r \hat{\bm{a}}_r \hat{\bm{b}}_r^\top$, 
where $\hat{\eta}_r = \hat{\bm{a}}_r^\top \hat{\bm{\Gamma}} \hat{\bm{b}}_r$.

\medskip
\qquad Compute the canonical functional loadings 
$\bm{\hat{A}}_r^X=\sum_{k=1}^{\kappa_X}\hat{\bm{a}}_r^{(k)}\hat{\psi}_k^X$
and $\bm{\hat{A}}_r^Y=\sum_{k=1}^{\kappa_Y}\hat{\bm{b}}_r^{(k)}\hat{\psi}_k^Y$.  
The canonical scores are
\(
\hat{z}_{ir}^X = \sum_{j=1}^p\frac{1}{N_i^X}\sum_{g=1}^{N_i^X}
\hat{A}_{rj}^X(T_{ig})\{x_{ijg}-\hat{\mu}_j^X(T_{ig})\}\) and
\(
\hat{z}_{ir}^Y = \sum_{m=1}^q\frac{1}{N_i^Y}\sum_{h=1}^{N_i^Y}
\hat{A}_{rm}^Y(T_{ih})\{y_{imh}-\hat{\mu}_m^Y(T_{ih})\}.
\)

\medskip
\qquad\textbf{Output:} 
$\bm{\hat{A}}_r^X$, $\bm{\hat{A}}_r^Y$, $\hat{z}_{ir}^X$, and $\hat{z}_{ir}^Y$, 
for $i=1,\ldots,n$.

\medskip
\quad End for.
\end{algorithm}

The tuning parameters \( \varrho_X \) and \( \varrho_Y \) in \eqref{opt_SVD} are selected separately for each component. 
Using the first component as an example, we split the \( n \) samples into 5 groups indexed by \( \mathcal{U}_g \), \( g = 1, \dots, 5 \), with \( \cup_{g=1}^5 \mathcal{U}_g = \{1, \dots, n\} \). 
Let \( \hat{\bm{a}}_1^{(-g)} \) and \( \hat{\bm{b}}_1^{(-g)} \) be the solutions to \eqref{opt_SVD} obtained by replacing \( \hat{\bm{\Gamma}} \) with 
\(
\frac{1}{n - |\mathcal{U}_g|} \sum_{i \notin \mathcal{U}_g} \hat{\bm{\Gamma}}_i,
\)
where \( \hat{\bm{\Gamma}}_i \) is a block matrix whose \((k_1, k_2)\)th block is \( \hat{\bm{\Lambda}}^{XY}_{i, k_1, k_2} \), and \( |\cdot| \) denotes set cardinality.  
The optimal \( \varrho_X \) and \( \varrho_Y \) are chosen to maximize
\(
\sum_{g=1}^5 
\big(\hat{\bm{a}}_1^{(-g)}\big)^\top
\left( \frac{1}{|\mathcal{U}_g|} \sum_{i \in \mathcal{U}_g} \hat{\bm{\Gamma}}_i \right)
\hat{\bm{b}}_1^{(-g)}.
\)

\section{Simulation}\label{sec:sim}
\noindent\textbf{Data Generation.}
For data generation, we consider the multivariate Karhunen--Lo\`eve expansion \citep{happ2018multivariate} for modeling $\bm{X}$ and $\bm{Y}$.
As such, the observed data are generated as
\begin{eqnarray*}
x_{ijg}&=&\sum_{r=1}^{20} A^X_{rj}(T_{ig}) z_{ir}^X+{\tau}^X_{ijg},\ i=1,\dots,n,\ j=1,\dots,p,\ g=1,\dots,N^X_i,\\
{y}_{imh}&=&\sum_{r=1}^{20} A^Y_{rm}(T_{ih}) z_{ir}^Y+{\tau}^Y_{imh},\ i=1,\dots,n,\ m=1,\dots,q,\ h=1,\dots,N^Y_i,
\end{eqnarray*}
where the eigenfunctions \( A^X_{rj} \) and \( A^Y_{rm} \) are constructed as combinations of 10 basis functions, with the nonzero loadings concentrated in 10 variables of both datasets; $(z_{ir}^X, z_{ir}^Y)^\top$ denote the random scores, which are independent across subjects $i$ and components $r$,  
and whose cross-covariance and correlation decay as $r$ increases; $\tau^X_{ijg}$s and $\tau^Y_{imh}$s are correlated random noise terms. 
The noise terms $\tau^X_{ijg}$ and $\tau^Y_{imh}$ are generated such that they are independent across different $i$ and across the two datasets, but are dependent across different features and time points within the same dataset.
The simulation of these components is detailed in Part~\ref{sec:simu_setup} of the Supplementary Materials. 
$\{T_{ig};g=1,\dots,N_{i}^X\}$ and $\{T_{ih};{h}=1,\dots,N_{i}^Y\}$, $i=1,\dots,n$, are random time points sampled from $\operatorname{Unif}([0,1])$, with $N_i^X$s and $N_i^Y$s independently sampled from $\{5,6,7,8\}$.

Under the above setting, $(z_{ir}^X,z_{ir}^Y)^\top$, $i=1,\dots,n$, are the $r$th FACD and CCA scores from the two datasets, with the corresponding functional loadings given by $A^X_{rj}$s and $A_{rm}^Y$s.
Given $x_{ijg}$s and $y_{imh}$s, our goal is to estimate $(z_{ir}^X, z_{ir}^Y)^\top$, $A^X_{rj}$s, and $A_{rm}^Y$s.

\noindent\textbf{Competing Methods.} We compare the proposed FACD with several existing CCA and dimension reduction methods: longitudinal canonical correlation analysis \citep[LCCA;][]{lee2023longitudinal}, sparse canonical correlation analysis \citep[SCCA;][]{witten2009extensions}, multivariate functional principal component analysis \citep[MFPCA;][]{happ2018multivariate}, and a sparse variant of functional CCA \citep{gorecki2017correlation}, referred to here as functional sparse canonical correlation analysis (FSCCA). 
Because some of these methods are not directly applicable to our simulated data, we make several necessary adaptations.

\begin{figure}[htbp]
\centering
\includegraphics[width=0.95\textwidth]{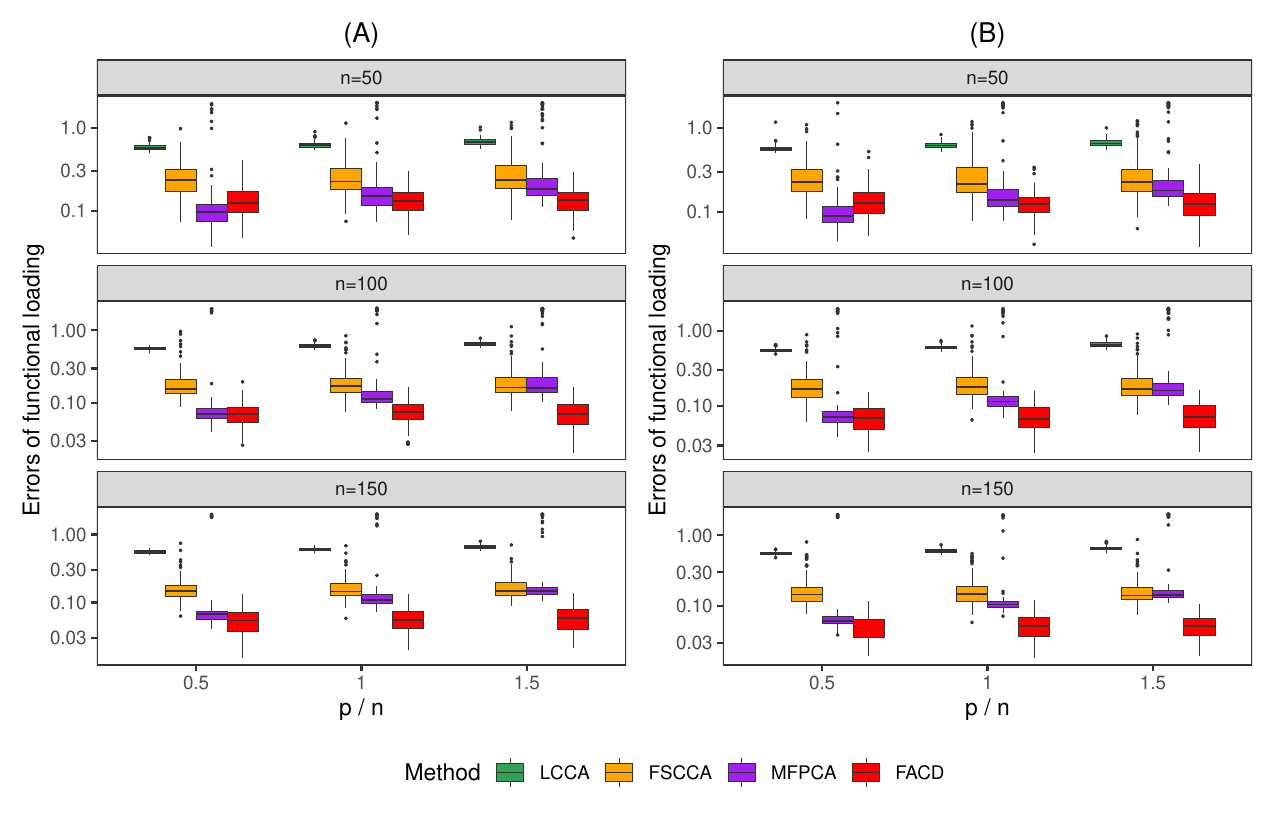}
\caption[Estimation errors for functional loadings]{
Boxplots of estimation errors for functional loadings from the first
(Panel A) and second (Panel B) datasets.
}
\label{err_functional_loading}
\end{figure}

FSCCA is a modified version of functional CCA designed to accommodate the high-dimensional setting in our simulation. Specifically, we project the longitudinal data \(x_{ijg}\) and \(y_{imh}\) onto orthonormal cubic polynomial basis functions to obtain the projected data. 
Unlike \citet{gorecki2017correlation}, who applied CCA to the projected data to estimate scores and functional loadings, we employ GSCCA \citep{lin2013group}, which groups the projected variables of the same features during sparse estimation, thereby facilitating joint feature selection across datasets. 
Moreover, SCCA handles high-dimensional features but is limited to cross-sectional data; thus, we apply it to the subject-wise means of the longitudinal observations. 
LCCA and MFPCA are implemented as originally proposed; however, LCCA uses linear bases to capture temporal patterns and lacks sparsity, whereas MFPCA is an unsupervised approach that independently extracts scores and loadings for each dataset.

\noindent\textbf{Comparison Measure.} We evaluate the performance of different methods using three criteria: 
(i) Estimation errors of functional loadings:
$\sum_{j=1}^p\|A^X_{rj}-\hat{A}^X_{rj}\|^2$ and $\sum_{m=1}^q\|A^Y_{rm}-\hat{A}^Y_{rm}\|^2$, where $\hat{A}^X_{rj}$ and $\hat{A}^Y_{rm}$ are the estimated functional loadings from different methods; 
(ii) Variable selection errors:
false positive rates $\frac{1}{p-10}\sum_{j=11}^{p}\mathbb{I}(\|\hat{A}^X_{rj}\|\neq 0)\times 100\%$ and $\frac{1}{q-10}{\sum_{m=11}^{q}}\mathbb{I}(\|\hat{A}^Y_{rm}\|\neq 0)\times 100\%$, and false negative rates $\frac{1}{10}\sum_{j=1}^{10}\mathbb{I}(\|\hat{A}_{rj}^X\|= 0)\times 100\%$ and $\frac{1}{10}{\sum_{m=1}^{10}}\mathbb{I}(\|\hat{A}_{rm}^Y\|= 0)\times 100\%$; 
(iii) Estimation accuracy of functional scores:
$\widehat{\operatorname{Cor}}(z_{ir}^X, \hat{z}_{ir}^X)$ and $\widehat{\operatorname{Cor}}(z_{ir}^Y, \hat{z}_{ir}^Y)$, where $\hat{z}_{ir}^X$ and $\hat{z}_{ir}^Y$ are the estimated scores, and $\widehat{\operatorname{Cor}}(\cdot, \cdot)$ denotes the empirical correlation.

\noindent\textbf{Simulation Results.}
We examine settings with different combinations of \( n \), \( p \), and \( q \), setting \( p = q \) for simplicity. 
We conduct 100 simulations for each scenario and assess the estimation performance of the first component. 
Figure~\ref{err_functional_loading} reports the estimation errors of functional loadings across methods. We observe that the performance of FACD is not sensitive to dimensionality when the true number of relevant variables is fixed. This phenomenon is consistent with Theorem~\ref{Theorem: consistency_diverge_dimension} in Section~\ref{Sec: sta_the} below.

Overall, our proposed FACD consistently outperforms LCCA and FSCCA and achieves performance comparable to or better than MFPCA. 
The advantage of FACD over MFPCA becomes more evident as the \( p/n \) ratio increases, underscoring the benefit of exploiting cross-covariance information to improve estimation in high-dimensional settings. 
In contrast, LCCA shows markedly higher errors, as it captures only linear longitudinal associations. 
Moreover, both LCCA and FSCCA rely on pre-specified basis functions, which may be inefficient for capturing the true underlying structure, leading to their inferior performance relative to FACD.

In Part~\ref{sec:linear} of the Supplementary Materials, we further examine the estimation errors in the case of linear functional loadings, i.e., when $A^X_{rj}$ and $A^Y_{rm}$ are linear functions as assumed in LCCA. Under this setting, we find that FACD continues to outperform LCCA and FSCCA. This advantage stems from FACD’s ability to adaptively identify efficient longitudinal patterns while incorporating sparsity in the estimation of high-dimensional structures, capabilities that are not accommodated by LCCA or FSCCA.

\begin{figure}[htbp]
\centering
\includegraphics[width=\textwidth]{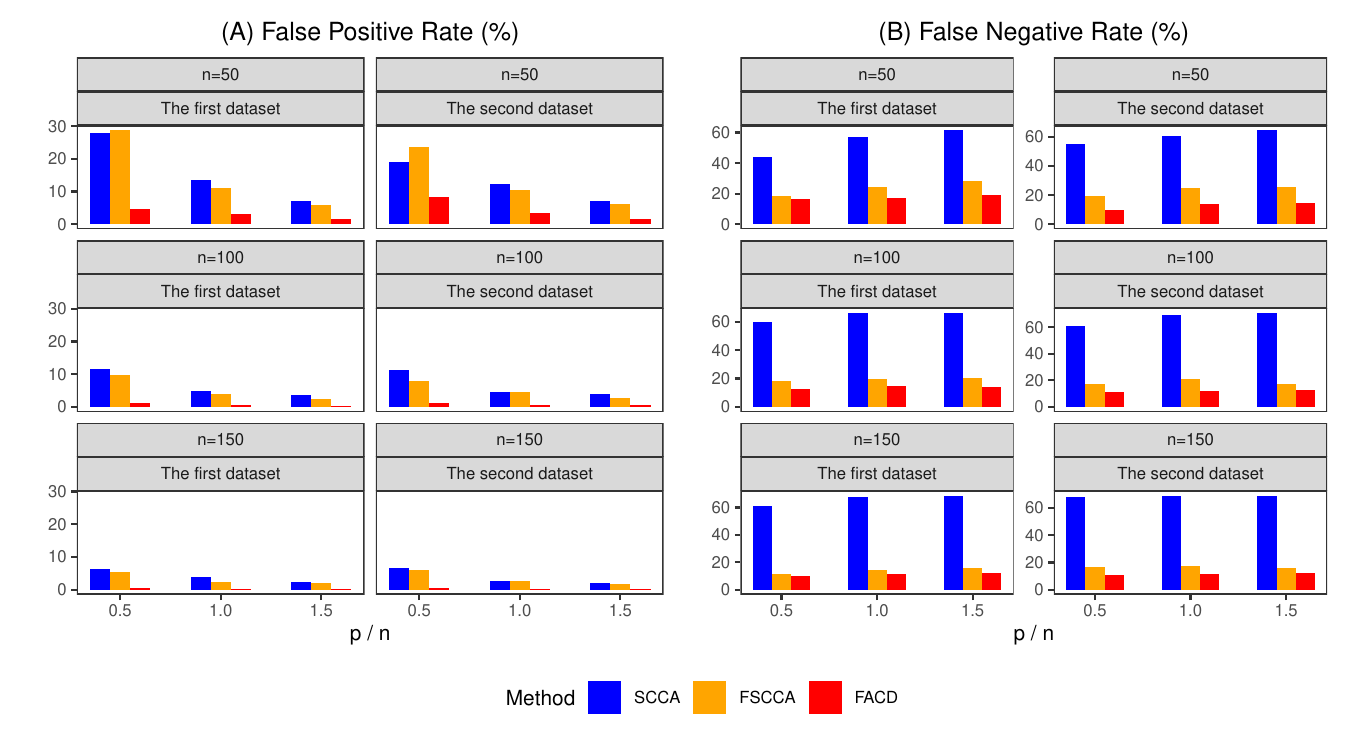}
\caption[False positive and false negative rates of variable selection]{
The false positive rates and false negative rates of variable selection.
}
\label{err_fal_var}
\end{figure}

Figure~\ref{err_fal_var} shows the false positive and false negative rates for variable selection across the three methods capable of this task.
FACD consistently attains markedly lower rates for both metrics.
SCCA performs the worst, as it averages over all time points before applying CCA, leading to a substantial loss of temporal information.
FSCCA improves upon SCCA by incorporating longitudinal signals but relies on pre-specified basis functions; when these bases are misaligned with the true temporal patterns, they capture inefficient signals and reduce selection accuracy.
In particular, FSCCA significantly increases the false positive rate compared with FACD, suggesting that inefficient basis representations in FSCCA may introduce redundant cross-view associations that inflate weak effects into significant ones.
In contrast, FACD employs data-adaptive bases that capture the true underlying longitudinal structure, resulting in more accurate variable selection.

Similar results can be found in Table~\ref{cor_table} in Part~\ref{sec: corre} of the Supplementary Materials, which compares the Pearson correlations between the estimated and true scores across methods. 
These results confirm that FACD most effectively recovers the top correlated structures between longitudinal datasets. 
{Furthermore, in Part~\ref{sec: component_covariance} of the Supplementary Materials, we also investigate the recovery of the second component and the overall cross-covariance structure using FACD. The results show that the recovery errors decrease as the sample size \(n\) increases, indicating the statistical consistency of FACD in recovering the cross-covariance structure over the leading components.}

\section{Data Analysis}\label{sec: data}

To illustrate the utility of FACD, we analyze data from a longitudinal multi-omic study profiling blood molecules \citep{contrepois2020molecular}. 
Blood samples from 36 participants were collected at baseline and at time points during the 0--60 minute period after exercise, following an overnight fast in a controlled session.
Plasma samples were analyzed by SWATH-MS for untargeted proteomics, immunoassays for targeted proteomics, LC-MS for untargeted metabolomics, and Lipidyzer for semi-targeted lipidomics, capturing key dynamic molecular responses to acute physical activity. 
The observation times follow the models \eqref{model_X_data} and \eqref{model_Y_data} and are not aligned across subjects or across omic layers.
Preprocessed data were obtained from \url{http://hmp2-data.stanford.edu/} and include 728, 710, 109, and 260 features for metabolites, lipids, targeted proteins, and untargeted proteins, respectively. 
We remove 54 metabolites due to their low prevalence across samples.
Untargeted protein markers are combined with targeted protein markers. 
All features were log-transformed and standardized across subjects and time points before being analyzed by FACD for each pair of omic datasets.

Given the high collinearity among lipid features, we group lipids into 17 classes according to structural and functional similarity (Table~\ref{tab:lipid_num_ori} in the Supplementary Materials). 
Within each group, principal component analysis was applied, and the top components explaining over 95\% of the variance were retained as representative features. 
For instance, the first principal component of the cholesteryl ester (CE) group is denoted as ``CE.pc.1,'' with subsequent components named similarly. 
This procedure yielded 122 aggregated lipid features for FACD analysis (Table~\ref{tab:lipid_num_pca} in the Supplementary Materials). 
Further preprocessing details are provided in Part~\ref{sec: lipid_proc} of the Supplementary Materials.

\noindent\textbf{Comparison of Existing Methods.} Figure~\ref{extracted_score}(A) in Part~\ref{sec:com_data_com} of the Supplementary Material displays the functional scores obtained from rank-one FACD applied to each pairwise combination of the three omics types.  
Notably, we observe that lipids and metabolites exhibit the strongest overall correlation (0.866), consistent with the integrative analysis of \citet{contrepois2020molecular}, which found stronger cross-omic links between lipids and metabolites than between proteins and other omic layers.  

We further compare FACD with LCCA, FSCCA, and MFPCA in extracting correlations between the proteomic and metabolomic data. 
The data are divided into six folds, using one fold for testing and the remaining five for training, with tuning parameters selected via five-fold cross-validation. 
Given the estimated loadings \(\hat{A}^X_{1j}\) and \(\hat{A}^Y_{1m}\) from the training data, we then compute the Pearson correlation between the scores extracted from the testing set using these loadings.  
As shown in Figure~\ref{extracted_score}(B) in Part~\ref{sec:com_data_com} of the Supplementary Material, FACD consistently achieves the highest correlations, demonstrating a superior ability to capture key associations between the longitudinal datasets.

\begin{figure}[htbp]
\centering
\includegraphics[width=\textwidth, trim=1cm 0cm -3.5cm 0cm]
{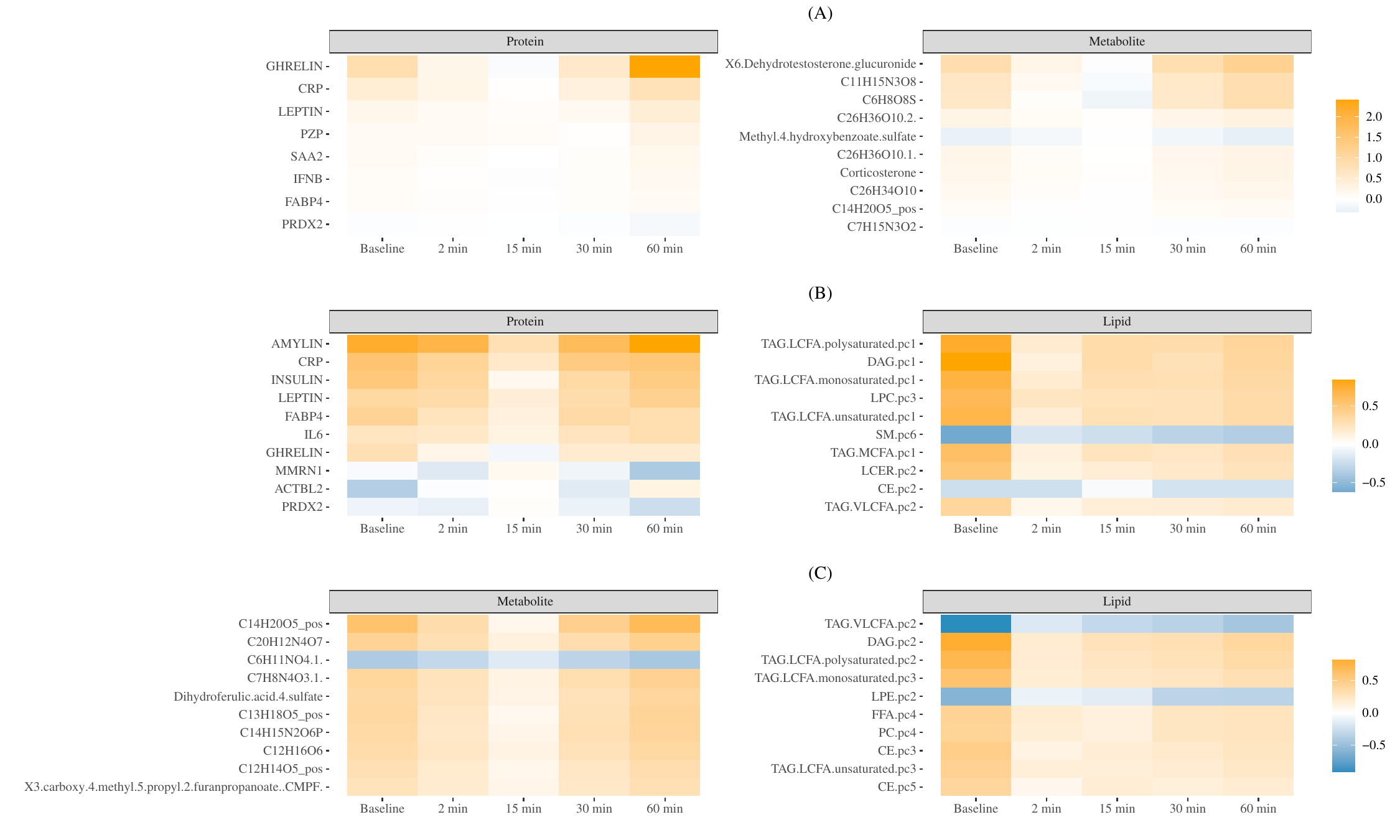}
\caption[Estimated functional loadings across omics datasets]{
The estimated functional loadings between any two omics in the
high-dimensional longitudinal data at baseline, early stage (2 min),
quarter stage (15 min), middle time (30 min), and end point (60 min).
The signs of the functional loadings are determined such that their
inner products with the constant function are positive.
The signs of lipid features depend on the signs used in the
principal component (PC) analysis, where the PC loadings are
arranged to have a positive inner product with the constant vector.
We only present the top 10 important features ranked by the
$L^2$ norm of their functional loadings.
}
\label{functional_loading}
\end{figure}

\noindent\textbf{Result.} 
Figure~\ref{functional_loading} presents the estimated functional loadings of the top features selected by the first FACD component, capturing dominant longitudinal associations across omics. 
The sparsity penalties in optimization \eqref{opt_SVD} result in the selection of 8 and 11 features between proteins and metabolites, 30 and 25 between proteins and lipids, and 24 and 29 between metabolites and lipids, respectively. 
These selected features are summarized in the CSV files provided in the Supplementary Materials.

The functional loadings reveal distinct temporal association patterns between omic layers. 
For instance, ghrelin, the ``hunger hormone,'' emerged as the leading protein marker associated with metabolites, with most associations occurring post-exercise. 
In contrast, lipid associations with other omics are primarily driven by baseline lipid levels, whereas protein and metabolite associations show strong activity at both pre- and post-exercise time points. 
These distinct temporal patterns highlight asynchronous biological roles across omic layers and demonstrate the value of longitudinal multi-omic integration for uncovering dynamic molecular interactions.

\begin{figure}[htbp]
\centering
\includegraphics[width=0.85\textwidth]{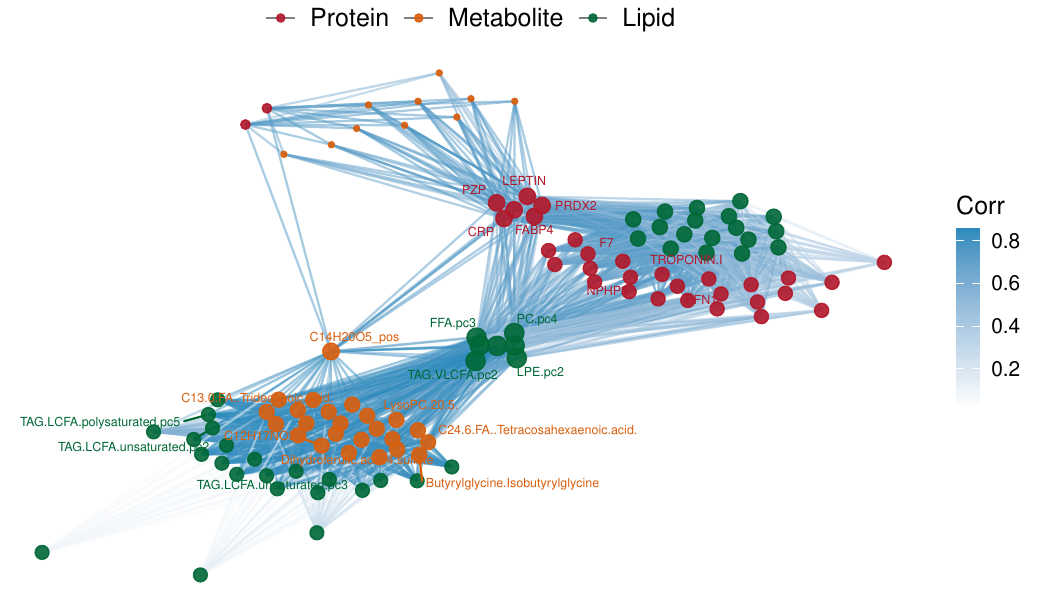}
\caption[Time-integrated correlations across omics datasets]{
The time-integrated correlations between features from any two omics
of high-dimensional longitudinal data, where larger node sizes
indicate higher connectivity and deeper edge colors represent
stronger correlations.
We highlight the top 10 features with the largest number of
connected edges for each omic.
This network is plotted using the R package \texttt{ggraph}
based on the correlation calculated by $\rho_{jm}$.
}
\label{association}
\end{figure}

Figure~\ref{association} illustrates a network of predominant time-aggregated associations across omics, derived from rank-one FACD reconstructions. 
Specifically, for any two datasets, the reconstructions are defined as 
\(\hat{X}_{ij}(\cdot) = \hat{A}_{1j}^X(\cdot)\,\hat{z}_{i1}^X\) and 
\(\hat{Y}_{im}(\cdot) = \hat{A}_{1m}^Y(\cdot)\,\hat{z}_{i1}^Y\), 
and the correlation between the \(j\)th and \(m\)th features is computed as  
\[
\rho_{jm} := 
\frac{\frac{1}{n}\sum_{i=1}^n \langle \hat{X}_{ij}, \hat{Y}_{im} \rangle}
{\sqrt{\frac{1}{n}\sum_{i=1}^n \|\hat{X}_{ij}\|^2} \cdot 
 \sqrt{\frac{1}{n}\sum_{i=1}^n \|\hat{Y}_{im}\|^2}},
\]
quantifying their overall temporal correlation based on the first FACD component.  
We find that the top protein markers identified exhibit clear and well-established biological relevance in the context of acute exercise.
For instance, leptin and ghrelin are key hormonal regulators of energy balance and appetite; troponin I is a cardiac-specific contractile protein reflecting transient cardiac strain; FABP4 (fatty acid-binding protein 4) mediates fatty acid transport and energy mobilization; and CRP (C-reactive protein) is a marker of systemic inflammation elevated under physiological stress.  
Together, these proteins reveal coordinated metabolic, cardiovascular, and inflammatory adaptations to acute exercise, underscoring FACD’s ability to uncover biologically meaningful cross-omic relationships.

{
\noindent\textbf{Sensitivity Analysis.} We conduct a sensitivity analysis to assess the stability of the estimated functional loadings. To this end, we repeatedly subsample observed time points from the original datasets and refit FACD, and then compare the resulting functional loadings with those estimated from the full dataset. This analysis helps assess the sensitivity of the estimated functional loadings to perturbations in the observed data. 

We consider two measures: the selection frequency, defined as the frequency that a feature selected from the full dataset is reselected based on a perturbed dataset, and the feature loading alignment,
\(
\frac{
\left|\left\langle \widehat A_j^{\mathrm{full}},\widehat A_j^{\mathrm{pert}}\right\rangle\right|
}{
\left\|\widehat A_j^{\mathrm{full}}\right\|\cdot
\left\|\widehat A_j^{\mathrm{pert}}\right\|
},
\)
where $\widehat A_j^{\mathrm{full}}$ and $\widehat A_j^{\mathrm{pert}}$ denote the estimated functional loadings of feature $j$ based on the full dataset and the perturbed dataset, respectively.
We report these measures for the top 10 selected features in the protein and metabolite modalities in Figure~\ref{fig:sensitivity} in Part~\ref{sec: sensitivity} of the Supplementary Materials. The results show that the selection frequencies are mostly above $0.9$, while the feature loading alignments are close to one. This indicates that both feature selection and functional loading estimation are stable under perturbations of the observed data.

}

\section{Statistical Consistency}\label{Sec: sta_the}

In this section, we establish the statistical convergence of FACD loadings in high-dimensional settings where \( p, q \gg n \). 
We assume that the numbers of observed time points, \( \{N_i^X : i = 1, \dots, n\} \) and \( \{N_i^Y : i = 1, \dots, n\} \), are fixed positive integers. 
Let \( N_X \) and \( N_Y \) denote their respective sample averages, i.e.,
\(
N_X := n^{-1}\sum_{i=1}^n N_i^X
\)
and
\(
N_Y := n^{-1}\sum_{i=1}^n N_i^Y.
\)
We further assume that \(\tau_{ijg}^X\)s, \(\tau_{imh}^Y\)s, \(T_{ig}\)s, \(T_{ih}\)s, and \( \{(\bm{\varepsilon}_i^X, \bm{\varepsilon}_i^Y)\} \) are mutually independent.  
Throughout, \( C \) and \( c \) denote generic positive constants independent of \( n, p, q, N_i^X, N_i^Y \), which may differ across contexts. 

We further introduce the following assumptions.

\begin{assumption}\label{A1} 
$\bm{\varepsilon}_i^X:=(\varepsilon^X_{i1},\dots,\varepsilon^X_{ip})^\top$ and 
$\bm{\varepsilon}_i^Y:=(\varepsilon^Y_{i1},\dots,\varepsilon^Y_{iq})^\top$ 
are independent and identically distributed across subjects \(i\). 
In addition, 
\(
\max_{1\leq j \leq p}\mathbb{E}\|\varepsilon^X_{ij}\|^8\leq C.
\)
Analogous conditions hold for $\varepsilon^Y_{im}$s.
Furthermore, we assume that the functional signals are supported on
$\mathcal{S}_X$ and $\mathcal{S}_Y$, respectively. Specifically,
\(
\varepsilon^X_{ij}(t)=0,\ j\notin\mathcal{S}_X,
\
\varepsilon^Y_{im}(t)=0,\ m\notin\mathcal{S}_Y,
\)
almost surely for almost every $t\in\mathcal{T}$.
\end{assumption}

\begin{assumption}\label{A3}
The noises ${\tau}^X_{ijg}$ and ${\tau}^Y_{imh}$
are mean-zero random variables that are independent
across datasets and subjects.
In addition,
\(
  \max_{1 \leq i \leq n,\; 1 \leq j \leq p,\;
  1 \leq g \leq N_i^X}
  \mathbb{E}\bigl({\tau}^X_{ijg}\bigr)^8
  \leq C
\)
and
\(
  \sup_{1\leq i\leq n,\;
  1\leq g_1\neq g_2\leq N_i^X}
  \sum_{j=1}^p
  \left|
  \operatorname{Cov}
  \bigl({\tau}^X_{ijg_1},
        {\tau}^X_{ijg_2}\bigr)
  \right|
  \leq C\sqrt{p}.
\)
Analogous conditions hold for ${\tau}^Y_{imh}$s.
\end{assumption}

\begin{assumption}\label{asum_time}
The time points $\{T_{ig} : i = 1, \dots, n,\ g = 1, \dots, N_i^X\}$ satisfy that 
$\{T_{ig} : g = 1, \dots, N_i^X\}$ are independently drawn from a distribution function $F$, where $F$ has a positive, bounded, and continuously differentiable density on $[0,1]$.
Analogous conditions hold for $\{T_{ih} : i = 1, \dots, n,\ h = 1, \dots, N_i^Y\}$. 
\end{assumption}

Assumptions~\ref{A1}--\ref{A3} control the uncertainty in the functional data and the noise.
In particular, Assumption~\ref{A1} imposes sparsity on the functional signals, requiring features outside the active sets to contain only measurement noise after removing their mean functions.
Assumption~\ref{A3} requires the sums of absolute temporal covariances of the measurement noise across features to grow no faster than $\sqrt{p}$ and $\sqrt{q}$, respectively, controlling the contribution of temporally dependent noise to the estimation of the cross-covariance structure.
Assumption~\ref{asum_time} is a common assumption on the observed time points in functional data analysis \citep{hsing2015theoretical,xiao2020asymptotic}.

\begin{assumption}\label{mean_assup}
For $l\geq 2$, the mean functions $\mu_{j}^X$ and $\mu_{m}^Y$ have continuous derivatives of order $l$, satisfying 
\(
\sup_{1\leq j \leq p,\, t \in \mathcal{T}} |\mu_{j}^X(t)| \leq C
\ \text{and}\
\sup_{1\leq m \leq q,\, t \in \mathcal{T}} |\mu_{m}^Y(t)| \leq C,
\)
and 
\(
\sup_{1\leq j \leq p}\|(\mu_{j}^X)^{(l)}\| \leq C
\quad \text{and}\quad
\sup_{1\leq m \leq q}\|(\mu_{m}^Y)^{(l)}\| \leq C.
\)
Moreover, $H_X(t_1,t_2)$ and $H_Y(t_1,t_2)$ are functions on $\mathcal{T} \times \mathcal{T}$ with continuous and bounded partial derivatives of total order $l$.
\end{assumption}

Assumption~\ref{mean_assup} imposes smoothness requirements on the mean and aggregated cross-covariance functions of the functional data, which control the estimation errors of $\hat{\mu}_j^X$, $\hat{\mu}_m^Y$, $\hat{H}_X$, and $\hat{H}_Y$ as $p, q \to \infty$.

As we implement the decomposition of $\bm{R}_{XY}(t_1,t_2)$ using the spectral decompositions of $H_X(t_1, t_2)$ and $H_Y(t_1, t_2)$, we require the following conditions.

\begin{assumption}\label{A5}
The eigenfunctions $\{\wtilde{\psi}_k^X : k\ge 1\}$ and $\{\wtilde{\psi}_k^Y : k\ge 1\}$ of 
$H_X(t_1, t_2)$ and $H_Y(t_1, t_2)$ are uniformly bounded. 
Moreover, there exist constants $c, C > 0$ and $a > 1$ such that the corresponding eigenvalues satisfy, for all $k\ge 1$,
$C\, k^{-a} \;\geq\; \zeta_{k}^X, \ \zeta_{k}^X-\zeta_{k+1}^X \;\geq\; c\, k^{-(a+1)}$, $C\, k^{-a} \;\geq\; \zeta_{k}^Y, \ \text{and}\ \zeta_{k}^Y-\zeta_{k+1}^Y \;\geq\; c\, k^{-(a+1)}$.
\end{assumption}

\begin{assumption}\label{A_r}
For $r\leq R$, the singular values $\{\lambda_r\}_{r\ge 1}$ of ${\bm{R}}_{XY}(t_1,t_2)$ satisfy $\lambda_R> 0$ and
\(
\frac{\inf_{1\leq r_1\neq r_2\leq R+1}\big|{\lambda}_{r_1}^2-{\lambda}_{r_2}^2\big|}{pq}>C.
\)
\end{assumption}

Assumption~\ref{A5} imposes polynomial decay rates on the eigenvalues of $H_X(t_1, t_2)$ and $H_Y(t_1, t_2)$ to control the errors arising from their spectral decompositions, consistent with conditions in the literature \citep{hall2007methodology}. Here, a larger $a$ indicates faster eigenvalue decay and hence lower model complexity.
Assumption~\ref{A_r} introduces a singular-gap condition for the $\mathbb{R}^{p \times q}$-valued kernel ${\bm{R}}_{XY}(t_1, t_2)$, ensuring the identifiability of its first $R$ singular components. Here, the denominator $pq$ accounts for the dimensional effects in the singular gap.

\begin{assumption}\label{A6}
Define $C_{\mu,X} 
= \sup_{f > 1,\ 1 \leq j \leq p,\ 1 \leq i \leq n,\ 1 \leq g \leq N_i^X}
\frac{\bigl(\mathbb{E}\,\bigl|{\mu}_{j}^X(T_{ig}) - \hat{\mu}_{j}^X(T_{ig})\bigr|^f\bigr)^{1/f}}
     {\sqrt{f \cdot \mathbb{E}\bigl({\mu}_{j}^X(T_{ig}) - \hat{\mu}_{j}^X(T_{ig})\bigr)^2}}$. We similarly define $C_{\mu,Y}$ and set $C_{\mu} = \max\{C_{\mu,X}, C_{\mu,Y}\}$. 
Define
\[
C_{\bm{\Gamma}} 
= \sup_{f > 2,\ 1 \leq j \leq p,\ 1 \leq m \leq q,\ k_1 \geq 1,\ k_2 \geq 1}
\frac{\bigl(\mathbb{E}\,\bigl|[\wtilde{\bm{\Lambda}}^{XY}_{k_1,k_2} - \hat{\bm{\Lambda}}^{XY}_{k_1,k_2}]_{j,m}\bigr|^f\bigr)^{1/f}}
     {f \cdot \sqrt{\mathbb{E}\Bigl({\bigl([\wtilde{\bm{\Lambda}}^{XY}_{k_1,k_2} - \hat{\bm{\Lambda}}^{XY}_{k_1,k_2}]_{j,m}\bigr)^2}\Bigr)}},
\]
where $[\cdot]_{j,m}$ denotes the $(j,m)$th entry of a matrix.
Let $s_X = |\mathcal{S}_X|$ and $s_Y = |\mathcal{S}_Y|$, where $\mathcal{S}_X$ and $\mathcal{S}_Y$ are defined in \eqref{sparse_condition} and \eqref{sparse_condition_Y}. Assume $s_X \leq C \sqrt{p}$ and $s_Y \leq C\sqrt{q}$, and as $n,p,q \rightarrow \infty$, 
\[
\sqrt{s_X}+\sqrt{s_Y} = o\!\left(\frac{n^{l/(2l+2)}}{C_{\mu} C_{\bm \Gamma} \log(p+q)}\right).
\]
\end{assumption}

The quantities $C_{\mu}$ and $C_{\bm{\Gamma}}$ quantify the tail behavior of the estimation errors for the mean functions $\mu_j^X$ and $\mu_m^Y$, and for the cross-covariance matrix $\bm{\Gamma}$, respectively. 
With this assumption, Assumption~\ref{A6} imposes restrictions on the growth rates of the sparsity levels $s_X$ and $s_Y$ as $n, p, q \to \infty$, ensuring that the penalized SVD in \eqref{opt_SVD} remains consistent.

Next, we analyze the estimation error of the functional loadings \( \wtilde{\bm{A}}^X_r \) and \( \wtilde{\bm{A}}^Y_r \). 
Since their signs are not identifiable, we align them such that 
\( \langle \wtilde{\bm{A}}^X_r, \hat{\bm{A}}^X_r \rangle_p \geq 0 \) and 
\( \langle \wtilde{\bm{A}}^Y_r, \hat{\bm{A}}^Y_r \rangle_q \geq 0 \).

\begin{theorem}\label{Theorem: consistency_diverge_dimension}
Suppose the conditions in Theorem~\ref{Theorem_appr} and Assumptions~\ref{A1}--\ref{A6}, as well as the tuning parameter conditions in Part~\ref{op_sec} of the Supplementary Materials, hold. Then, as $n,p,q\rightarrow\infty$,
\begin{align}\label{rate_A}
&\max\{\|\wtilde{\bm{A}}^X_r- \hat{\bm{A}}^X_r\|_p,\ 
\|\wtilde{\bm{A}}^Y_r- \hat{\bm{A}}^Y_r\|_q\}\nonumber\\
=& \mathcal{O}_p\!\bigg[\left\{C_{\bm{\Gamma}}\cdot\big(\text{Error}_{\mu}+\text{Error}_{H}\big)\cdot\max\big\{\log(p),\log(q)\big\}\cdot\big\{\sqrt{s_X}+\sqrt{s_Y}\big\}\bigg\}^{\frac{a-1}{3a+3}}\right],
\end{align}
where
\(
\text{Error}_{\mu}
= C_{\mu}\cdot\sqrt{\log(p+q)}\cdot 
\Big\{(n N_X)^{-l/(2l+1)} + (n N_Y)^{-l/(2l+1)} + n^{-1/2}\Big\}
\)
and
\(
\text{Error}_{H}
= (n N_X^2)^{-l/(2l+2)}+\allowbreak (n N_Y^2)^{-l/(2l+2)} + n^{-1/2}.
\)
\end{theorem}

Theorem~\ref{Theorem: consistency_diverge_dimension} establishes the convergence rates of $\hat{\bm{A}}^X_r$ and $\hat{\bm{A}}^Y_r$ as $n, p, q \to \infty$. 
Here, $\text{Error}_{\mu}$ quantifies the error arising from high-dimensional mean estimation, whereas $\text{Error}_{H}$ represents the covariance estimation error when the true mean is known. 
Note that the rates in $\text{Error}_{\mu}$ and $\text{Error}_{H}$ depend on \( n \), \( N_X \), and \( N_Y \). 
The terms \( (n N_X)^{-l/(2l+1)} \), \( (n N_Y)^{-l/(2l+1)} \), \( (n N_X^2)^{-l/(2l+2)} \), and \( (n N_Y^2)^{-l/(2l+2)} \) represent standard nonparametric convergence rates \citep{stone1982optimal}, while the \( n^{-1/2} \) term reflects the parametric rate from independent functional samples. These rates all tend to $0$ as $n$ goes to infinity, regardless of the temporal sampling density.
If \( C_{\mu} \) is bounded and \( N_X, N_Y \) are sufficiently large (e.g., \( \min\{N_X, N_Y\} \geq C n^{1/(2l)} \)), then \( \text{Error}_{\mu} + \text{Error}_{H} \) in \eqref{rate_A} is dominated by the order \( n^{-1/2} \).

Note that $\hat{\bm{A}}^X_r$ and $\hat{\bm{A}}^Y_r$ are consistent as $n \to \infty$ under the sparsity Assumption~\ref{A6}, where the convergence rates depend on the ambient dimensions only through a $\log(p+q)$ factor as $p$ and $q$ increase. This rate is similar to the Lasso-type rate \citep{wainwright2019high}, suggesting that high dimensionality does not affect the convergence of the functional loadings.

\section{Discussion}\label{sec:dis}
In this article, we introduce the Functional-Aggregated Cross-covariance Decomposition (FACD), a novel framework for identifying key features and their longitudinal associations between paired high-dimensional datasets. 
At its core, FACD employs a functional-aggregation strategy that extracts data-adaptive bases from high-dimensional cross-covariances, enabling efficient decomposition of the cross-covariance structure via a tractable sparse singular value decomposition. 
We establish theoretical guarantees for FACD, demonstrating statistical convergence in high-dimensional settings, and validate its empirical performance through extensive simulations. 
Finally, the application of FACD to longitudinal multi-omic data illustrates its potential to drive scientific discovery in biological research.

The FACD framework provides an effective approach for characterizing associations between high-dimensional or functional datasets \citep{witten2009extensions, parkhomenko2009sparse, yang2011functional, lin2013group,wang2026regionalization}, where cross-covariances are often highly complex and non-invertible. In such settings, conventional CCA procedures for multivariate functional data may fail due to their reliance on invertible covariance matrices/operators and the stability of correlation estimates. In contrast, FACD operates directly on the cross-covariance structure without requiring explicit inversion. Furthermore, by deriving functional bases from the cross-covariance itself, FACD more accurately characterizes the underlying associations than approaches based on prespecified bases commonly used in the literature.

{
Although FACD accommodates sparse and irregular longitudinal observations, the current framework assumes a noninformative observation-time mechanism, under which the observation schedule is independent of the underlying longitudinal processes. In practice, observation times may be nonuniform or clustered around particular visits, in which case alternative numerical integration schemes, such as trapezoidal weighting, may improve the approximation of temporal integrals (see Part~\ref{sec:trapezoidal weighting} of the Supplementary Materials for a demonstration). More importantly, visit times, dropout, or missingness may themselves be informative. For example, patients with more severe disease may be observed more frequently or may drop out after reaching a clinical endpoint. Such mechanisms may violate the current sampling assumptions and potentially introduce bias into the estimated marginal and cross-covariance structures. Extending FACD to accommodate informative observation processes, for example through joint modeling of the longitudinal and observation processes, therefore represents an important direction for future research.
}

\bibliographystyle{apalike}
\bibliography{refbib}

\appendix

\section{Mathematical Foundation of FACD}

\paragraph*{Notation} For a kernel function \( \bm{K}(t_1, t_2) \in \mathbb{R}^{p \times q} \), define its associated integral operator \( \mathcal{K} \) from \( L^2(\mathbb{R}^q) \) to \( L^2(\mathbb{R}^p) \) by
\(
(\mathcal{K}\bm{f})(t_1) = \int_{\mathcal{T}} \bm{K}(t_1, t_2) \bm{f}(t_2)\ \mathrm{d}t_2,\quad t_1\in\mathcal T,\ \forall \bm{f} \in L^2(\mathbb{R}^q).
\)
Additionally, we use the notations \( (\cdot)^* \), \( (\cdot)^{-1} \), \( \cdot \succeq 0 \), \( \cdot \circ \cdot \), and \( (\cdot)^{1/2} \) to denote the adjoint, inverse, non-negative definiteness, composition, and square root of matrices or operators, respectively. 
Define the norm \( \|\mathcal{A}\|_{\infty} := \sup_{\{\bm{f} \in L^2(\mathbb{R}^p) : \|\bm{f}\|_p \leq 1\}} \|\mathcal{A}\bm{f}\|_q \) for an operator \( \mathcal{A} \) mapping from \( L^2(\mathbb{R}^p) \) to \( L^2(\mathbb{R}^q) \). 
We say that \( \mathcal{A}\) is compact if there exist \( \lambda_1 \geq \lambda_2 \geq \cdots \geq 0 \) and orthonormal functions \( \{\bm{f}_r \in L^2(\mathbb{R}^p);\, r \geq 1\} \) and \( \{\bm{g}_r \in L^2(\mathbb{R}^q);\, r \geq 1\} \) such that  
\(
\lim_{R \to \infty} \left\| \mathcal{A} - \sum_{r=1}^{R} \lambda_r\, \bm{f}_r \otimes \bm{g}_r \right\|_{\infty} = 0,
\)
where \( \otimes \) denotes the tensor product, defining an operator \( \bm{f}_r \otimes \bm{g}_r \) by 
\(
(\bm{f}_r \otimes \bm{g}_r)\bm{h} = \bm{f}_r \langle \bm{g}_r, \bm{h} \rangle_q
\)
for all \( \bm{h} \in L^2(\mathbb{R}^q) \).  
If \( \mathcal{A} \) is an infinite-dimensional compact operator, we write  
\(
\mathcal{A} = \sum_{r=1}^{\infty} \lambda_r\, \bm{f}_r \otimes \bm{g}_r,
\)
where \( \lambda_r \) is the \( r \)th singular value and \( \bm{f}_r, \bm{g}_r \) are the corresponding singular functions.  
See Section~4.3 of \citet{hsing2015theoretical} for details.

\

\subsection{Non-invertibility of Functional CCA}\label{FCCA_irre}
We denote the covariance of \( \bm{X} \) and \( \bm{Y} \), as well as their cross-covariance functions, as
\begin{eqnarray*}
\bm{R}_X(t_1, t_2) &:=& \operatorname{Cov}\big\{\bm{X}(t_1), \bm{X}(t_2)\big\} \in \mathbb{R}^{p \times p}, \quad t_1, t_2 \in \mathcal{T}, \\
\bm{R}_Y(t_1, t_2) &:=& \operatorname{Cov}\big\{\bm{Y}(t_1), \bm{Y}(t_2)\big\} \in \mathbb{R}^{q \times q}, \quad t_1, t_2 \in \mathcal{T}, \\
\bm{R}_{XY}(t_1, t_2) &:=& \operatorname{Cov}\big\{\bm{X}(t_1), \bm{Y}(t_2)\big\} \in \mathbb{R}^{p \times q}, \quad t_1, t_2 \in \mathcal{T}.
\end{eqnarray*}
Their associated integral operators are referred to as covariance operators, denoted by \( \mathcal{R}_X \), \( \mathcal{R}_Y \), and \( \mathcal{R}_{XY} \), respectively. These operators are considered the extensions of (cross-) covariance matrices from multivariate data to multivariate random functions.

In this section, we consider measuring the association between 
\[
\int_0^1 \bm{A}^\top_X(t)\,\bm{X}(t)\,\mathrm{d}t 
\quad \text{and} \quad 
\int_0^1 \bm{A}^\top_Y(t)\,\bm{Y}(t)\,\mathrm{d}t,
\]
using the Pearson correlation. 
\begin{eqnarray}\label{CCA_criterion}
\frac{\operatorname{Cov}\bigg(\int_0^1 \bm{A}^\top_X(t) \bm{X}(t)\ \mathrm{d}t, \int_0^1 \bm{A}^\top_Y(t) \bm{Y}(t)\ \mathrm{d}t\bigg)}{\sqrt{\operatorname{Var}\bigg(\int_0^1 \bm{A}^\top_X(t) \bm{X}(t)\ \mathrm{d}t\bigg)\cdot \operatorname{Var}\bigg(\int_0^1 \bm{A}^\top_Y(t) \bm{Y}(t)\ \mathrm{d}t\bigg)}},
\end{eqnarray}
which leads to the CCA between $\bm{X}$ and $\bm{Y}$. 
By the definitions of $\mathcal{R}_{XY}$, $\mathcal{R}_X$, and $\mathcal{R}_Y$, we represent the association by
\begin{gather*}
\rho^2\bigg(\int_0^1 \bm{A}^\top_X(t) \bm{X}(t)\ \mathrm{d}t, \int_0^1 \bm{A}^\top_Y(t) \bm{Y}(t)\ \mathrm{d}t\bigg) = 
\frac{\langle \bm{A}_X, \mathcal{R}_{XY} \bm{A}_Y \rangle^2_p}{\langle \bm{A}_X, \mathcal{R}_{X} \bm{A}_X \rangle_p \cdot \langle \bm{A}_Y, \mathcal{R}_{Y} \bm{A}_Y \rangle_q},
\end{gather*}
Since $\mathcal{R}_{X} \succeq 0$ and $\mathcal{R}_{Y} \succeq 0$, we utilize their square roots to define $\bm{B}_X := \mathcal{R}_{X}^{1/2} \bm{A}_X$ and $\bm{B}_Y := \mathcal{R}_{Y}^{1/2} \bm{A}_Y$. 
If the inverses of $\mathcal{R}_{Y}$ and $\mathcal{R}_{X}$ exist, we then calculate the operator 
\[
\mathcal{G}_{XY} := \mathcal{R}_{X}^{-1/2} \circ \mathcal{R}_{XY} \circ \mathcal{R}_{Y}^{-1/2},
\]
and solve the CCA by finding $\bm{B}_X$ and $\bm{B}_Y$ that maximize 
\begin{eqnarray*}
    \langle \bm{B}_X, \mathcal{G}_{XY} \bm{B}_Y \rangle_p\ \text{subject to}\ \|\bm{B}_X\|_p = \|\bm{B}_Y\|_q = 1.
\end{eqnarray*}
The above procedure is analogous to solving CCA for multivariate data. 

However, when $\bm{X}$ and $\bm{Y}$ are infinite-dimensional functions, 
the covariance operators $\mathcal{R}_{X}$ and $\mathcal{R}_{Y}$ are 
typically compact operators on infinite-dimensional functional spaces. 
By Theorem~4.1.4 in \citet{hsing2015theoretical}, these operators are not invertible. 
Therefore, the estimation of functional loadings using 
\eqref{CCA_criterion} is invalid.

\

\subsection{SVD for Canonical Cross-Covariance Analysis}\label{sec: prd}

\begin{theoremS}\label{theo: svd}
Define 
\( \bm{R}_{XY}(t_1, t_2) := \operatorname{Cov}\{\bm{X}(t_1), \bm{Y}(t_2)\} \in \mathbb{R}^{p \times q} \) for \( t_1, t_2 \in \mathcal{T} \). 
Assume that \( \bm{R}_{XY}(t_1, t_2) \) is continuous with respect to \( t_1 \) and \( t_2 \). Then, $\bm{R}_{XY}(t_1, t_2)$ has the following decomposition:
\begin{eqnarray*}
    \bm{R}_{XY}(t_1,t_2)=\sum_{r=1}^{\infty}\lambda_r\wtilde{\bm{A}}^X_{r}(t_1) \big\{\wtilde{\bm{A}}^Y_{r}(t_2)\big\}^\top,
\end{eqnarray*}
where $\lambda_1\geq \lambda_2\geq \cdots\geq 0$ are the singular values, and $\wtilde{\bm{A}}^X_{r}$ and $\wtilde{\bm{A}}^Y_{r}$ are singular functions, respectively. In addition,
\[
\max_{\bm{A}_X\in \mathbb{S}_p,\ \bm{A}_Y\in \mathbb{S}_q}
\operatorname{Cov}^2\!\bigg(\int_0^1 \bm{A}^\top_X(t)\bm{X}(t)\, \mathrm{d}t,\ 
\int_0^1 \bm{A}^\top_Y(t)\bm{Y}(t)\, \mathrm{d}t\bigg)
=\lambda_1^2,
\]
where the maximum is attained at $\bm{A}_X=\pm\wtilde{\bm{A}}^X_{1}$ and $\bm{A}_Y=\pm\wtilde{\bm{A}}^Y_{1}$.
\end{theoremS}
\begin{proof}
By the Stone–Weierstrass Theorem, we can show that \( \mathcal{R}_{XY} \) is a compact operator. The proof is similar to Theorem 4.6.2 in \citet{hsing2015theoretical}.
Therefore, we have the following decomposition 
\begin{eqnarray*}
    \mathcal{R}_{XY}=\sum_{r=1}^{\infty}\lambda_r\wtilde{\bm{A}}^X_{r}\otimes \wtilde{\bm{A}}^Y_{r}\ \text{or equivalently,}\quad 
    \bm{R}_{XY}(t_1,t_2)=\sum_{r=1}^{\infty}\lambda_r\wtilde{\bm{A}}^X_{r}(t_1) \big\{\wtilde{\bm{A}}^Y_{r}(t_2)\big\}^\top,
\end{eqnarray*}
by Theorem 4.3.1 in \citet{hsing2015theoretical}, where {\(\lambda_1\ge \lambda_2\ge \cdots \ge 0\)} and
{\(\{\wtilde{\bm{A}}^X_r\}_{r\ge 1}\subset L^2(\mathbb{R}^p)\), \(\{\wtilde{\bm{A}}^Y_r\}_{r\ge 1}\subset L^2(\mathbb{R}^q)\) are orthonormal systems.}
Accordingly, by Fubini's theorem and exchanging the sum/integrals,
\begin{align*}
    &\operatorname{Cov}\bigg(\int_0^1 \bm{A}^\top_X(t)\bm{X}(t)\,\mathrm{d}t,\int_0^1 \bm{A}^\top_Y(t)\bm{Y}(t)\,\mathrm{d}t\bigg) \\
    =& \int_0^1\int_0^1 \bm{A}^\top_X(t_1) \bm{R}_{XY}(t_1,t_2) \bm{A}_Y(t_2)\,\mathrm{d}t_1\mathrm{d}t_2 \\
    =& \int_0^1\int_0^1 \bm{A}^\top_X(t_1) \bigg(\sum_{r=1}^{\infty} \lambda_r \wtilde{\bm{A}}^X_{r}(t_1) \big\{\wtilde{\bm{A}}^Y_{r}(t_2)\big\}^\top \bigg) \bm{A}_Y(t_2)\,\mathrm{d}t_1\mathrm{d}t_2 \\
    =& \sum_{r=1}^{\infty} \lambda_r \int_0^1 \bm{A}^\top_X(t_1) \wtilde{\bm{A}}^X_{r}(t_1)\,\mathrm{d}t_1 \cdot \int_0^1 \big\{\wtilde{\bm{A}}^Y_{r}(t_2)\big\}^\top \bm{A}_Y(t_2)\,\mathrm{d}t_2.
\end{align*}
For $\bm{A}_X \in \mathbb{S}_p$ and $\bm{A}_Y \in \mathbb{S}_q$, we have
\begin{align*}
    \sum_{r \geq 1} \bigg(\int_0^1 \bm{A}^\top_X(t_1) \wtilde{\bm{A}}^X_{r}(t_1)\,\mathrm{d}t_1\bigg)^2 &\leq \|\bm{A}_X\|_p^2 = 1, \\
    \sum_{r \geq 1} \bigg(\int_0^1 \bm{A}^\top_Y(t_2) \wtilde{\bm{A}}^Y_{r}(t_2)\,\mathrm{d}t_2\bigg)^2 &\leq \|\bm{A}_Y\|_q^2 = 1.
\end{align*}
Thus,
\begin{align*}
    &\Bigg|\operatorname{Cov}\bigg(\int_0^1 \bm{A}^\top_X(t)\bm{X}(t)\,\mathrm{d}t,\int_0^1 \bm{A}^\top_Y(t)\bm{Y}(t)\,\mathrm{d}t\bigg)\Bigg| \\
    =& \Bigg|\sum_{r=1}^{\infty} \lambda_r \int_0^1 \bm{A}^\top_X(t_1) \wtilde{\bm{A}}^X_{r}(t_1)\,\mathrm{d}t_1 \cdot \int_0^1 \big\{\wtilde{\bm{A}}^Y_{r}(t_2)\big\}^\top \bm{A}_Y(t_2)\,\mathrm{d}t_2\Bigg| \\
    \leq& \sum_{r=1}^{\infty} \lambda_r \Bigg|\int_0^1 \bm{A}^\top_X(t_1) \wtilde{\bm{A}}^X_{r}(t_1)\,\mathrm{d}t_1\Bigg|
    \cdot \Bigg|\int_0^1 \big\{\wtilde{\bm{A}}^Y_{r}(t_2)\big\}^\top \bm{A}_Y(t_2)\,\mathrm{d}t_2\Bigg| \\
    \leq& \lambda_1 \sum_{r=1}^{\infty} \Bigg|\int_0^1 \bm{A}^\top_X(t_1) \wtilde{\bm{A}}^X_{r}(t_1)\,\mathrm{d}t_1\Bigg|
    \cdot \Bigg|\int_0^1 \big\{\wtilde{\bm{A}}^Y_{r}(t_2)\big\}^\top \bm{A}_Y(t_2)\,\mathrm{d}t_2\Bigg| \\
    \leq & \lambda_1\cdot \sqrt{\sum_{r \geq 1} \bigg(\int_0^1 \bm{A}^\top_X(t_1) \wtilde{\bm{A}}^X_{r}(t_1)\,\mathrm{d}t_1\bigg)^2 \cdot 
    \sum_{r \geq 1} \bigg(\int_0^1 \bm{A}^\top_Y(t_2) \wtilde{\bm{A}}^Y_{r}(t_2)\,\mathrm{d}t_2\bigg)^2} \\
    \leq & \lambda_1.
\end{align*}
As a result,
\begin{align*}
    \max_{\bm{A}_X \in \mathbb{S}_p, \bm{A}_Y \in \mathbb{S}_q} 
    \Bigg|\operatorname{Cov}\bigg(\int_0^1 \bm{A}^\top_X(t)\bm{X}(t)\,\mathrm{d}t,\int_0^1 \bm{A}^\top_Y(t)\bm{Y}(t)\,\mathrm{d}t\bigg)\Bigg|^2 = \lambda_1^2,
\end{align*}
where the maximum is attained at $\bm{A}_X =\pm \wtilde{\bm{A}}^X_{1}$ and $\bm{A}_Y =\pm \wtilde{\bm{A}}^Y_{1}$, {(with the signs chosen so that the covariance is \(\lambda_1\)).}
\end{proof}

\

\subsection{Proof of Theorem \ref{theo: svd_mar}}
\begin{proof}
First, note that the subspaces $\mathcal{H}_1$ and $\mathcal{H}_2$ of $L^2(\mathbb{R})$ are separable Hilbert spaces.
Since $\{\wtilde{A}_{rj}^X; j=1, \dots, p,  r\geq 1\}$ and $\{\wtilde{A}_{rm}^Y; m=1, \dots, q, r\geq 1\}$ are contained in $\mathcal{H}_1$ and $\mathcal{H}_2$, respectively, we can write:
\begin{align*}
    \wtilde{A}_{rj}^X &= \sum_{k_1 \geq 1} \langle \wtilde{A}_{rj}^X, \psi_{k_1}^X \rangle \psi_{k_1}^X, \\
    \wtilde{A}_{rm}^Y &= \sum_{k_2 \geq 1} \langle \wtilde{A}_{rm}^Y, \psi_{k_2}^Y \rangle \psi_{k_2}^Y,
\end{align*}
where \(\{\psi_{k_1}^X\}_{k_1\ge1}\) and \(\{\psi_{k_2}^Y\}_{k_2\ge1}\) are orthonormal bases of \(\mathcal{H}_1\) and \(\mathcal{H}_2\), respectively.

Let ${R}^{XY}_{jm}(t_1, t_2)$ be the $(j, m)$th element of $\bm{R}_{XY}(t_1, t_2)$.
Then,
\begin{align*}
    {R}^{XY}_{jm}(t_1, t_2) 
    &= \sum_{r=1}^\infty \lambda_r \wtilde{A}_{rj}^X(t_1) \wtilde{A}_{rm}^Y(t_2) \\
    &= \sum_{r=1}^\infty \lambda_r 
       \bigg\{\sum_{k_1 \geq 1} \langle \wtilde{A}_{rj}^X, \psi_{k_1}^X \rangle \psi_{k_1}^X(t_1)\bigg\}
       \bigg\{\sum_{k_2 \geq 1} \langle \wtilde{A}_{rm}^Y, \psi_{k_2}^Y \rangle \psi_{k_2}^Y(t_2)\bigg\} \\
    &= \sum_{k_1 \geq 1} \sum_{k_2 \geq 1} 
       \bigg(\sum_{r=1}^\infty \lambda_r \langle \wtilde{A}_{rj}^X, \psi_{k_1}^X \rangle \langle \wtilde{A}_{rm}^Y, \psi_{k_2}^Y \rangle \bigg) 
       \psi_{k_1}^X(t_1) \psi_{k_2}^Y(t_2).
\end{align*}
Moreover, by the orthonormal expansions above and the definition of the coefficients,
\begin{align*}
    \sum_{r=1}^\infty \lambda_r \langle \wtilde{A}_{rj}^X, \psi_{k_1}^X \rangle \langle \wtilde{A}_{rm}^Y, \psi_{k_2}^Y \rangle
    &= \iint_{0}^{1} {R}^{XY}_{jm}(t_1, t_2)\,\psi_{k_1}^X(t_1)\,\psi_{k_2}^Y(t_2)\,\mathrm{d}t_1\mathrm{d}t_2,
\end{align*}
Therefore,
\begin{align*}
    {R}^{XY}_{jm}(t_1, t_2)
    &= \sum_{k_1 \geq 1} \sum_{k_2 \geq 1} 
       \bigg( \iint_0^1 {R}^{XY}_{jm}(t_1, t_2)\,\psi_{k_1}^X(t_1)\,\psi_{k_2}^Y(t_2) \, \mathrm{d}t_1 \mathrm{d}t_2 \bigg) 
       \psi_{k_1}^X(t_1) \psi_{k_2}^Y(t_2)\\
    &=: \sum_{k_1=1}^{\infty}\sum_{k_2=1}^{\infty}{\Lambda^{XY}_{jm;k_1,k_2}}\,\psi_{k_1}^X(t_1)\,\psi_{k_2}^Y(t_2),
\end{align*}
with \(\Lambda^{XY}_{jm;k_1,k_2}:=\iint_0^1 {R}^{XY}_{jm}(t_1, t_2)\,\psi_{k_1}^X(t_1)\,\psi_{k_2}^Y(t_2)\,\mathrm{d}t_1\mathrm{d}t_2\).
This completes the proof.
\end{proof}

\

\subsection{Proof of Theorem \ref{The_opt_bas}}
\begin{proof}
For any complete orthonormal bases
$\{\psi_{k_1}^X; k_1 \geq 1\}$ for $\mathcal{H}_1$
and $\{\psi_{k_2}^Y; k_2 \geq 1\}$ for $\mathcal{H}_2$,
note that:
\begin{align*}
    H_X(t_1, t_2)
    &= \operatorname{tr}\{\bm{L}_X(t_1, t_2)\}/(pq) \\
    &= \frac{1}{pq}
    \operatorname{tr}\bigg\{
    \int_{0}^1
    \bm{R}_{XY}(t_1, t)
    \bm{R}^\top_{XY}(t_2, t)
    \, \mathrm{d}t
    \bigg\} \\
    &= \frac{1}{pq}
    \operatorname{tr}\bigg\{
    \int_{0}^1
    \bigg(
    \sum_{k_1=1}^\infty
    \sum_{k_2=1}^\infty
    \bm{\Lambda}^{XY}_{k_1,k_2}
    \psi_{k_1}^X(t_1)
    \psi_{k_2}^Y(t)
    \bigg) \\
    &\qquad\qquad\qquad \times
    \bigg(
    \sum_{k_1'=1}^\infty
    \sum_{k_2'=1}^\infty
    (\bm{\Lambda}^{XY}_{k_1',k_2'})^\top
    \psi_{k_1'}^X(t_2)
    \psi_{k_2'}^Y(t)
    \bigg)
    \, \mathrm{d}t
    \bigg\} \\
    &= \frac{1}{pq}
    \sum_{k_1=1}^\infty
    \sum_{k_1'=1}^\infty
    \operatorname{tr}\bigg\{
    \sum_{k=1}^\infty
    \bm{\Lambda}^{XY}_{k_1,k}
    (\bm{\Lambda}^{XY}_{k_1',k})^\top
    \bigg\}
    \psi_{k_1}^X(t_1)
    \psi_{k_1'}^X(t_2) \\
    &:= \frac{1}{pq}
    \sum_{k_1=1}^\infty
    \sum_{k_1'=1}^\infty
    \operatorname{tr}(\bm{D}_{k_1,k_1'})
    \psi_{k_1}^X(t_1)
    \psi_{k_1'}^X(t_2),
\end{align*}
by Theorem~\ref{theo: svd_mar}, where
$\bm{\Lambda}^{XY}_{k_1,k_2}
\in\mathbb{R}^{p\times q}$
and
\[
\bm{D}_{k_1,k_1'}
:=
\sum_{k\ge1}
\bm{\Lambda}^{XY}_{k_1,k}
(\bm{\Lambda}^{XY}_{k_1',k})^\top
\in\mathbb{R}^{p\times p}.
\]

On the other hand,
$H_X(t_1,t_2)
=
\sum_{k\ge1}
\zeta_k^X
\wtilde{\psi}_k^X(t_1)
\wtilde{\psi}_k^X(t_2)$.
For $\zeta_k^X>0$, we have:
\begin{align*}
    \wtilde{\psi}_k^X(t_1)
    &=
    \frac{1}{\zeta_k^X}
    \int_0^1
    H_X(t_1,t_2)
    \wtilde{\psi}_k^X(t_2)
    \, \mathrm{d}t_2 \\
    &=
    \frac{1}{\zeta_k^X}
    \int_0^1
    \frac{1}{pq}
    \sum_{k_1=1}^\infty
    \sum_{k_1'=1}^\infty
    \operatorname{tr}(\bm{D}_{k_1,k_1'})
    \psi_{k_1}^X(t_1)
    \psi_{k_1'}^X(t_2)
    \wtilde{\psi}_k^X(t_2)
    \, \mathrm{d}t_2 \\
    &=
    \frac{1}{\zeta_k^X}
    \frac{1}{pq}
    \sum_{k_1=1}^\infty
    \sum_{k_1'=1}^\infty
    \operatorname{tr}(\bm{D}_{k_1,k_1'})
    \psi_{k_1}^X(t_1)
    \int_0^1
    \psi_{k_1'}^X(t_2)
    \wtilde{\psi}_k^X(t_2)
    \, \mathrm{d}t_2.
\end{align*}
Thus,
$\wtilde{\psi}_k^X\in\mathcal{H}_1$
when $\zeta_k^X>0$,
since it is a linear combination of
$\{\psi_{k_1}^X\}$ in $\mathcal{H}_1$.

Let
$\wtilde{\mathcal{H}}_1
:=
\operatorname{span}
\{\wtilde{\psi}_k^X;
k\geq1,\ \zeta_k^X>0\}$,
and denote
$\wtilde{\mathcal{H}}_1^\perp$
as the orthogonal complement of
$\wtilde{\mathcal{H}}_1$
in $\mathcal{H}_1$
under the $L^2$ inner product.
For any
$\psi\in\wtilde{\mathcal{H}}_1^\perp$,
we have
\begin{align*}
    \int_0^1
    H_X(t_1,t_2)\psi(t_2)
    \, \mathrm{d}t_2
    &=
    \sum_{k\ge1}
    \zeta_k^X
    \wtilde{\psi}_k^X(t_1)
    \int_0^1
    \wtilde{\psi}_k^X(t_2)
    \psi(t_2)
    \, \mathrm{d}t_2
    =0.
\end{align*}

Moreover, by the spectral decomposition
of $\bm{L}_X$, we have
\begin{align*}
    0
    &=
    \int_0^1\int_0^1
    \psi(t_1)
    H_X(t_1,t_2)
    \psi(t_2)
    \, \mathrm{d}t_1
    \mathrm{d}t_2 \\
    &=
    \frac{1}{pq}
    \sum_{r:\lambda_r>0}
    \lambda_r^2
    \sum_{j=1}^p
    \left|
    \left\langle
    \psi,\wtilde{A}_{rj}^X
    \right\rangle
    \right|^2.
\end{align*}
Since $\lambda_r>0$, it follows that
$\langle\psi,\wtilde{A}_{rj}^X\rangle=0$
for all $j$ and all $r$ with
$\lambda_r>0$.
By the definition of $\mathcal{H}_1$,
we obtain $\psi=0$.
Hence,
$\wtilde{\mathcal{H}}_1^\perp=\{0\}$,
which implies
\[
\mathcal{H}_1
=
\wtilde{\mathcal{H}}_1
=
\operatorname{span}
\{\wtilde{\psi}_k^X;
k\ge1,\ \zeta_k^X>0\}.
\]

By a similar argument, we can prove that
\[
\mathcal{H}_2
=
\operatorname{span}
\{\wtilde{\psi}_k^Y;
k\ge1,\ \zeta_k^Y>0\}.
\]
\end{proof}

\

\subsection{Proof of Theorem~\ref{Theorem_appr}}

\begin{proof}
For any complete orthonormal bases $\{\psi_{k_1}^X; k_1 \geq 1\}$ {such that $\mathcal{H}_1\subset \operatorname{span}\{{\psi}_k^X; k \geq 1\}$}, note that:
\begin{align*}
    \bm{L}_X(t_1, t_2)
    &= \int_{0}^1 \bm{R}_{XY}(t_1, t) \bm{R}^\top_{XY}(t_2, t) \, \mathrm{d}t \\
    &= \int_{0}^1 
       \bigg(\sum_{k_1=1}^\infty \sum_{k_2=1}^\infty \bm{\Lambda}^{XY}_{k_1, k_2} \psi_{k_1}^X(t_1) \psi_{k_2}^Y(t)\bigg)
       \bigg(\sum_{k_1'=1}^\infty \sum_{k_2'=1}^\infty (\bm{\Lambda}^{XY}_{k_1', k_2'})^\top \psi_{k_1'}^X(t_2) \psi_{k_2'}^Y(t)\bigg)\, \mathrm{d}t \\
    &= \sum_{k_1=1}^\infty \sum_{k_1'=1}^\infty 
       \bigg\{\sum_{k=1}^\infty \bm{\Lambda}^{XY}_{k_1, k} (\bm{\Lambda}^{XY}_{k_1', k})^\top\bigg\} 
       \psi_{k_1}^X(t_1) \psi_{k_1'}^X(t_2) \\
    &:= \sum_{k_1=1}^\infty \sum_{k_1'=1}^\infty \bm{D}_{k_1, k_1'}\,
       \psi_{k_1}^X(t_1) \psi_{k_1'}^X(t_2),
\end{align*}
by Theorem~\ref{theo: svd_mar}, where {\(\bm{D}_{k_1,k_1'}:=\sum_{k\ge1}\bm{\Lambda}^{XY}_{k_1,k}(\bm{\Lambda}^{XY}_{k_1',k})^\top\).}

Since $\mathcal{H}_1\subset \operatorname{span}\{\wtilde{\psi}_k^X; k \geq 1\}$ due to Theorem~\ref{The_opt_bas}, we obtain
\begin{align*}
    \bm{L}_X(t_1, t_2) 
    &= \sum_{k_1=1}^\infty \sum_{k_1'=1}^\infty {\bm{D}}_{k_1, k_1'}\,
       \wtilde{\psi}_{k_1}^X(t_1) \wtilde{\psi}_{k_1'}^X(t_2),
\end{align*}
where ${\bm{D}}_{k_1, k_1'}$ denotes $\sum_{k=1}^\infty \wtilde{\bm{\Lambda}}^{XY}_{k_1, k} (\wtilde{\bm{\Lambda}}^{XY}_{k_1', k})^\top$.

With the same approach, we obtain
\begin{align*}
    \bm{L}_Y(t_1, t_2)
    &= \int_{0}^1\bm{R}^\top_{XY}(t,t_1)\bm{R}_{XY}(t,t_2)\ \mathrm{d}t \\
    &= \int_{0}^1 
       \bigg(\sum_{k_1'=1}^\infty \sum_{k_2'=1}^\infty (\bm{\Lambda}^{XY}_{k_1', k_2'})^\top \psi_{k_1'}^X(t) \psi_{k_2'}^Y(t_1)\bigg)
       \bigg(\sum_{k_1=1}^\infty \sum_{k_2=1}^\infty \bm{\Lambda}^{XY}_{k_1, k_2} \psi_{k_1}^X(t) \psi_{k_2}^Y(t_2)\bigg)\, \mathrm{d}t \\
    &= \sum_{k_2=1}^\infty \sum_{k_2'=1}^\infty 
       \bigg\{\sum_{k=1}^\infty (\bm{\Lambda}^{XY}_{k, k_2'})^\top\bm{\Lambda}^{XY}_{k, k_2} \bigg\} 
       \psi_{k_2'}^Y(t_1) \psi_{k_2}^Y(t_2) \\
    &:= \sum_{k_2=1}^\infty \sum_{k_2'=1}^\infty \bm{E}_{k_2', k_2}\,
       \psi_{k_2'}^Y(t_1) \psi_{k_2}^Y(t_2),
\end{align*}
where {\(\bm{E}_{k_2',k_2}:=\sum_{k\ge1}(\bm{\Lambda}^{XY}_{k,k_2'})^\top\bm{\Lambda}^{XY}_{k,k_2}\).}
Since $\mathcal{H}_2\subset \operatorname{span}\{\wtilde{\psi}_k^Y; k \geq 1\}$ due to Theorem~\ref{The_opt_bas}, we obtain
\begin{align*}
    \bm{L}_Y(t_1, t_2) 
    &= \sum_{k_2=1}^\infty \sum_{k_2'=1}^\infty \bm{E}_{k_2', k_2}\,
       \wtilde{\psi}_{k_2'}^Y(t_1) \wtilde{\psi}_{k_2}^Y(t_2).
\end{align*}
In the following, we simplify the notation $\wtilde{\bm{\Lambda}}^{XY}_{k_1, k_2}$ to ${\bm{\Lambda}}^{XY}_{k_1, k_2}$.

Define
\begin{align*}
    {\bm{L}}^{(\kappa_X,\kappa_Y)}_X(t_1,t_2) &= \int_{0}^1{\bm{R}}^{(\kappa_X,\kappa_Y)}_{XY}(t_1,t)\big\{{\bm{R}}^{(\kappa_X,\kappa_Y)}_{XY}(t_2,t)\big\}^\top\ \mathrm{d}t,\\
    {\bm{L}}^{(\kappa_X,\kappa_Y)}_Y(t_1,t_2) &= \int_{0}^1\big\{{\bm{R}}^{(\kappa_X,\kappa_Y)}_{XY}(t,t_1)\big\}^\top{\bm{R}}^{(\kappa_X,\kappa_Y)}_{XY}(t,t_2)\ \mathrm{d}t,
\end{align*}
where ${\bm{R}}_{XY}^{(\kappa_X,\kappa_Y)}(t_1,t_2)= \sum_{k_1=1}^{\kappa_X}\sum_{k_2=1}^{\kappa_Y}{\bm{\Lambda}}^{XY}_{k_1,k_2}\wtilde{\psi}_{k_1}^X(t_1) \wtilde{\psi}_{k_2}^Y(t_2)$.
Notice that 
\begin{align*}
    {\bm{L}}^{(\kappa_X,\kappa_Y)}_X(t_1, t_2) 
    &=  \int_{0}^1{\bm{R}}^{(\kappa_X,\kappa_Y)}_{XY}(t_1,t){\bm{R}}^{(\kappa_X,\kappa_Y)\top}_{XY}(t_2,t)\ \mathrm{d}t\\ 
   & = 
   \sum_{k_1=1}^{\kappa_X} \sum_{k_1'=1}^{\kappa_X} \sum_{k=1}^{\kappa_Y} {\bm{\Lambda}}^{XY}_{k_1, k} ({\bm{\Lambda}}^{XY}_{k_1', k})^\top
       \wtilde{\psi}_{k_1}^X(t_1) \wtilde{\psi}_{k_1'}^X(t_2).
\end{align*}
Based on the above {derivations}, we have
\begin{align*}
   & \bm{L}_X(t_1, t_2) - {\bm{L}}^{(\kappa_X,\kappa_Y)}_X(t_1, t_2) \\
    =& \sum_{\substack{k_1\ge 1,\,k_1'\ge 1:\\ k_1>\kappa_X\ \text{or}\ k_1'>\kappa_X}} {\bm{D}}_{k_1, k_1'}\,
       \wtilde{\psi}_{k_1}^X(t_1) \wtilde{\psi}_{k_1'}^X(t_2)
       + \sum_{k_1=1}^{\kappa_X} \sum_{k_1'=1}^{\kappa_X} \sum_{k>\kappa_Y} {\bm{\Lambda}}^{XY}_{k_1, k} ({\bm{\Lambda}}^{XY}_{k_1', k})^\top\,
       \wtilde{\psi}_{k_1}^X(t_1) \wtilde{\psi}_{k_1'}^X(t_2),
\end{align*}
{where \({\bm{D}}_{k_1,k_1'}=\sum_{k\ge 1}{\bm{\Lambda}}^{XY}_{k_1,k}({\bm{\Lambda}}^{XY}_{k_1',k})^\top\).}
Thus, using the orthonormality of \(\{\wtilde{\psi}^X_{k_1}(t_1)\wtilde{\psi}^X_{k_1'}(t_2)\}_{k_1,k_1'\ge1}\) in \(L^2([0,1]^2)\), we have
\begin{align*}
    &\iint_0^1 \bigg\|\bm{L}_X(t_1, t_2) - {\bm{L}}^{(\kappa_X,\kappa_Y)}_X(t_1, t_2)\bigg\|_F^2 \, \mathrm{d}t_1 \mathrm{d}t_2 \\
    =& {{\iint_0^1 \bigg\|\sum_{\substack{k_1\ge 1,\,k_1'\ge 1:\\ k_1>\kappa_X\ \text{or}\ k_1'>\kappa_X}} {\bm{D}}_{k_1, k_1'}
       \wtilde{\psi}_{k_1}^X(t_1) \wtilde{\psi}_{k_1'}^X(t_2)+
       \sum_{k_1=1}^{\kappa_X} \sum_{k_1'=1}^{\kappa_X} \sum_{k>\kappa_Y} {\bm{\Lambda}}^{XY}_{k_1, k} ({\bm{\Lambda}}^{XY}_{k_1', k})^\top
       \wtilde{\psi}_{k_1}^X(t_1) \wtilde{\psi}_{k_1'}^X(t_2)\bigg\|_F^2 \, \mathrm{d}t_1 \mathrm{d}t_2}} \\
    =& {\sum_{\substack{k_1\ge 1,\,k_1'\ge 1:\\ k_1>\kappa_X\ \text{or}\ k_1'>\kappa_X}} \big\|{\bm{D}}_{k_1, k_1'}\big\|_F^2}
    \;+\;
    \sum_{k_1=1}^{\kappa_X} \sum_{k_1'=1}^{\kappa_X}  
    \bigg\|\sum_{k>\kappa_Y}{\bm{\Lambda}}^{XY}_{k_1, k} ({\bm{\Lambda}}^{XY}_{k_1', k})^\top \bigg\|_F^2.
\end{align*}
By the properties of the Frobenius norm, 
\begingroup
\setlength{\arraycolsep}{2pt}
\begin{eqnarray*}
  &&  \sum_{\substack{k_1\ge 1,\,k_1'\ge 1:\\ k_1>\kappa_X\ \text{or}\ k_1'>\kappa_X}} \big\|{\bm{D}}_{k_1, k_1'}\big\|_F^2\\
    &\leq& \sum_{k_1>\kappa_X} \sum_{k_1'\ge 1} \bigg\| \sum_{k=1}^\infty {\bm{\Lambda}}^{XY}_{k_1, k} ({\bm{\Lambda}}^{XY}_{k_1', k})^\top\bigg\|_F^2
    \;+\;\sum_{k_1=1}^{\kappa_X} \sum_{k_1'>\kappa_X} \bigg\| \sum_{k=1}^\infty {\bm{\Lambda}}^{XY}_{k_1, k} ({\bm{\Lambda}}^{XY}_{k_1', k})^\top\bigg\|_F^2\\
    &\leq & \sum_{k_1>\kappa_X} \sum_{k_1'\ge 1} \bigg(\sum_{k=1}^\infty\big\|  {\bm{\Lambda}}^{XY}_{k_1, k} ({\bm{\Lambda}}^{XY}_{k_1', k})^\top\big\|_F\bigg)^2
    \;+\;\sum_{k_1=1}^{\kappa_X} \sum_{k_1'>\kappa_X} \bigg(\sum_{k=1}^\infty\big\|  {\bm{\Lambda}}^{XY}_{k_1, k} ({\bm{\Lambda}}^{XY}_{k_1', k})^\top\big\|_F\bigg)^2\\
    &\leq &\sum_{k_1>\kappa_X} \sum_{k_1'\ge 1} \bigg(\sum_{k=1}^\infty\big\|  {\bm{\Lambda}}^{XY}_{k_1, k} \big\|_F\cdot\big\|{\bm{\Lambda}}^{XY}_{k_1', k}\big\|_F\bigg)^2
    \;+\;\sum_{k_1=1}^{\kappa_X} \sum_{k_1'>\kappa_X} \bigg(\sum_{k=1}^\infty\big\|  {\bm{\Lambda}}^{XY}_{k_1, k} \big\|_F\cdot\big\|{\bm{\Lambda}}^{XY}_{k_1', k}\big\|_F\bigg)^2\\
    &\leq &\sum_{k_1>\kappa_X} \sum_{k_1'\ge 1} \bigg(\sum_{k=1}^\infty\big\|  {\bm{\Lambda}}^{XY}_{k_1, k} \big\|_F^2\bigg)\cdot\bigg(\sum_{k=1}^\infty\big\|{\bm{\Lambda}}^{XY}_{k_1', k}\big\|_F^2\bigg)
    \;+\;\sum_{k_1=1}^{\kappa_X} \sum_{k_1'>\kappa_X} \bigg(\sum_{k=1}^\infty\big\|  {\bm{\Lambda}}^{XY}_{k_1, k} \big\|_F^2\bigg)\cdot\bigg(\sum_{k=1}^\infty\big\|{\bm{\Lambda}}^{XY}_{k_1', k}\big\|_F^2\bigg)\\
    &=& \bigg(\sum_{k_1>\kappa_X} \sum_{k=1}^\infty\big\|  {\bm{\Lambda}}^{XY}_{k_1, k} \big\|_F^2\bigg)\bigg(\sum_{k_1'\ge 1} \sum_{k=1}^\infty\big\|  {\bm{\Lambda}}^{XY}_{k_1', k} \big\|_F^2\bigg)
    \;+\;\bigg(\sum_{k_1=1}^{\kappa_X} \sum_{k=1}^\infty\big\|  {\bm{\Lambda}}^{XY}_{k_1, k} \big\|_F^2\bigg)\bigg(\sum_{k_1'>\kappa_X} \sum_{k=1}^\infty\big\|  {\bm{\Lambda}}^{XY}_{k_1', k} \big\|_F^2\bigg)\\
    &\leq& 2\bigg(\sum_{k_1>\kappa_X} \sum_{k=1}^\infty\big\|  {\bm{\Lambda}}^{XY}_{k_1, k} \big\|_F^2\bigg)\bigg(\sum_{k_1\ge 1} \sum_{k=1}^\infty\big\|  {\bm{\Lambda}}^{XY}_{k_1, k} \big\|_F^2\bigg)\\
    &=& 2\bigg(\sum_{k_1>\kappa_X} \sum_{k=1}^\infty\operatorname{tr}\big\{{\bm{\Lambda}}^{XY}_{k_1, k}({\bm{\Lambda}}^{XY}_{k_1, k})^\top \big\}\bigg)
    \bigg(\sum_{k_1\ge 1} \sum_{k=1}^\infty\operatorname{tr}\big\{{\bm{\Lambda}}^{XY}_{k_1, k}({\bm{\Lambda}}^{XY}_{k_1, k})^\top \big\}\bigg).
\end{eqnarray*}
\endgroup
Here, we used \(\|AB^\top\|_F\le \|A\|_F\|B\|_F\), \(\|B^\top\|_F=\|B\|_F\), and the Cauchy--Schwarz inequality.

Similarly, we have
\begin{eqnarray*}
    \sum_{k_1=1}^{\kappa_X} \sum_{k_1'=1}^{\kappa_X}  \bigg\|\sum_{k>\kappa_Y}{\bm{\Lambda}}^{XY}_{k_1, k} ({\bm{\Lambda}}^{XY}_{k_1^{'}, k})^\top \bigg\|_F^2
   &\leq & 
  \sum_{k_1=1}^{\kappa_X} \sum_{k_1'=1}^{\kappa_X}  \bigg(\sum_{k>\kappa_Y}\big\|{\bm{\Lambda}}^{XY}_{k_1, k} ({\bm{\Lambda}}^{XY}_{k_1^{'}, k})^\top \big\|_F\bigg)^2\\
   &\leq & 
  \sum_{k_1=1}^{\kappa_X} \sum_{k_1'=1}^{\kappa_X}  \bigg(\sum_{k>\kappa_Y}\big\|{\bm{\Lambda}}^{XY}_{k_1, k} \big\|_F\big\|{\bm{\Lambda}}^{XY}_{k_1^{'}, k} \big\|_F\bigg)^2\\
   &\leq & 
  \bigg(\sum_{k_1=1}^{\kappa_X} \sum_{k>\kappa_Y}\big\|{\bm{\Lambda}}^{XY}_{k_1, k} \big\|_F^2\bigg)^2\\
  &\leq &
  \bigg( \sum_{k>\kappa_Y}\sum_{k_1\geq 1}\big\|{\bm{\Lambda}}^{XY}_{k_1, k} \big\|_F^2\bigg)^2\\
  &= &\bigg(\sum_{k>\kappa_Y}\sum_{k_1=1}^\infty \operatorname{tr}\big\{ ({\bm{\Lambda}}^{XY}_{k_1, k})^\top{\bm{\Lambda}}^{XY}_{k_1, k} \big\}\bigg)^2.
\end{eqnarray*}
Note that
\begin{align*}
    \frac{1}{pq} \sum_{k=1}^\infty\operatorname{tr}\big\{ {\bm{\Lambda}}^{XY}_{k_1, k}({\bm{\Lambda}}^{XY}_{k_1, k})^\top \big\}
    &=\frac{1}{pq}\operatorname{tr}({\bm{D}}_{k_1, k_1})\\
    &= \frac{1}{pq} \operatorname{tr}\bigg(\iint_0^1 \bm{L}_X(t_1, t_2)\, \wtilde{\psi}_{k_1}^X(t_1)\, \wtilde{\psi}_{k_1}^X(t_2) \, \mathrm{d}t_1 \mathrm{d}t_2\bigg) \\
    &=\iint_0^1 H_X(t_1,t_2)\, \wtilde{\psi}_{k_1}^X(t_1)\, \wtilde{\psi}_{k_1}^X(t_2) \, \mathrm{d}t_1 \mathrm{d}t_2\\
    &= \zeta_{k_1}^X,
\end{align*}
and
\begin{align*}
     \frac{1}{pq}\sum_{k_1=1}^\infty \operatorname{tr}\big\{ ({\bm{\Lambda}}^{XY}_{k_1, k})^\top{\bm{\Lambda}}^{XY}_{k_1, k} \big\}
     &= \frac{1}{pq}\operatorname{tr}({\bm{E}}_{k, k})\\
    &= \frac{1}{pq} \operatorname{tr}\bigg(\iint_0^1 \bm{L}_Y(t_1, t_2)\, \wtilde{\psi}_{k}^Y(t_1)\, \wtilde{\psi}_{k}^Y(t_2) \, \mathrm{d}t_1 \mathrm{d}t_2\bigg) \\
    &=\iint_0^1 H_Y(t_1,t_2)\, \wtilde{\psi}_{k}^Y(t_1)\, \wtilde{\psi}_{k}^Y(t_2) \, \mathrm{d}t_1 \mathrm{d}t_2\\
    &= \zeta_{k}^Y.
\end{align*}
Combining the above inequalities yields
\begin{align}\label{inter_eq}
    \iint_0^1 \bigg\|\bm{L}_X(t_1, t_2) - {\bm{L}}^{(\kappa_X,\kappa_Y)}_X(t_1, t_2)\bigg\|_F^2 \, \mathrm{d}t_1 \mathrm{d}t_2 
    &\leq 2\bigg(pq \sum_{k_1>\kappa_X} \zeta_{k_1}^X\bigg)\cdot \bigg(pq \sum_{k_1\geq 1} \zeta_{k_1}^X\bigg)  + \bigg(pq \sum_{k>\kappa_Y} \zeta_{k}^Y\bigg)^2.
\end{align}

Moreover, notice that
\[
\sum_{k_1\ge1}\zeta_{k_1}^X=\int_0^1 H_X(t,t)\,\mathrm{d}t,
\qquad
\sum_{k\ge1}\zeta_{k}^Y=\int_0^1 H_Y(t,t)\,\mathrm{d}t,
\]
In our setting, using \(H_X(t_1,t_2)=\operatorname{tr}\{\bm{L}_X(t_1,t_2)\}/(pq)\), we have
\begin{align*}
\int_0^1 H_X(t,t)\,\mathrm{d}t
&=\frac{1}{pq}\int_0^1 \operatorname{tr}\{\bm{L}_X(t,t)\}\,\mathrm{d}t
= \frac{1}{pq}\int_0^1 \operatorname{tr}\bigg\{\int_0^1 \bm{R}_{XY}(t,s)\bm{R}_{XY}^\top(t,s)\,\mathrm{d}s\bigg\}\,\mathrm{d}t \\
&= \frac{1}{pq}\iint_0^1 \operatorname{tr}\{\bm{R}_{XY}(t,s)\bm{R}_{XY}^\top(t,s)\}\,\mathrm{d}s\,\mathrm{d}t
= \frac{1}{pq}\iint_0^1 \|\bm{R}_{XY}(t,s)\|_F^2\,\mathrm{d}s\,\mathrm{d}t.
\end{align*}
Since \(\max_{1\le j\le p,\,1\le m\le q}\int_0^1\int_0^1\{R^{XY}_{jm}(t,s)\}^2\ \mathrm{d}t\mathrm{d}s\le M\) for some constant \(M>0\), we have \[\int_0^1\int_0^1\|\bm{R}_{XY}(t,s)\|_F^2\ \mathrm{d}t\mathrm{d}s=\sum_{j=1}^p\sum_{m=1}^q \int_0^1\int_0^1\{R^{XY}_{jm}(t,s)\}^2 \ \mathrm{d}t\mathrm{d}s\le pq\,M.\]
Therefore,
\[
\sum_{k_1\ge1}\zeta_{k_1}^X=\int_0^1 H_X(t,t)\,\mathrm{d}t
\le \frac{1}{pq}\iint_0^1 \|\bm{R}_{XY}(t,s)\|_F^2\,\mathrm{d}s\,\mathrm{d}t
\le M
=: C.
\]
A symmetric calculation gives \(\int_0^1 H_Y(t,t)\,\mathrm{d}t=\frac{1}{pq}\iint_0^1\|\bm{R}_{XY}(t,s)\|_F^2\,\mathrm{d}s\,\mathrm{d}t\), and thus
\(\sum_{k\ge1}\zeta_k^Y=\int_0^1 H_Y(t,t)\,\mathrm{d}t\le C\) as well.
Then \eqref{inter_eq} leads to
\begin{align*}
    \iint_0^1 \bigg\|\bm{L}_X(t_1, t_2) - {\bm{L}}^{(\kappa_X,\kappa_Y)}_X(t_1, t_2)\bigg\|_F^2 \, \mathrm{d}t_1 \mathrm{d}t_2 
    \leq C\cdot (pq)^2\bigg(\sum_{k_1>\kappa_X} \zeta_{k_1}^X + \sum_{k>\kappa_Y} \zeta_{k}^Y\bigg),
\end{align*}
for some constant \(C>0\).
We similarly {obtain} that
\begin{align*}
    \iint_0^1 \bigg\|\bm{L}_Y(t_1, t_2) - {\bm{L}}^{(\kappa_X,\kappa_Y)}_Y(t_1, t_2)\bigg\|_F^2 \, \mathrm{d}t_1 \mathrm{d}t_2 
    \leq C\cdot (pq)^2\bigg(\sum_{k_1>\kappa_X} \zeta_{k_1}^X + \sum_{k>\kappa_Y} \zeta_{k}^Y\bigg),
\end{align*}
{by symmetry of the above argument.}

Finally, notice that $\wtilde{\bm{A}}^X_r$ and $\bm{\wtilde{A}}_{r,\text{appr}}^{X}$ are the eigenfunctions of $\bm{L}_X(t_1,t_2)$ and $\bm{L}^{(\kappa_X,\kappa_Y)}_X(t_1,t_2)$, respectively. {It follows} that  
\begin{eqnarray}\label{eigen_gap}
    \|\wtilde{\bm{A}}^X_r- \bm{\wtilde{A}}_{r,\text{appr}}^{X}\|_p
    \leq C\cdot \frac{\|\mathcal{L}_X-{\mathcal{L}}^{(\kappa_X,\kappa_Y)}_X\|_{\infty}}{\inf_{r^{'}\neq r}|\lambda_r^2-\lambda_{r^{'}}^2|},
\end{eqnarray}
according to Lemma 4.3 in \citet{bosq2000linear}, where $C$ is a constant, and $\mathcal{L}_X$ and ${\mathcal{L}}^{(\kappa_X,\kappa_Y)}_X$ are the integral operators induced by $\bm{L}_X(t_1,t_2)$ and ${\bm{L}}^{(\kappa_X,\kappa_Y)}_X(t_1,t_2)$, respectively.

Notice that
\begin{eqnarray*}
  && \|\mathcal{L}_X-{\mathcal{L}}^{(\kappa_X,\kappa_Y)}_X\|_{\infty}^2\\
   &=&\sup_{\bm{f}\in \mathcal{S}_p}\bigg\|\int_{0}^1 \big\{\bm{L}_X(\cdot,t_2)-{\bm{L}}^{(\kappa_X,\kappa_Y)}_X(\cdot,t_2)\big\}  \bm{f}(t_2)\ \mathrm{d}t_2\bigg\|_p^2\\
   &\le& \sup_{\bm{f}\in \mathcal{S}_p}\int_{0}^1\bigg(\int_{0}^1 \big\|\bm{L}_X(t_1,t_2)-{\bm{L}}^{(\kappa_X,\kappa_Y)}_X(t_1,t_2)\big\|_F\,\|\bm{f}(t_2)\|_E\,\mathrm{d}t_2\bigg)^2\mathrm{d}t_1\\
   &\le& \sup_{\bm{f}\in \mathcal{S}_p}\int_{0}^1\bigg(\int_{0}^1 \big\|\bm{L}_X(t_1,t_2)-{\bm{L}}^{(\kappa_X,\kappa_Y)}_X(t_1,t_2)\big\|_F^2\,\mathrm{d}t_2\bigg)
   \bigg(\int_{0}^1 \|\bm{f}(t_2)\|^2_E\,\mathrm{d}t_2\bigg)\mathrm{d}t_1\\
   &=& \iint_0^1 \big\|\bm{L}_X(t_1,t_2)-{\bm{L}}^{(\kappa_X,\kappa_Y)}_X(t_1,t_2)\big\|_F^2\,\mathrm{d}t_1\mathrm{d}t_2\\
   &\le& C\cdot (pq)^2\bigg(\sum_{k_1>\kappa_X}\zeta_{k_1}^X+\sum_{k>\kappa_Y}\zeta_k^Y\bigg),
\end{eqnarray*}
where the second inequality uses the Cauchy--Schwarz inequality and \(\int_0^1\|\bm{f}(t)\|_E^2\  \mathrm{d}t=1\) for \(\bm{f}\in\mathcal{S}_p\), and the last inequality follows from the previously established bound on \(\iint\|\bm{L}_X-{\bm{L}}^{(\kappa_X,\kappa_Y)}_X\|_F^2\).

Combining the above two inequalities, we have
\begin{eqnarray*}
   \|\wtilde{\bm{A}}^X_r- \bm{\wtilde{A}}_{r,\text{appr}}^{X}\|_p
   &\leq& \frac{C\cdot pq}{\inf_{r^{'}\neq r}|\lambda_r^2-\lambda_{r^{'}}^2|}
  \sqrt{{\sum_{k_1>\kappa_X} \zeta_{k_1}^X +  \sum_{k>\kappa_Y} \zeta_{k}^Y}},
\end{eqnarray*}
We similarly {derive} that
\begin{eqnarray*}
   \|\wtilde{\bm{A}}^Y_r- \bm{\wtilde{A}}_{r,\text{appr}}^{Y}\|_q
   &\leq& \frac{C\cdot pq}{\inf_{r^{'}\neq r}|\lambda_r^2-\lambda_{r^{'}}^2|}
  \sqrt{{\sum_{k_1>\kappa_X} \zeta_{k_1}^X +  \sum_{k>\kappa_Y} \zeta_{k}^Y}}.
\end{eqnarray*}
\end{proof}

\begin{remark}
We can control the error between singular values in a similar way:
By Theorem~4.2.8 in \citet{hsing2015theoretical},
\begin{eqnarray}\label{eigen_value_gap}
    |\lambda_r^2-\eta_r^2|\leq \|\mathcal{L}_X-{\mathcal{L}}^{(\kappa_X,\kappa_Y)}_X\|_{\infty}
   \leq C\cdot pq\cdot\sqrt{\sum_{k_1>\kappa_X} \zeta_{k_1}^X + \sum_{k_2>\kappa_Y} \zeta_{k_2}^Y},
\end{eqnarray}
for any $r\geq 1$.
    
\end{remark}

\

\
\section{Additional Results}

{
\subsection{Algorithms}\label{algo_sec}

It is worth noting that optimization problem~\eqref{opt_SVD} is equivalent to
\begin{align*}
\min_{\substack{
\bm a\in\mathbb R^{p\kappa_X},\,
\bm b\in\mathbb R^{q\kappa_Y}
}}
\quad &
\bigl\|\widehat{\bm\Gamma}-\bm a\bm b^\top\bigr\|_F^2
+
\varrho_X
\sum_{j=1}^p
\sqrt{
\sum_{k=1}^{\kappa_X}
\bigl(a_j^{(k)}\bigr)^2
}
+
\varrho_Y
\sum_{m=1}^q
\sqrt{
\sum_{k=1}^{\kappa_Y}
\bigl(b_m^{(k)}\bigr)^2
}
\\
\text{subject to}
\quad &
\bm a^\top\bm a= 1,
\qquad
\bm b^\top\bm b= 1,
\end{align*}
where \(a_j^{(k)}\) and \(b_m^{(k)}\) are the \(j\)th and \(m\)th elements of \(\bm a^{(k)}\) and \(\bm b^{(k)}\), respectively. Define the feature-specific coefficient groups
\(
(\bm a)_j
=
\bigl(
a_j^{(1)},\ldots,a_j^{(\kappa_X)}
\bigr)^\top,
\ j=1,\ldots,p,
\)
and
\(
(\bm b)_m
=
\bigl(
b_m^{(1)},\ldots,b_m^{(\kappa_Y)}
\bigr)^\top,
\ m=1,\ldots,q.
\)
The two penalty terms have the form of group-Lasso penalties
\citep{hastie2009elements}, with each group corresponding to one feature.
We therefore employ an iterative matrix factorization algorithm in the
spirit of \citet{lin2013group} to solve~\eqref{opt_SVD}.

Let
\(
\widehat{\bm\Gamma}
=
\bm U\bm D\bm V^\top
\)
be the singular value decomposition of
\(\widehat{\bm\Gamma}\). We initialize the algorithm using the leading
left and right singular vectors,
\(
\widehat{\bm a}_{1,(0)}
=
\bm U_{\cdot 1},
\
\widehat{\bm b}_{1,(0)}
=
\bm V_{\cdot 1},
\)
which correspond to the leading directions of the unpenalized matrix
decomposition. The complete procedure is given in
Algorithm~\ref{algo:matrix_fac}.

\begin{algorithm}[ht!]
\caption{Iterative Matrix Factorization}
\label{algo:matrix_fac}
\begin{algorithmic}[1]
\State \textbf{Input:}
Matrix \(\widehat{\bm\Gamma}\), tuning parameters
\(\varrho_X\) and \(\varrho_Y\), initial vectors
\(\widehat{\bm a}_{1,(0)}\) and
\(\widehat{\bm b}_{1,(0)}\), convergence tolerance
\(\epsilon\), and maximum number of iterations \(U_{\max}\).
\State Set \(u=0\).
\Repeat
    \State Set \(u=u+1\).

    \State \textbf{Right-to-left multiplication:}
    \(
    \widehat{\bm a}_1
    =
    \widehat{\bm\Gamma}
    \widehat{\bm b}_{1,(u-1)}.
    \)

    \State \textbf{Left thresholding:}
    For \(j=1,\ldots,p\), update
    \[
    (\widehat{\bm a}_1)_j
    =
    (\widehat{\bm a}_1)_j
    \max\left\{
    1-
    \frac{\varrho_X}
    {2\|(\widehat{\bm a}_1)_j\|_E},
    0
    \right\},
    \]
    where
    \((\widehat{\bm a}_1)_j=\bm 0\) if
    \(\|(\widehat{\bm a}_1)_j\|_E=0\).

    \State \textbf{Left normalization:}
    If \(\|\widehat{\bm a}_1\|_E\neq0\), set
    \(
    \widehat{\bm a}_{1,(u)}
    =
    \frac{\widehat{\bm a}_1}
    {\|\widehat{\bm a}_1\|_E};
    \)
    otherwise, set
    \(\widehat{\bm a}_{1,(u)}=\bm 0\).

    \State \textbf{Left-to-right multiplication:}
    \(
    \widehat{\bm b}_1
    =
    \widehat{\bm\Gamma}^\top
    \widehat{\bm a}_{1,(u)}.
    \)

    \State \textbf{Right thresholding:}
    For \(m=1,\ldots,q\), update
    \[
    (\widehat{\bm b}_1)_m
    =
    (\widehat{\bm b}_1)_m
    \max\left\{
    1-
    \frac{\varrho_Y}
    {2\|(\widehat{\bm b}_1)_m\|_E},
    0
    \right\},
    \]
    where
    \((\widehat{\bm b}_1)_m=\bm 0\) if
    \(\|(\widehat{\bm b}_1)_m\|_E=0\).

    \State \textbf{Right normalization:}
    If \(\|\widehat{\bm b}_1\|_E\neq0\), set
    \(
    \widehat{\bm b}_{1,(u)}
    =
    \frac{\widehat{\bm b}_1}
    {\|\widehat{\bm b}_1\|_E};
    \)
    otherwise, set
    \(\widehat{\bm b}_{1,(u)}=\bm 0\).

    \State Compute
    \[
    e_u
    =
    \frac{
    \|\widehat{\bm a}_{1,(u)}
    -
    \widehat{\bm a}_{1,(u-1)}\|_E
    }{
    \|\widehat{\bm a}_{1,(u-1)}\|_E
    }
    +
    \frac{
    \|\widehat{\bm b}_{1,(u)}
    -
    \widehat{\bm b}_{1,(u-1)}\|_E
    }{
    \|\widehat{\bm b}_{1,(u-1)}\|_E
    },
    \]
    with a ratio set to zero when its denominator is zero.
\Until{\(e_u\leq\epsilon\) or \(u=U_{\max}\).}
\State \textbf{Output:}
The final vectors
\(\widehat{\bm a}_{1,(u)}\) and
\(\widehat{\bm b}_{1,(u)}\).
\end{algorithmic}
\end{algorithm}

We evaluate Algorithm~\ref{algo:matrix_fac} using 100 independently
generated datasets for each setting with
\(
n=100,
\
q=p,
\
\frac{p}{n}\in\{0.5,1,1.5\},
\)
following the simulation setup in Part~\ref{sec:simu_setup}. We set
\(
\varrho_X=\varrho_Y=0.02,
\
\epsilon=10^{-3},
\
U_{\max}=200.
\)
For each dataset, the solution obtained from the SVD initialization is
denoted by
\(
\bigl(
\widehat{\bm a}_{\mathrm{SVD}},
\widehat{\bm b}_{\mathrm{SVD}}
\bigr).
\)
We record the total fitting time and the number of alternating iterations
required for this run to converge.

To assess sensitivity to initialization, we construct 20 perturbed SVD
initializations for each dataset. For \(g=1,\ldots,20\), we independently
generate standard Gaussian vectors \(\bm z_{a,g}\) and \(\bm z_{b,g}\),
and define
\(
\bm a_g^{(0)}
=
\frac{
\widehat{\bm a}_{1,(0)}
+
0.1\,\bm z_{a,g}/\|\bm z_{a,g}\|_E
}{
\left\|
\widehat{\bm a}_{1,(0)}
+
0.1\,\bm z_{a,g}/\|\bm z_{a,g}\|_E
\right\|_E
},
\)
and
\(
\bm b_g^{(0)}
=
\frac{
\widehat{\bm b}_{1,(0)}
+
0.1\,\bm z_{b,g}/\|\bm z_{b,g}\|_E
}{
\left\|
\widehat{\bm b}_{1,(0)}
+
0.1\,\bm z_{b,g}/\|\bm z_{b,g}\|_E
\right\|_E
}.
\)
Among the converged and nondegenerate solutions obtained from these
perturbed initializations, we select the one with the smallest penalized
objective value in~\eqref{opt_SVD}, denoted by
\(
\bigl(
\widehat{\bm a}_{\mathrm{pert}},
\widehat{\bm b}_{\mathrm{pert}}
\bigr).
\)
Initialization stability is measured by the rank-one relative difference
\[
\mathrm{RD}
=
\frac{
\left\|
\widehat{\bm a}_{\mathrm{SVD}}
\widehat{\bm b}_{\mathrm{SVD}}^\top
-
\widehat{\bm a}_{\mathrm{pert}}
\widehat{\bm b}_{\mathrm{pert}}^\top
\right\|_F
}{
\left\|
\widehat{\bm a}_{\mathrm{SVD}}
\widehat{\bm b}_{\mathrm{SVD}}^\top
\right\|_F
}
\cdot 100\%.
\]
This rank-one comparison is invariant to simultaneous sign changes in
the two loading vectors.

Figure~\ref{FACD_com}
summarizes the distributions of the runtime, number of iterations, and
rank-one relative difference across the 100 simulation replications.
All experiments are conducted on a high-performance computing system with 40 CPU cores and 208 GB of RAM.

\begin{figure}[htbp]
\begin{center}
\includegraphics[width=0.8\textwidth]{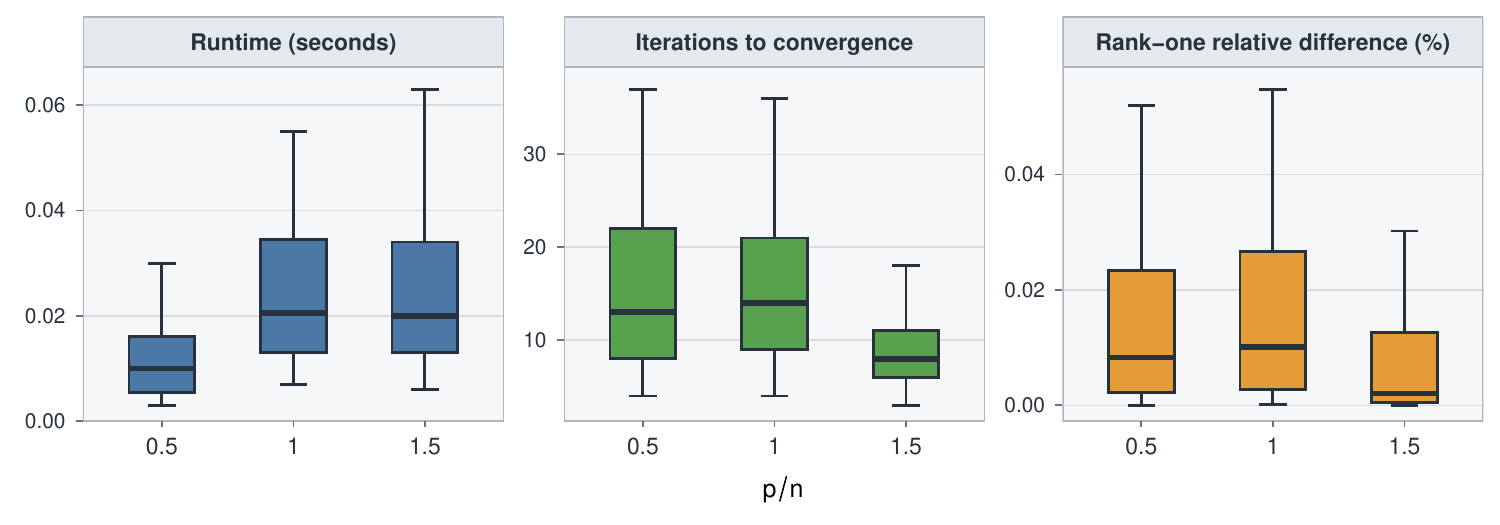}
\end{center}
\caption{
Boxplots of the runtime, number of iterations to convergence, and rank-one relative difference for solving~\eqref{opt_SVD} over 100 simulation replications. We consider settings with \(n=100\), \(q=p\), and \(p/n\in\{0.5,1,1.5\}\), and set \(\varrho_X=\varrho_Y=0.02\) throughout. The rank-one relative difference is defined as
\(
{
\left\|
\widehat{\bm a}_{\mathrm{SVD}}
\widehat{\bm b}_{\mathrm{SVD}}^\top
-
\widehat{\bm a}_{\mathrm{pert}}
\widehat{\bm b}_{\mathrm{pert}}^\top
\right\|_{\mathrm F}
} / {
\left\|
\widehat{\bm a}_{\mathrm{SVD}}
\widehat{\bm b}_{\mathrm{SVD}}^\top
\right\|_{\mathrm F}
}
\cdot 100\%,
\)
where
\((\widehat{\bm a}_{\mathrm{SVD}},\widehat{\bm b}_{\mathrm{SVD}})\)
denotes the solution obtained using the SVD initialization, and
\((\widehat{\bm a}_{\mathrm{pert}},\widehat{\bm b}_{\mathrm{pert}})\)
denotes the best converged solution among those obtained using perturbed SVD initializations.
All experiments were conducted on a high-performance computing system with 40 CPU cores and 208 GB of RAM.
}
\label{FACD_com}
\end{figure}
}

\

\subsection{Simulation Setup}\label{sec:simu_setup}
We generate the functional loadings as
$A^X_{rj}=\sum_{k=1}^{10} a_{r,j,k}\psi_k$ and
$A^Y_{rm}=\sum_{k=1}^{10} b_{r,m,k}\psi_k$,
for $r=1,\dots,R$, $j=1,\dots,p$, and $m=1,\dots,q$, respectively,
where $\{\psi_k; k\leq 10\}$ are the first ten non-constant Fourier basis functions.
Define
$$\bm{a}_r := (a_{r,1,1}, a_{r,1,2}, \ldots, a_{r,1,10}, \dots, a_{r,p,1}, a_{r,p,2}, \ldots, a_{r,p,10})^\top \in \mathbb{R}^{10p}$$
and
$\bm{b}_r := (b_{r,1,1}, b_{r,1,2}, \ldots, b_{r,1,10}, \dots, b_{r,q,1}, b_{r,q,2}, \ldots, b_{r,q,10})^\top \in \mathbb{R}^{10q}$,
which are sparse vectors and are orthonormal across different $r$’s.

Specifically, we first independently generate
$a_{r,j,k} \sim \operatorname{Unif}\big([-\exp\{1.5(11-k)\}, -\exp(11-k)] \cup [\exp(11-k), \exp\{1.5(11-k)\}]\big)$
for $r\leq R$, $j\leq 10$, and $k\leq 10$, and set $a_{r,j,k}=0$ for $j>10$; the coefficients $b_{r,m,k}$ are generated similarly.
We then orthonormalize $\bm{a}_r$s and $\bm{b}_r$s using the Gram--Schmidt process.
This setup ensures that the functional loadings are orthogonal across different $r$’s, i.e.,
$\sum_{j=1}^p \langle A^X_{r_1 j}, A^X_{r_2 j} \rangle = \mathbb{I}(r_1 = r_2)$ and
$\sum_{m=1}^q \langle A^Y_{r_1 m}, A^Y_{r_2 m} \rangle = \mathbb{I}(r_1 = r_2)$,
and that the nonzero functional loadings are concentrated in the first 10 variables in both datasets.

We generate the random scores $(z_{ir}^X,z_{ir}^Y)^\top$, $i=1,\dots,n$ independently from a mean-zero Gaussian distribution with covariance matrix 
$$\begin{pmatrix}
  \sigma_r^2 & \sigma_r^2\rho_r \\
  \sigma_r^2\rho_r & \sigma_r^2 \\
\end{pmatrix}$$ for $r=1,\dots, R$ independently. Here, the variance $\sigma_r^2$ and the correlation $\rho_r$ are taken as $0.5\cdot r ^ {-2}$ and $0.8\cdot r^{-2}$.
The noise terms $\tau^X_{ijg}$s and $\tau^Y_{imh}$s are generated in a way such that they are independent across different $i$ but dependent across different features and times.
Specifically, we set $\tau_{ijg}^X=\sum_{k=1}^3\vartheta_{ijk}\psi_k(T_{ig})+\varepsilon_{ijg}$, $i=1,\dots,n$, $j=1,\dots,p$, and $g=1,\dots,N_i^X$, where $\bm{\vartheta_{ik}}:=(\vartheta_{i1k},\dots,\vartheta_{ipk})^\top\in \mathbb{R}^p$ are random vectors independent across $i$'s and $k$'s and drawn from a mean-zero Gaussian distribution with the covariance matrix $\bm{\Sigma}_{\vartheta}$, where $\left[\bm{\Sigma}_{\vartheta}\right]_{j_1,j_2}=\frac{1}{6}\sigma_{\text{main}}^2\exp\big\{-(j_1-j_2)^2/10\big\}$ for $1\leq j_1,j_2\leq p$, and  $\varepsilon_{ijg}$s are independent mean zero Gaussian random variables with a standard deviation of $\sigma_{\text{main}}\times 10\%$. Here, $\sigma_{\text{main}}$ is the standard deviation of the main signal level of the model, defined by $$\sqrt{\frac{1}{10}\sum_{j=1}^{10}\frac{1}{\sum_{i=1}^nN_i^X}\sum_{i=1}^n\sum_{g=1}^{N_i^X}(\sum_{r=1}^RA^X_{rj}(T_{ig}) z_{ir}^X)^2}.$$
The noise terms ${\tau}_{imh}^Y$ are drawn similarly. 

\begin{figure}[ht!]
\begin{center}
\includegraphics[scale = 0.7]{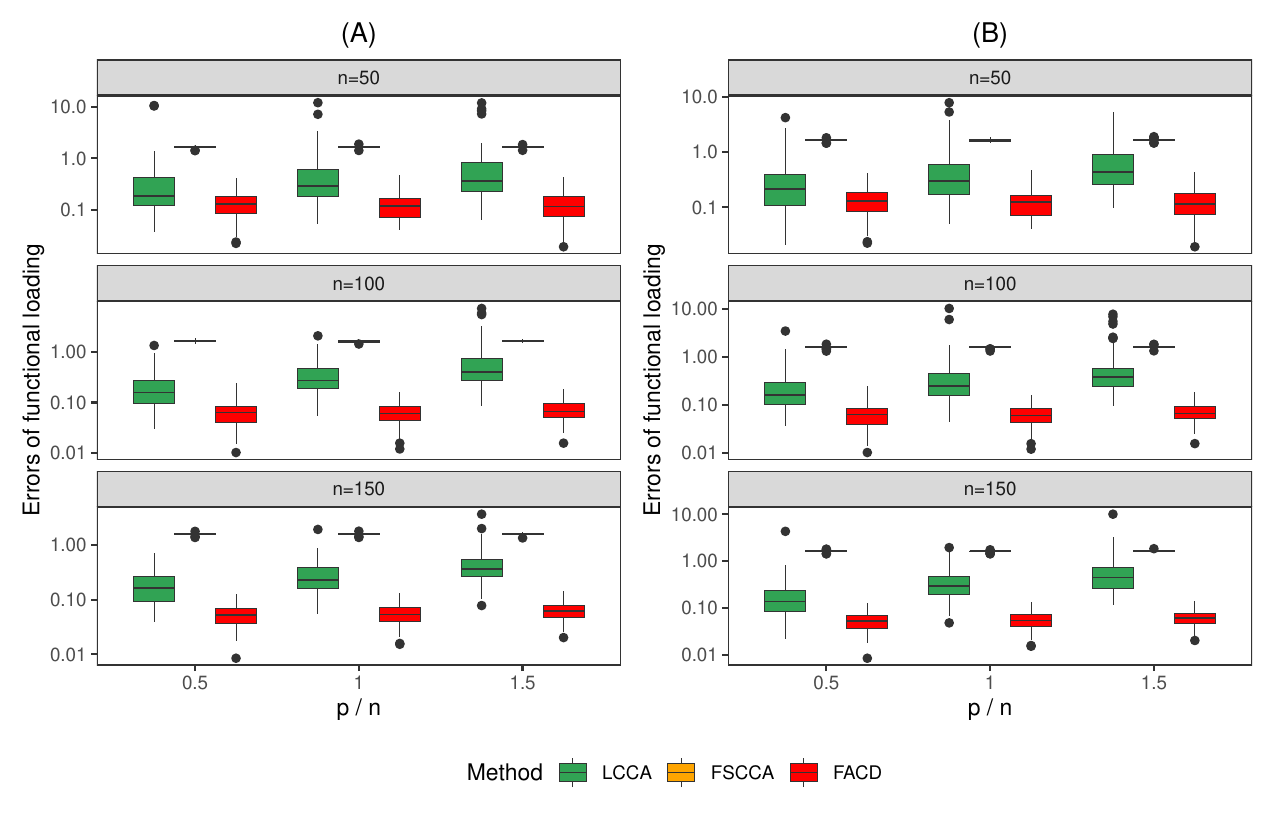}
\end{center}
\caption{Boxplot of estimation errors for functional loadings from the first (Panel A) and second (Panel B) datasets.} \label{err_functional_loading_1}
\end{figure}

\

\subsection{Simulations for Linear Associations}\label{sec:linear}

In this section, we illustrate the simulation results for functional loadings that are linear functions. 
To this end, we modify $\{\psi_k : k \leq 10\}$ to be constant and linear functions for the generation of functional loadings, given by 
\[
A^X_{rj}(t) = a_{rj1} + a_{rj2}\, t 
\quad \text{and} \quad 
A^Y_{rm}(t) = b_{rm1} + b_{rm2}\, t.
\]
The vectors $\bm{a}_r := (a_{r11}, a_{r12}, \dots, a_{rp1}, a_{rp2})^\top \in \mathbb{R}^{2p}$ 
and $\bm{b}_r := (b_{r11}, b_{r12}, \dots, b_{rq1}, b_{rq2})^\top \in \mathbb{R}^{2q}$, 
along with the other settings, are generated in the same manner as described in 
Part~\ref{sec:simu_setup}. The estimation errors for the linear functional loadings are given in Figure~\ref{err_functional_loading_1}. 

\

\subsection{Correlations between Estimated and True Functional Scores.}\label{sec: corre}

We measure the estimation accuracy of functional scores using $\widehat{\operatorname{Cor}}(z_{ir}^X, \hat{z}_{ir}^X)$ and $\widehat{\operatorname{Cor}}(z_{ir}^Y, \hat{z}_{ir}^Y)$, where $\hat{z}_{ir}^X$s and $\hat{z}_{ir}^Y$s are the estimated scores and $\widehat{\operatorname{Cor}}(\cdot, \cdot)$ denotes the empirical correlation. The correlation is given in Table~\ref{cor_table}.

\begin{table}[h]
\caption{Correlations between the estimated and true functional scores.}\label{cor_table}
\centering
\renewcommand{\arraystretch}{1.3}
\setlength\tabcolsep{6pt}
\begin{tabular}{c|c|ccc|ccc}
  \hline
\multicolumn{2}{c|}{ \multirow{2}{*}{Correlation}}
   &  \multicolumn{3}{c|}{The first dataset} &  \multicolumn{3}{c}{The second dataset}\\
 \multicolumn{2}{c|}{}   & $p=0.5n$ & $p=n$ & $p=1.5n$ &  $p=0.5n$ & $p=n$ & $p=1.5n$  \\ 
  \hline
  \multirow{5}{*}{$n=50$} &
LCCA & 0.23 & 0.22 & 0.19 & 0.24 & 0.21 & 0.20 \\ 
   & SCCA & 0.18 & 0.18 & 0.18 & 0.15 & 0.16 & 0.18 \\ 
   & MFPCA & 0.92 & 0.87 & 0.84 & 0.95 & 0.88 & 0.84 \\ 
   & FSCCA & 0.92 & 0.92 & 0.91 & 0.92 & 0.92 & 0.92 \\ 
   & FACD & 0.94 & 0.94 & 0.94 & 0.94 & 0.95 & 0.94 \\ 
       \hline
         \multirow{5}{*}{$n=100$} &
 LCCA & 0.16 & 0.18 & 0.15 & 0.15 & 0.18 & 0.15 \\ 
   & SCCA & 0.11 & 0.11 & 0.12 & 0.10 & 0.12 & 0.11 \\ 
   & MFPCA & 0.93 & 0.85 & 0.81 & 0.89 & 0.86 & 0.82 \\ 
   & FSCCA & 0.93 & 0.93 & 0.92 & 0.92 & 0.92 & 0.92 \\ 
 & FACD & 0.95 & 0.95 & 0.95 & 0.95 & 0.95 & 0.95 \\  
       \hline
         \multirow{5}{*}{$n=150$} &
LCCA & 0.16 & 0.13 & 0.12 & 0.17 & 0.14 & 0.12 \\ 
   & SCCA & 0.11 & 0.09 & 0.10 & 0.10 & 0.09 & 0.09 \\ 
 & MFPCA & 0.92 & 0.89 & 0.89 & 0.92 & 0.93 & 0.87 \\ 
   & FSCCA & 0.93 & 0.93 & 0.93 & 0.92 & 0.93 & 0.93 \\ 
 & FACD & 0.95 & 0.95 & 0.95 & 0.95 & 0.95 & 0.95 \\ 
       \hline
\end{tabular}
\end{table}

\

{
\subsection{Recovery of FACD Components and Overall Cross-covariance}
\label{sec: component_covariance}

We further investigate the recovery of the leading FACD components and the overall cross-covariance structure as the sample size increases. We consider sample sizes \(n\in\{50,100,150,\allowbreak 200\}\) and set \(p=q=1.5n\). For each setting, we generate 100 independent datasets using the same data-generating mechanism as in Part~\ref{sec:simu_setup}, and then fit FACD to these datasets.

Let \(\bm A_r^X=(A_{r1}^X,\ldots,A_{rp}^X)^\top\) and \(\bm A_r^Y=(A_{r1}^Y,\ldots,A_{rq}^Y)^\top\) denote the true loading functions associated with the \(r\)th component, and let \(z_{ir}^X\) and \(z_{ir}^Y\) denote the corresponding subject-specific scores. We first evaluate the recovery accuracy of the first two components. For \(r=1,2\), the recovery errors on the \(X\)- and \(Y\)-sides are defined as
\begin{align*}
\operatorname{Err}_{X,r}
&=
\frac{1}{n}
\sum_{i=1}^n
\sum_{j=1}^p
\left\|
z_{ir}^X A_{rj}^X
-
\widehat{z}_{ir}^X\widehat{A}_{rj}^X
\right\|_{L^2}^2,\\
\operatorname{Err}_{Y,r}
&=
\frac{1}{n}
\sum_{i=1}^n
\sum_{m=1}^q
\left\|
z_{ir}^Y A_{rm}^Y
-
\widehat{z}_{ir}^Y\widehat{A}_{rm}^Y
\right\|_{L^2}^2.
\end{align*}
We then report the combined component recovery error
\[
\operatorname{Err}_{\mathrm{comp},r}
=
\operatorname{Err}_{X,r}
+
\operatorname{Err}_{Y,r}.
\]

We next evaluate the recovery of the overall cross-covariance structure. Let \(\bm R_{XY}\) denote the true cross-covariance, with its \((j,m)\)th element defined as $R_{jm}^{XY}(s,t)$. The corresponding FACD estimate is denoted by
\[
\widehat{\bm R}_{XY}
=
\left\{
\widehat R_{jm}^{XY}(s,t):
j=1,\ldots,p,\;
m=1,\ldots,q
\right\},
\]
with
\(
\widehat R_{jm}^{XY}(s,t)
=
\sum_{r=1}^{\widehat R}
\widehat\eta_{r}
\widehat A_{rj}^X(s)
\widehat A_{rm}^Y(t).
\)
To select the estimated rank, we first order the estimated singular values such that
\[
\left|\widehat\eta_{1}\right|
\geq
\cdots
\geq
\left|\widehat\eta_{R_{\max}}\right|,
\qquad
R_{\max}=5,
\]
and apply the adjacent singular-value ratio criterion
\[
\widehat R
=
\underset{1\leq r<R_{\max}}{\arg\max}
\frac{\left|\widehat\eta_{r}\right|}
{\left|\widehat\eta_{r+1}\right|}.
\]
The overall absolute cross-covariance recovery error is then defined as
\[
\operatorname{Err}_{\mathrm{cov}}
=
\sum_{j=1}^{p}\sum_{m=1}^{q}
\int_0^1\int_0^1
\left\{
R_{jm}^{XY}(s,t)
-
\widehat R_{jm}^{XY}(s,t)
\right\}^2
\,\mathrm{d}s\,\mathrm{d}t.
\]

Figure~\ref{fig:FACD_recovery} presents the mean recovery errors over the 100 replications, together with their \(95\%\) confidence intervals. We observe that the errors decrease as the sample size increases in all three cases, providing empirical evidence for the statistical consistency of FACD in recovering both the leading components and the overall cross-covariance structure.

\begin{figure}[t]
\centering
\includegraphics[width=\textwidth]{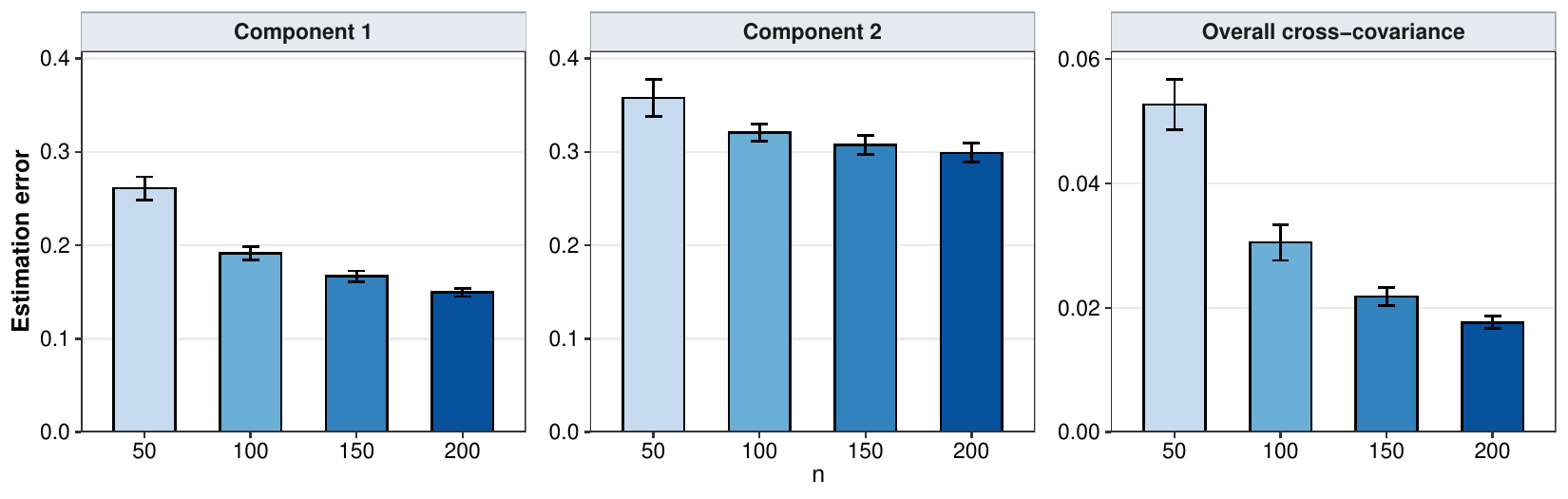}
\caption{Recovery of the first two FACD components and the overall cross-covariance structure. Bars represent the mean errors over 100 replications, and error bars represent \(95\%\) confidence intervals.}
\label{fig:FACD_recovery}
\end{figure}
}

\

\subsection{Processing of Lipid Data}\label{sec: lipid_proc}

The lipid features can be grouped into 13 types:
Cholesteryl Ester (CE),
Ceramide (CER),
Diacylglycerol (DAG),
Dihydroceramide (DCER),
Free Fatty Acid (FFA),
Hexosylceramide (HCER),
Lactosylceramide (LCER),
Lysophosphatidylcholine (LPC), 
Lysophosphatidyletha-nolamine (LPE),
Phosphatidylcholine (PC),
Phosphatidylethanolamine (PE),
Sphingomyelin (SM),
Triacylglycerol (TAG). For TAG, lipid features are further categorized into mid-chain fatty acid (MCFA, $\leq$ 12 carbons), very long-chain fatty acid (VLCFA, $\geq $22 carbons), and long-chain fatty acid (LCFA, between 13 and 21 carbons) that is subcategorized based on their degree of saturation. 
The number of lipid features in each group is summarized in Tables~\ref{tab:lipid_num_ori}.

\begin{table}[htbp]
\caption{Numbers of lipid features in each lipid group}
\label{tab:lipid_num_ori}
\footnotesize
\centering
\begin{tabular}{c|cccccc}
\hline
Cluster & {CE} & {CER} & {DAG} & {DCER} & {FFA} & {HCER} \\ 
\hline
Number & 25 & 9 & 35 & 4 & 26 & 7 \\ 
\hline
\hline
Cluster & {LCER} & {LPC} & {LPE} & {PC} & {PE} & {SM}  \\ 
\hline
Number & 6 & 11 & 6 & 51 & 38 & 12  \\ 
\hline
\hline
\multirow{2}{*}{Cluster} & \multirow{2}{*}{{TAG MCFA}} & {TAG LCFA}  & {TAG LCFA}  & {TAG LCFA}  & \multirow{2}{*}{{TAG VLCFA}} & \\
& & {unsaturated} & {monosaturated} & {polysaturated} & &\\
\hline
Number & 12 & 35 & 63 & 325 & 45 &\\ 
\hline
\end{tabular}
\end{table}

For each lipid group, we use samples from all subjects and time points to perform principal component analysis (PCA) and keep the top principal components explaining 95\% of the variance. The signs of the principal component scores are arranged such that the principal component loadings have a positive inner product with the constant vector. The number of features after PCA is summarized in Tables~\ref{tab:lipid_num_pca}. 

\begin{table}[htbp]
\caption{Numbers of Principal Components in Each Cluster of Lipid}
\label{tab:lipid_num_pca}
\footnotesize
\centering
\begin{tabular}{c|cccccc}
\hline
Cluster & {CE} & {CER} & {DAG} & {DCER} & {FFA} & {HCER} \\ 
\hline
Number & 10 & 6 & 15 & 3 & 10 & 5 \\ 
\hline
\hline
Cluster & {LCER} & {LPC} & {LPE} & {PC} & {PE} & {SM}  \\ 
\hline
Number & 4 & 6 & 3 & 22 & 13 & 6  \\ 
\hline
\hline
\multirow{2}{*}{Cluster} & \multirow{2}{*}{{TAG MCFA}} & {TAG LCFA}  & {TAG LCFA}  & {TAG LCFA}  & \multirow{2}{*}{{TAG VLCFA}} & \\
& & {unsaturated} & {monosaturated} & {polysaturated} & &\\
\hline
Number & 2 & 3 & 4 & 6 & 4 &\\ 
\hline
\end{tabular}
\end{table}

\

\subsection{Comparison Across Competing Methods in Data Analysis}\label{sec:com_data_com}

\begin{figure}[htbp]
\begin{center}
\includegraphics[width=0.85\textwidth]{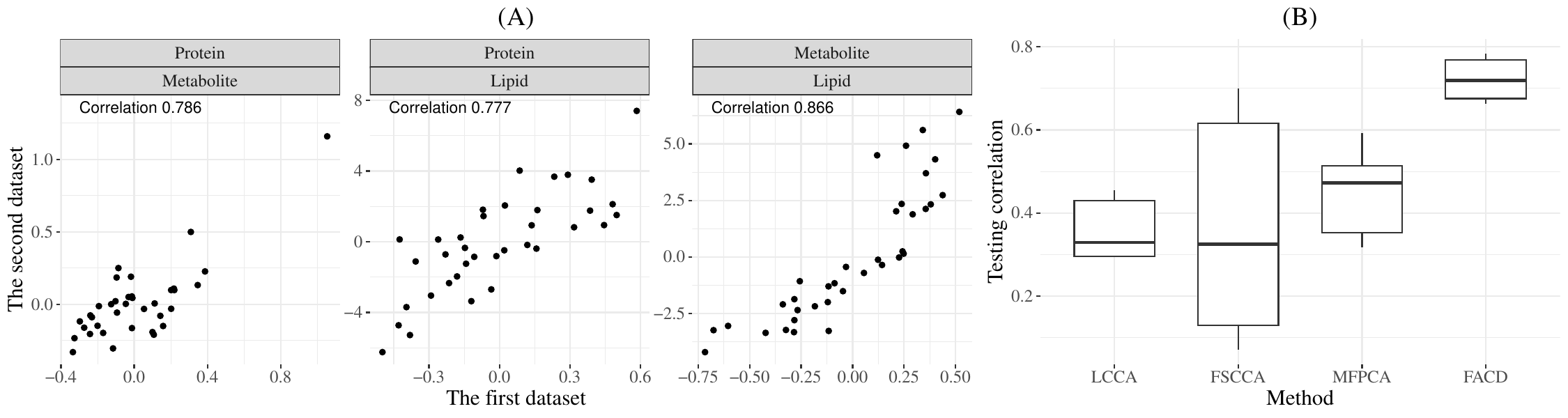}
\end{center}
\caption{(A) The extracted first canonical scores of the participants between any two omics types of high-dimensional longitudinal data, with correlations calculated using the Pearson correlation coefficient. (B) Boxplot of Pearson correlations between the scores of proteome and metabolome for each testing fold.}\label{extracted_score} 
\end{figure}

{
\subsection{Sensitivity Analysis}
\label{sec: sensitivity}

We assess the stability of the estimated functional loadings under perturbations of the observed time points of datasets. The analysis is based on the protein and metabolite modalities considered in the real-data application and focuses on the first FACD component. Let $\widehat A_j^{\mathrm{full}}$ denote the functional loading of feature $j$ estimated from the full dataset, and let $\mathcal S^{\mathrm{full}}$ denote the corresponding set of selected features.

We generate $B=500$ perturbed datasets through repeated time-point subsampling. For each repetition $b=1,\ldots,B$ and each subject, we independently determine whether its longitudinal observations are perturbed, with probability $0.1$. If a subject is selected for perturbation, we randomly retain $80\%$ of its available observation times without replacement, subject to retaining at least three time points. Otherwise, all of its original observation times are retained. The same retained time points are used for the protein and metabolite measurements of a given subject, and all subjects are retained throughout the analysis. Given the resampled dataset, we refit FACD to each perturbed dataset using the same tuning and estimation procedure as in the original analysis, and denote the resulting loading of feature $j$ in repetition $b$ by $\widehat A_j^{(b)}$ and the corresponding selected feature set by $\mathcal S^{(b)}$.

We evaluate stability from two complementary perspectives. First, for each feature $j\in\mathcal S^{\mathrm{full}}$, we compute its selection frequency
\[
\text{SF}
=
\frac{1}{B}
\sum_{b=1}^B
\mathbb I\bigl\{j\in\mathcal S^{(b)}\bigr\},
\]
which measures how frequently a feature selected from the full dataset is reselected after perturbing the observation times. Second, conditional on feature $j$ being reselected in repetition $b$, we compute the feature loading alignment
\[
\operatorname{Align}_j^{(b)}
=
\frac{
\left|
\left\langle
\widehat A_j^{\mathrm{full}},
\widehat A_j^{(b)}
\right\rangle
\right|
}{
\left\|
\widehat A_j^{\mathrm{full}}
\right\|
\cdot\left\|
\widehat A_j^{(b)}
\right\|
}.
\]
The absolute value accounts for the sign indeterminacy of the functional loadings. Hence, an alignment close to one indicates that the estimated loading retains essentially the same functional shape after perturbation, up to an overall sign change.

For visualization, we consider the top 10 selected features within each modality, ranked according to the magnitude of their full-data functional loadings. For each feature, we display the empirical distribution of $\operatorname{Align}_j^{(b)}$ across repetitions in which the feature is reselected, together with its selection frequency. 
These results are reported in Figure~\ref{fig:sensitivity}

\begin{figure}
    \centering
    \includegraphics[width=0.9\linewidth]{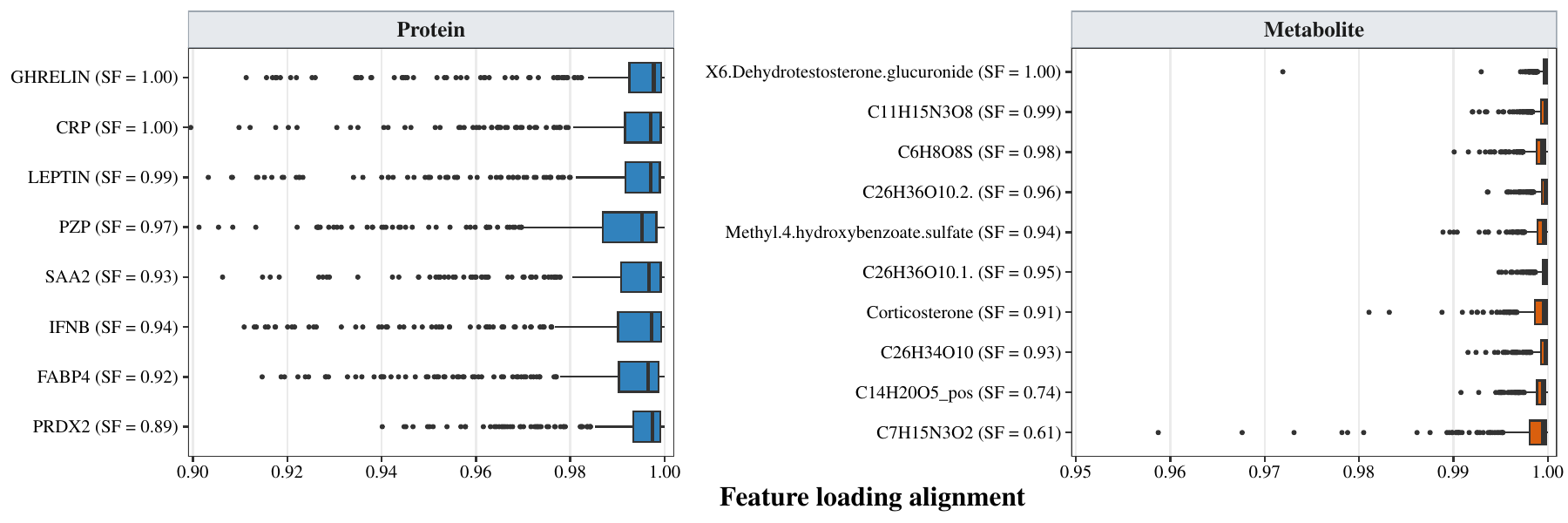}
    \caption{Sensitivity analysis of the estimated functional loadings under time-point subsampling. The boxplots show the loading alignment between the full-data estimates and those obtained from repeatedly perturbed datasets for the top 10 selected features in the protein and metabolite modalities. The selection frequency (SF) of each feature is reported in parentheses. Both the selection frequencies and feature loading alignments are computed based on 500 repeated time-point subsamples.}
    \label{fig:sensitivity}
\end{figure}

\subsection{Trapezoidal Weighting under Nonuniform Observation Times}
\label{sec:trapezoidal weighting}

We further investigate the effect of nonuniform observation times on the functional-aggregation step of FACD. Recall that, for the $X$ modality, FACD constructs
\[
U_i^X(T_{ig_1},T_{ig_2})
=
\frac{1}{pq}
\sum_{j=1}^{p}
\sum_{m=1}^{q}
\frac{1}{N_i^Y}
\sum_{h=1}^{N_i^Y}
G_{i,jm}^{XY}(T_{ig_1},T_{ih})
G_{i,jm}^{XY}(T_{ig_2},T_{ih}),
\]
where the empirical average over the observed time points provides a simple approximation to the corresponding temporal aggregation. This construction is particularly natural when the observation times are approximately uniformly distributed. When the observation times are highly nonuniform or clustered, however, equally weighting all observed time points may place relatively more emphasis on densely observed regions of the time domain.

To adapt FACD to this approximation, we additionally consider a trapezoidal-weighted version of the functional aggregation. For subject $i$, let
\[
T_{i1}<\cdots<T_{iN_i^Y}
\]
denote the ordered observation times for the $Y$ modality. Define the trapezoidal weights as
\[
w_{i1}
=
\frac{T_{i2}-T_{i1}}{2},
\qquad
w_{ih}
=
\frac{T_{i,h+1}-T_{i,h-1}}{2},
\quad
h=2,\ldots,N_i^Y-1,
\]
and
\[
w_{iN_i^Y}
=
\frac{T_{iN_i^Y}-T_{i,N_i^Y-1}}{2}.
\]
We further normalize these weights by
\[
\widetilde w_{ih}
=
\frac{w_{ih}}
{\sum_{l=1}^{N_i^Y}w_{il}},
\qquad
h=1,\ldots,N_i^Y,
\]
so that $\sum_{h=1}^{N_i^Y}\widetilde w_{ih}=1$. We then replace the equally weighted empirical average in $U_i^X$ by
\[
U_{i,\mathrm{trap}}^X(T_{ig_1},T_{ig_2})
=
\frac{1}{pq}
\sum_{j=1}^{p}
\sum_{m=1}^{q}
\sum_{h=1}^{N_i^Y}
\widetilde w_{ih}
G_{i,jm}^{XY}(T_{ig_1},T_{ih})
G_{i,jm}^{XY}(T_{ig_2},T_{ih}).
\]
An analogous modification is applied to $U_i^Y$. Compared with equal weighting, the trapezoidal construction accounts for the local spacing between adjacent observation times and therefore reduces the relative contribution of observations concentrated within densely sampled time regions.
}

\
\section{Statistical Convergence of FACD}
For two sequences of non-negative real values $\{a_n\}$ and $\{b_n\}$, we say $a_n \lesssim b_n$ or $b_n \gtrsim a_n$ if there exists a constant $C > 0$ such that $a_n \leq C b_n$ for all $n$. Similarly, this notation is extended to sequences indexed by $p$ and $q$, such as $a_p \lesssim b_p$ and $a_q \lesssim b_q$. In the following, we abuse $\langle\cdot,\cdot\rangle$ to denote the inner-product of two vectors.

We recall that $\bm{\Gamma}$ is a block matrix with the $(k_1,k_2)$th block as $\wtilde{\bm{\Lambda}}^{XY}_{k_1,k_2}$ for $k_1=1,\dots,\kappa_X$ and $k_2=1,\dots,\kappa_Y$, where 
$\wtilde{\bm{\Lambda}}^{XY}_{k_1,k_2}=\int_0^1\int_0^1 \bm{R}_{XY}(t_1,t_2)\wtilde\psi_{k_1}^X(t_1)\wtilde\psi_{k_2}^Y(t_2)\ \mathrm{d}t_1\mathrm{d}t_2$. Let 
\begin{gather*}
   \bm{\Gamma} =  \sum_{r=1}^{R}
   \eta_{r}\cdot \bm{a}_{r}
\bm{b}_{r}^{\top},
\end{gather*}
where $R$ is the rank of $\bm{\Gamma}$, and $\eta_{r}$,  $\bm{a}_{r}$, and 
$\bm{b}_{r}$ are the $r$th singular value/vectors of $\bm{\Gamma}$, respectively. Denote 
$\bm{a}_{r}=((\bm{a}_{r}^{(1)})^\top,\dots,(\bm{a}_{r}^{(\kappa_X)})^\top)^\top\in \mathbb{R}^{p\cdot\kappa_X}$ and $\bm{b}_{r}=((\bm{b}_{r}^{(1)})^\top,\dots,(\bm{b}_{r}^{(\kappa_Y)})^\top)^\top\in \mathbb{R}^{q\cdot\kappa_Y}$. In the following, $(\bm{a})_j$ or $(\bm{b})_m$ indicate the elements in $\bm{a}\in \mathbb{R}^{p\cdot\kappa_X}$ or $\bm{b}\in \mathbb{R}^{q\cdot\kappa_Y}$ corresponding to the $j$th or $m$th feature, respectively. Besides, $a_{1j}^{(k_1)}$ denotes the element of $\bm{a}_1$ corresponding to the $j$th feature and the $k_1$th component, and $b_{1m}^{(k_2)}$ is defined analogously. 

To control errors, we require that the estimated eigenfunctions 
$\hat\psi_{k}^X$ and $\hat\psi_{k}^Y$, $k \geq 1$, are obtained under the constraint
\(
\sup_{t \in \mathcal{T},\, k \geq 1} \max\bigl\{ |\hat\psi_{k}^X(t)|,\ |\hat\psi_{k}^Y(t)| \bigr\} \leq C.
\)
Define $\operatorname{dist}(\cdot,\cdot)$ as the sine distance,
\(
\operatorname{dist}(\bm{u},\bm{v}) = \sqrt{1 - \frac{\langle \bm{u}, \bm{v}\rangle^2}{\|\bm{u}\|^2_E \|\bm{v}\|^2_E}}.
\)
We further impose constraints on the estimated vectors $\hat{\bm{a}}_r$ and $\hat{\bm{b}}_r$ obtained from the optimization problem~\eqref{opt_SVD} in the main text, namely
\(
\max\left\{ \operatorname{dist}(\hat{\bm{a}}_r, \bm{a}_r),\ 
\operatorname{dist}(\hat{\bm{b}}_r, \bm{b}_r)\right\} 
\leq 1 - \frac{\eta_{r+1}}{\eta_r} - c,
\)
where $c$ is a positive constant satisfying $c < 1$.  
These boundedness requirements are imposed for technical convenience but are not necessary in practice.

\begin{proof}[Proof of Theorem~\ref{Theorem: consistency_diverge_dimension}]
We only focus on the error bound of $\|\wtilde{\bm{A}}^X_r- \hat{\bm{A}}^X_r\|_p$, while those of $\|\wtilde{\bm{A}}^Y_r- \hat{\bm{A}}^Y_r\|_q$ can be proven similarly.
Notice that
\begin{eqnarray}
 \|\wtilde{\bm{A}}^X_r- \hat{\bm{A}}^X_r\|_p^2&\lesssim &
    \bigg\|\wtilde{\bm{A}}^X_r-\sum_{k=1}^{\kappa_X}{\bm{a}}^{(k)}_{r}\cdot\wtilde{\psi}_{k}^X\bigg\|_p^2+
    \bigg\|\sum_{k=1}^{\kappa_X}{\bm{a}}^{(k)}_{r}\cdot\wtilde{\psi}_{k}^X-\sum_{k=1}^{\kappa_X}\hat{\bm{a}}^{(k)}_{r}\cdot\hat{\psi}_{k}^X\bigg\|_p^2\nonumber\\
    &:=&(1) + (2).\label{bound_init}
\end{eqnarray}
\begin{itemize}
    \item  Notice that $\wtilde{\bm{A}}^X_r$ and $\sum_{k=1}^{\kappa_X}{\bm{a}}^{(k)}_{r}\cdot\wtilde{\psi}_{k}^X$ are the $r$th eigenfunctions of $\bm{L}_X(t_1,t_2)$ and $\bm{L}^{(\kappa_X,\kappa_Y)}_X(t_1,t_2)$, respectively. Without loss of generality, we assume $\langle \wtilde{\bm{A}}^X_r,\sum_{k=1}^{\kappa_X}{\bm{a}}^{(k)}_{r}\cdot\wtilde{\psi}_{k}^X\rangle_p\geq 0$.
By Theorem \ref{Theorem_appr} and Assumption~\ref{A_r}, we have
\begin{eqnarray*}
    (1)&\lesssim&
     {\sum_{k_1>\kappa_X} \zeta_{k_1}^X + \sum_{k_2>\kappa_Y} \zeta_{k_2}^Y}.
\end{eqnarray*}
By Assumption~\ref{A5}, $\zeta_{k_1}^X\leq C\cdot k_1^{-a}$ and $\zeta_{k_2}^Y\leq C\cdot k_2^{-a}$. Then, {for \(a>1\),}
\begin{eqnarray}\label{term_(1)_diverge}
  (1)&\lesssim&   \sum_{k_1>\kappa_X} \zeta_{k_1}^X + \sum_{k_2>\kappa_Y} \zeta_{k_2}^Y\nonumber\\
  &\leq&  {C\bigg(\sum_{k_1>\kappa_X} k_1^{-a}+\sum_{k_2>\kappa_Y} k_2^{-a}\bigg)}
  \ {\leq\ C\bigg(\frac{\kappa_X^{1-a}}{a-1}+\frac{\kappa_Y^{1-a}}{a-1}\bigg).}
\end{eqnarray}

\item Notice that
\begin{eqnarray}
   (2)&\lesssim&  \bigg\|\sum_{k=1}^{\kappa_X}{\bm{a}}^{(k)}_{r}\cdot\wtilde{\psi}_{k}^X-\sum_{k=1}^{\kappa_X}{\bm{a}}^{(k)}_{r}\cdot\hat{\psi}_{k}^X\bigg\|_p^2
   +  \bigg\|\sum_{k=1}^{\kappa_X}{\bm{a}}^{(k)}_{r}\cdot\hat{\psi}_{k}^X-\sum_{k=1}^{\kappa_X}\hat{\bm{a}}^{(k)}_{r}\cdot\hat{\psi}_{k}^X\bigg\|_p^2\nonumber\\
   &= &
   \bigg\|\sum_{k=1}^{\kappa_X}{\bm{a}}^{(k)}_{r}\cdot(\wtilde{\psi}_{k}^X-\hat{\psi}_{k}^X)\bigg\|_p^2
   +  \bigg\|\sum_{k=1}^{\kappa_X}({\bm{a}}^{(k)}_{r}-\hat{\bm{a}}^{(k)}_{r})\cdot\hat{\psi}_{k}^X\bigg\|_p^2\nonumber\\
    &\leq & \bigg(\sum_{k=1}^{\kappa_X}{\|{\bm{a}}^{(k)}_{r}\|_E^2}\bigg)\bigg(\sum_{k=1}^{\kappa_X}{\|\wtilde{\psi}_{k}^X-\hat{\psi}_{k}^X\|^2}\bigg)
    + {\big\|{\bm{a}}_{r}-\hat{\bm{a}}_{r}\big\|_E^2}\nonumber\\
    &\lesssim & \kappa_X\cdot \max_{k\leq \kappa_X}{\|\wtilde{\psi}_{k}^X-\hat{\psi}_{k}^X\|^2}
    + {\big\|{\bm{a}}_{r}-\hat{\bm{a}}_{r}\big\|^2_E}.\label{bound_(2)}
\end{eqnarray}
\end{itemize}

In the following, we examine the error bounds of $\max_{k\leq \kappa_X}\|\wtilde{\psi}_{k}^X-\hat{\psi}_{k}^X\|^2$ and $\big\|{\bm{a}}_{r}-\hat{\bm{a}}_{r}\big\|_E^2$, respectively.

Without loss of generality, $\langle \wtilde{\psi}_{k}^X,\hat{\psi}_{k}^X\rangle\geq 0$ {for all} $k\leq \kappa_X$. Note that $\wtilde{\psi}_{k}^X$ and $\hat{\psi}_{k}^X$ are the eigenfunctions of $H_X(t_1,t_2)$ and $\hat{H}_X(t_1,t_2)$, respectively. By Lemma 4.3 in \citet{bosq2000linear} and Assumption \ref{A5}, we obtain 
\begin{eqnarray*}
   \max_{k\leq \kappa_X}\|\wtilde{\psi}_{k}^X-\hat{\psi}_{k}^X\|^2
   &\lesssim& {\max_{k\le \kappa_X}\frac{\|\hat{H}_X-H_X\|_{\infty}^2}{\inf_{\ell\neq k}|\zeta_k^X-\zeta_\ell^X|^2}}
   \ \lesssim\  \kappa_X^{2a+2}\cdot {\big\|\hat{H}_X - H_X\big\|_{\infty}^2}.
\end{eqnarray*}
where $\|\hat{H}_X-H_X\|_{\infty}$ {denotes} the operator norm of the {integral} operator induced by the kernel {\(\hat{H}_X-H_X\)}.
{Moreover, since \(\|T\|_{\infty}\le \|T\|\), we have \(\|\hat{H}_X-H_X\|_{\infty}^2\le \|\hat{H}_X-H_X\|^2\). Hence the above display also implies the same rate with \(\|\hat{H}_X-H_X\|^2\) in place of \(\|\hat{H}_X-H_X\|_{\infty}^2\).}
Utilizing Lemma~\ref{A4}, we have
\begin{eqnarray}\label{equ_phi}
   \mathbb{E}\max_{k\leq \kappa_X}\|\wtilde{\psi}_{k}^X-\hat{\psi}_{k}^X\|^2\lesssim \kappa_X^{2a+2}\cdot\bigg\{\text{Error}^2_{\mu}+\text{Error}^2_{H}\bigg\}.
\end{eqnarray}
Similarly,
\begin{eqnarray}\label{equ_phi_Y}
    \mathbb{E}\max_{k\leq \kappa_Y}\|\wtilde{\psi}_{k}^Y-\hat{\psi}_{k}^Y\|^2\lesssim\kappa_Y^{2a+2}\cdot \bigg\{\text{Error}^2_{\mu}+\text{Error}^2_{H}\bigg\}.
\end{eqnarray}

In the following, we focus on the error rate of $\|\bm{a}_r - \hat{\bm{a}}_r\|$.
We first consider $r = 1$.
Assuming $\sup_{1\leq j\leq p} \|(\bm{\Gamma}\bm{\hat{b}}_{1}-\hat{\bm{\Gamma}}\bm{\hat{b}}_{1})_j\|_E\lesssim \varrho_X$ and $\sup_{1\leq m\leq q} \|\bm(\bm{\Gamma}^\top\bm{\hat{a}}_{1}-\bm{\hat{\Gamma}}^\top\bm{\hat{a}}_{1})_m\|_E\lesssim \varrho_Y$,
{Lemmas~\ref{Le: Iterative_bound} and \ref{Le: Iterative_bound_2}} indicate that
\begin{eqnarray*}
\operatorname{dist}(\hat{\bm{a}}_1,\bm{a}_1)  & \leq& C\cdot
\frac{\varrho_X}{\eta_1}\cdot\sqrt{s_X}+\frac{\eta_2}{\eta_1}\cdot\operatorname{dist}(\hat{\bm{b}}_1,\bm{b}_1)+\operatorname{dist}^2(\bm{b}_1,\bm{\hat{b}}_1),\\ 
\operatorname{dist}(\hat{\bm{b}}_{1},\bm{b}_{1}) &\leq& C\cdot \frac{\varrho_Y}{\eta_1}\cdot\sqrt{s_Y}+  \operatorname{dist}^2(\bm{a}_1,\bm{\hat{a}}_1)+\frac{\eta_2}{\eta_1} \cdot \operatorname{dist}(\bm{a}_1,\bm{\hat{a}}_1).
\end{eqnarray*} 
Then,
\begin{eqnarray*}
\operatorname{dist}(\hat{\bm{a}}_1,\bm{a}_1) 
&\leq& C\cdot
\frac{\varrho_X}{\eta_1}\cdot\sqrt{s_X}+\operatorname{dist}(\hat{\bm{b}}_1,\bm{b}_1)\cdot \bigg(\frac{\eta_2}{\eta_1}+\operatorname{dist}(\hat{\bm{b}}_1,\bm{b}_1)\bigg)  \\
&\leq& C\cdot
\frac{\varrho_X}{\eta_1}\cdot\sqrt{s_X}+{\operatorname{dist}(\hat{\bm{b}}_1,\bm{b}_1),}
\end{eqnarray*}
{where the last inequality uses \(\eta_2/\eta_1+\operatorname{dist}(\hat{\bm{b}}_1,\bm{b}_1)\le 1\).}
Consequently,
\begin{eqnarray*}
\operatorname{dist}(\hat{\bm{a}}_1,\bm{a}_1) 
&\lesssim& 
\frac{\varrho_X}{\eta_1}\cdot\sqrt{s_X}+
\frac{\varrho_Y}{\eta_1}\cdot\sqrt{s_Y}+  \operatorname{dist}^2(\bm{a}_1,\bm{\hat{a}}_1)+\frac{\eta_2}{\eta_1} \cdot \operatorname{dist}(\bm{a}_1,\bm{\hat{a}}_1).
\end{eqnarray*}
Rearranging the above terms, we obtain
\begin{eqnarray*}
\left(1 - \frac{\eta_2}{\eta_1} - \operatorname{dist}(\bm{a}_{1}, \hat{\bm{a}}_{1})\right)
\cdot \operatorname{dist}(\hat{\bm{a}}_{1}, \bm{a}_{1})
\;\;\leq\;\;
C \left(
\frac{\varrho_X}{\eta_1}\sqrt{s_X} \;+\; \frac{\varrho_Y}{\eta_1}\sqrt{s_Y}
\right).
\end{eqnarray*}
As {\(\operatorname{dist}(\hat{\bm{a}}_1,\bm{a}_1)+\eta_2/\eta_1 \leq 1-c\)}, 
\begin{eqnarray*}
\operatorname{dist}(\hat{\bm{a}}_{1},\bm{a}_{1})  &\lesssim& \frac{\varrho_X}{\eta_1}\cdot\sqrt{s_X}+ \frac{\varrho_Y}{\eta_1}\cdot\sqrt{s_Y}.
\end{eqnarray*}
Again by Lemma \ref{Le: Iterative_bound_2}, we have
\begin{eqnarray*}
\operatorname{dist}(\hat{\bm{b}}_{1},\bm{b}_{1})  \lesssim \frac{\varrho_X}{\eta_1}\cdot\sqrt{s_X}+ \frac{\varrho_Y}{\eta_1}\cdot\sqrt{s_Y}.
\end{eqnarray*}
Without loss of generality, we assume that $\langle \bm{\hat{a}}_1,\bm{a}_1\rangle \geq 0$ and $\langle \bm{\hat{b}}_1,\bm{b}_1\rangle \geq 0$, which in turn {implies}
$$
{\|\hat{\bm{a}}_{1}-\bm{a}_{1}\|_E\leq \sqrt{2}\,\operatorname{dist}(\hat{\bm{a}}_{1},\bm{a}_{1})
\ \ \text{and}\ \ 
\|\hat{\bm{b}}_{1}-\bm{b}_{1}\|_E\leq \sqrt{2}\,\operatorname{dist}(\hat{\bm{b}}_{1},\bm{b}_{1}).}
$$
Therefore,
\begin{eqnarray*}
    \max\big\{\|\hat{\bm{a}}_{1}-\bm{a}_{1}\|_E,\|\hat{\bm{b}}_{1}-\bm{b}_{1}\|_E\big\}
    \leq C\cdot\bigg(\frac{\varrho_X}{\eta_1}\cdot\sqrt{s_X}+ \frac{\varrho_Y}{\eta_1}\cdot\sqrt{s_Y}\bigg).
\end{eqnarray*}
The above derivation indicates that
\begin{eqnarray}\label{imp_inq}
& &\mathbb{P}\bigg\{ \max\big\{\|\hat{\bm{a}}_{1}-\bm{a}_{1}\|_E,\|\hat{\bm{b}}_{1}-\bm{b}_{1}\|_E\big\}> C\cdot \bigg(\frac{\varrho_X}{\eta_1}\cdot\sqrt{s_X}+ \frac{\varrho_Y}{\eta_1}\cdot\sqrt{s_Y}\bigg)\bigg\}\nonumber\\
&\leq& 
    \mathbb{P}\bigg\{\sup_{1\leq j\leq p} \|(\bm{\Gamma}\bm{\hat{b}}_{1}-\hat{\bm{\Gamma}}\bm{\hat{b}}_{1})_j\|_E\gtrsim \varrho_X\ \text{or}\ 
    \sup_{1\leq m\leq q} {\|(\bm{\Gamma}^\top\bm{\hat{a}}_{1}-\bm{\hat{\Gamma}}^\top\bm{\hat{a}}_{1})_m\|_E}\gtrsim \varrho_Y\bigg\}.
\end{eqnarray}

We next consider bounding the right-hand side of \eqref{imp_inq}.
First note that
\begin{eqnarray*}
&&\|(\bm{\Gamma}\bm{\hat{b}}_{1}-\hat{\bm{\Gamma}}\bm{\hat{b}}_{1})_j\|_E\\
&= & \sqrt{\sum_{k_1=1}^{\kappa_X}\bigg(\sum_{k_2=1}^{\kappa_Y}\sum_{m=1}^q ({\Lambda}^{XY}_{jm,k_1,k_2}- \hat{{\Lambda}}^{XY}_{jm,k_1,k_2})\hat{b}_{1m}^{(k_2)}\bigg)^2}\\
&\leq& {\sqrt{\sum_{k_1=1}^{\kappa_X}\bigg(\sum_{k_2=1}^{\kappa_Y}\sum_{m=1}^q |{\Lambda}^{XY}_{jm,k_1,k_2}- \hat{{\Lambda}}^{XY}_{jm,k_1,k_2}|^2\bigg)\bigg(\sum_{k_2=1}^{\kappa_Y}\sum_{m=1}^q |\hat{b}_{1m}^{(k_2)}|^2\bigg)}}\\
&=& {\|\hat{\bm b}_1\|\cdot \sqrt{\sum_{k_1=1}^{\kappa_X}\sum_{k_2=1}^{\kappa_Y}\sum_{m=1}^q |{\Lambda}^{XY}_{jm,k_1,k_2}- \hat{{\Lambda}}^{XY}_{jm,k_1,k_2}|^2}}\\
&=& \sqrt{\sum_{k_1=1}^{\kappa_X}\sum_{k_2=1}^{\kappa_Y}\sum_{m=1}^q |{\Lambda}^{XY}_{jm,k_1,k_2}- \hat{{\Lambda}}^{XY}_{jm,k_1,k_2}|^2}\,,
\end{eqnarray*}
{where the last equality uses \(\|\hat{\bm b}_1\|_E=1\).}
By the definition of $C_{\bm{\Gamma}}$, 
$\big|{\Lambda}^{XY}_{jm,k_1,k_2}- \hat{{\Lambda}}^{XY}_{jm,k_1,k_2}\big|$
is a sub-exponential random variable with its sub-exponential norm ($\psi_1$-norm) bounded by 
\[
\big\|{\Lambda}^{XY}_{jm,k_1,k_2}- \hat{{\Lambda}}^{XY}_{jm,k_1,k_2}\big\|_{\psi_1}
\ \le\
C_{\bm{\Gamma}}\cdot C \cdot\sqrt{\mathbb{E}\big|{\Lambda}^{XY}_{jm,k_1,k_2}- \hat{{\Lambda}}^{XY}_{jm,k_1,k_2}\big|^2},
\]
see Proposition 2.7.1 in \citet{vershynin2018high}.
Let $\Delta_{k_1,k_2,m}:={\Lambda}^{XY}_{jm,k_1,k_2}-\hat{{\Lambda}}^{XY}_{jm,k_1,k_2}$ and define
\[
S:=\Big(\sum_{k_1=1}^{\kappa_X}\sum_{k_2=1}^{\kappa_Y}\sum_{m=1}^q |\Delta_{k_1,k_2,m}|^2\Big)^{1/2}.
\]
Proposition 2.7.1 in \citet{vershynin2018high} indicates that:
there exists an absolute constant $C_0>0$ such that
\[
\|X\|_{\psi_1}\ \le\ C_0\sup_{u\ge 2}\frac{\big(\mathbb{E}|X|^u\big)^{1/u}}{u}.
\]

Fix any $u\ge 2$. By Minkowski's inequality, we have
\[
\Big(\mathbb{E}S^{u}\Big)^{1/u}
=
\Big(\mathbb{E}\Big(\sum_{k_1,k_2,m}|\Delta_{k_1,k_2,m}|^2\Big)^{u/2}\Big)^{1/u}
\ \le\
\Big(\sum_{k_1,k_2,m}\big(\mathbb{E}|\Delta_{k_1,k_2,m}|^{u}\big)^{2/u}\Big)^{1/2}.
\]
Since each $\Delta_{k_1,k_2,m}$ is sub-exponential, the standard moment growth bound implies
\[
\big(\mathbb{E}|\Delta_{k_1,k_2,m}|^{u}\big)^{1/u}\ \le\ C_1\,u\,\|\Delta_{k_1,k_2,m}\|_{\psi_1},
\qquad u\ge 2,
\]
for an absolute constant $C_1>0$. Hence,
\[
\Big(\mathbb{E}S^{u}\Big)^{1/u}
\ \le\
C_1u\Big(\sum_{k_1,k_2,m}\|\Delta_{k_1,k_2,m}\|_{\psi_1}^2\Big)^{1/2}.
\]
Dividing by $u$ and taking $\sup_{u\ge 2}$, we obtain
\[
\|S\|_{\psi_1}
\ \le\
C\Big(\sum_{k_1=1}^{\kappa_X}\sum_{k_2=1}^{\kappa_Y}\sum_{m=1}^q
\|\Delta_{k_1,k_2,m}\|_{\psi_1}^2\Big)^{1/2},
\]
for an absolute constant $C>0$.
Consequently, for any $x>0$,
\begin{eqnarray*}
&&\mathbb{P}\big\{\|(\bm{\Gamma}\bm{\hat{b}}_{1}-\hat{\bm{\Gamma}}\bm{\hat{b}}_{1})_j\|_E\geq x\big\}\\
&\leq& \mathbb{P}\bigg\{\sqrt{\sum_{k_1=1}^{\kappa_X}\sum_{k_2=1}^{\kappa_Y}\sum_{m=1}^q 
\big|{\Lambda}^{XY}_{jm,k_1,k_2}- \hat{{\Lambda}}^{XY}_{jm,k_1,k_2}\big|^2}\geq x\bigg\}\\
&\leq& 2\exp\bigg(
-\frac{x}{C_{\bm{\Gamma}}\cdot C\cdot 
\sqrt{\sum_{k_1=1}^{\kappa_X}\sum_{k_2=1}^{\kappa_Y}\sum_{m=1}^q 
\mathbb{E}\big|{\Lambda}^{XY}_{jm,k_1,k_2}- \hat{{\Lambda}}^{XY}_{jm,k_1,k_2}\big|^2}}
\bigg),
\end{eqnarray*}
by Proposition 2.7.1 in \citet{vershynin2018high}.
Accordingly, by the union bound,
\begin{eqnarray*}
& & \mathbb{P}\bigg\{\sup_{1\leq j\leq p}\|(\bm{\Gamma}\bm{\hat{b}}_{1}-\hat{\bm{\Gamma}}\bm{\hat{b}}_{1})_j\|_E\gtrsim   \varrho_X\bigg\}\\
 &\leq& 
 2p\exp\bigg(
-\frac{\varrho_X}{C_{\bm{\Gamma}}\cdot C\cdot 
\max_{1\le j\le p}\sqrt{\sum_{k_1=1}^{\kappa_X}\sum_{k_2=1}^{\kappa_Y}\sum_{m=1}^q 
\mathbb{E}\big|{\Lambda}^{XY}_{jm,k_1,k_2}- \hat{{\Lambda}}^{XY}_{jm,k_1,k_2}\big|^2}}
\bigg).
\end{eqnarray*}

Utilizing Lemma \ref{Le: bound_gamma} and \eqref{equ_phi} and \eqref{equ_phi_Y},
\begin{eqnarray}\label{error_gamma}
&&\max_{1\le j\le p}\sqrt{\sum_{k_1=1}^{\kappa_X}\sum_{k_2=1}^{\kappa_Y}\sum_{m=1}^q 
\mathbb{E}\big|{\Lambda}^{XY}_{jm,k_1,k_2}- \hat{{\Lambda}}^{XY}_{jm,k_1,k_2}\big|^2}\nonumber\\
&=&  \sqrt{q\kappa_X\kappa_Y} \cdot \mathcal{O} \big(\text{Error}_{\mu}+\text{Error}_{H}\big)\cdot \max\{\kappa_X^{a+1},\kappa_Y^{a+1}\}.
\end{eqnarray}
Let
\[
\varrho_X
:= C\cdot C_{\bm{\Gamma}}\cdot
\Big(\sqrt{q}\big(\text{Error}_{\mu}+\text{Error}_{H}\big)\max\{\kappa_X^{a+2},\kappa_Y^{a+2}\}\Big)\cdot \log p,
\]
{with} a sufficiently large constant \(C>0\). Then, by \eqref{error_gamma},
\begin{eqnarray*}
& & \mathbb{P}\bigg\{\sup_{1\leq j\leq p}\|(\bm{\Gamma}\bm{\hat{b}}_{1}-\hat{\bm{\Gamma}}\bm{\hat{b}}_{1})_j\|_E\gtrsim  \varrho_X\bigg\}
\ \le\
2p\exp\big(-C\log p\big)
\ =\ 2p^{1-C}
\ =\ o(1),
\end{eqnarray*}
as \(p\to\infty\).
We similarly obtain 
\begin{eqnarray*}
& & \mathbb{P}\bigg\{\sup_{1\leq m\leq q} \|(\bm{\Gamma}^\top\bm{\hat{a}}_{1}-\hat{\bm{\Gamma}}^\top\bm{\hat{a}}_{1})_m\|_E\gtrsim \varrho_Y\bigg\}= o(1), 
\end{eqnarray*}
when 
\[
\varrho_Y
:= C\cdot C_{\bm{\Gamma}}\cdot
\Big(\sqrt{p}\big(\text{Error}_{\mu}+\text{Error}_{H}\big)\max\{\kappa_X^{a+2},\kappa_Y^{a+2}\}\Big)\cdot \log q,
\]
with a sufficiently large $C$.
Therefore,
\begin{eqnarray*}
    \mathbb{P}\bigg\{\sup_{1\leq j\leq p} \|(\bm{\Gamma}\bm{\hat{b}}_{1}-\hat{\bm{\Gamma}}^{}\bm{\hat{b}}_{1})_j\|_E\gtrsim \varrho_X\ \text{or}\ 
    \sup_{1\leq m\leq q} \|(\bm{\Gamma}^\top\bm{\hat{a}}_{1}-\hat{\bm{\Gamma}}^\top\bm{\hat{a}}_{1})_m\|_E\gtrsim \varrho_Y\bigg\}=o(1),
\end{eqnarray*}
as $p,q\to \infty$.
Due to \eqref{imp_inq}, {it follows that}
\begin{eqnarray*}
 &&   \max\big\{\|\hat{\bm{a}}_{1}-\bm{a}_{1}\|_E^2,\|\hat{\bm{b}}_{1}-\bm{b}_{1}\|_E^2\big\}
 =\mathcal{O}_p\bigg(\bigg(\frac{\varrho_X}{\eta_1}\cdot\sqrt{s_X}+ \frac{\varrho_Y}{\eta_1}\cdot\sqrt{s_Y}\bigg)^2\bigg)\\
&\le& {\frac{2}{\eta_1^2}\Big(\varrho_X^2 s_X+\varrho_Y^2 s_Y\Big)}\\
&\le& {\frac{2}{\eta_1^2}\Big(\varrho_X^2 s_X+\varrho_Y^2 s_Y\Big)}\\
&=& {\mathcal{O}_p\bigg(
\frac{C_{\bm{\Gamma}}^2}{\eta_1^2}\cdot
\big(\text{Error}_{\mu}+\text{Error}_{H}\big)^2\cdot
\max\{\kappa_X^{2a+4},\kappa_Y^{2a+4}\}\cdot
\Big[q\big(\log p\big)^2\cdot s_X
\;+\;
p\big(\log q\big)^2\cdot s_Y\Big]
\bigg)}\\
&=& {\max\{\kappa_X^{2a+4},\kappa_Y^{2a+4}\}\cdot 
\mathcal{O}_p\bigg[
C_{\bm{\Gamma}}^2\cdot\big(\text{Error}_{\mu}+\text{Error}_{H}\big)^2\cdot
\max\bigg\{\frac{\sqrt{q}\log p}{\eta_1},\frac{\sqrt{p}\log q}{\eta_1}\bigg\}^2\cdot
\big(s_X+s_Y\big)
\bigg].}
\end{eqnarray*}
Combining the above inequality with \eqref{bound_(2)} and \eqref{equ_phi}, we have
\begin{eqnarray*}
    (2)&=&\kappa_X^{2a+3}\cdot\mathcal{O}_p\big(\text{Error}_{\mu}^2+\text{Error}_{H}^2\big)
    +\max\{\kappa_X^{2a+4},\kappa_Y^{2a+4}\}\\
    &\cdot& 
\mathcal{O}_p\bigg[
C_{\bm{\Gamma}}^2\cdot\big(\text{Error}_{\mu}+\text{Error}_{H}\big)^2\cdot
\max\bigg\{\frac{\sqrt q\log p}{\eta_1},\frac{\sqrt p\log q}{\eta_1}\bigg\}^2\cdot
\big(s_X+s_Y\big)
\bigg].
\end{eqnarray*}

Next we relate \(\eta_1\) to \(\lambda_1\). By \eqref{eigen_value_gap}, we have \(\eta_1^2=\lambda_1^2-pq\cdot o(1)\) as \(\kappa_X\) and \(\kappa_Y\) go to \(\infty\). Hence
\begin{eqnarray*}
 \frac{1}{\eta_1}
 &=&
 \frac{1}{\lambda_1}+\frac{\lambda_1-\eta_1}{\eta_1\lambda_1}
 \ \le\
 \frac{1}{\lambda_1}+\frac{\lambda_1^2-\eta_1^2}{\eta_1\lambda_1(\lambda_1+\eta_1)}
 \ =\
 \frac{1}{\lambda_1}
 \ +\ 
 \frac{pq\cdot o(1)}{\eta_1\lambda_1(\lambda_1+\eta_1)}
 \ =\
 \frac{1}{\lambda_1}+o\!\left(\frac{1}{\lambda_1}\right).
\end{eqnarray*}
In particular, \(\eta_1\asymp \lambda_1\) and thus \(1/\eta_1\lesssim 1/\lambda_1\).
Moreover, under Assumption~\ref{A_r}, and the ordering \(\lambda_1^2\ge \lambda_2^2\), we have
\(
\lambda_1^2-\lambda_2^2 \ \gtrsim\ pq,
\)
so $\lambda_1
    \ \geq\
    \sqrt{\lambda_1^2-\lambda_2^2}
    \ \gtrsim\ \sqrt{pq}.$
Therefore,
\[
\max\bigg\{\frac{\sqrt q}{\eta_1},\frac{\sqrt p}{\eta_1}\bigg\}
\ \lesssim\
\max\bigg\{\frac{\sqrt q}{\lambda_1},\frac{\sqrt p}{\lambda_1}\bigg\}
\ \lesssim\
1.
\]
Plugging this bound back into the previous display yields
\begin{eqnarray}\label{bound_(2)_final_2}
    (2)
    &=&
    \kappa_X^{2a+3}\cdot\mathcal{O}_p\big(\text{Error}_{\mu}^2+\text{Error}_{H}^2\big)+\max\{\kappa_X^{2a+4},\kappa_Y^{2a+4}\}\cdot\mathcal{O}_p\bigg[
    C_{\bm{\Gamma}}^2\cdot\big(\text{Error}_{\mu}+\text{Error}_{H}\big)^2\nonumber\\
    &\cdot&
    \max\bigg\{\frac{\sqrt{q}}{\lambda_1}\log(p),\frac{\sqrt{p}}{\lambda_1}\log(q)\bigg\}^2
    \cdot (s_X+s_Y)
    \bigg]
    \nonumber\\
    &\le&
    \max\{\kappa_X^{2a+4},\kappa_Y^{2a+4}\}\cdot\mathcal{O}_p\bigg[
    C_{\bm{\Gamma}}^2\cdot\big(\text{Error}_{\mu}+\text{Error}_{H}\big)^2\nonumber\\
    &\cdot&
    \max\big\{\log(p),\log(q)\big\}^2
    \cdot (s_X+s_Y)
    \bigg].
\end{eqnarray}

Combining \eqref{term_(1)_diverge} and \eqref{bound_(2)_final_2} with \eqref{bound_init}, we have
\begin{eqnarray*}
&&\|\wtilde{\bm{A}}^X_1- \hat{\bm{A}}^X_1\|_p^2\\
&\leq&    C\bigg(\frac{\kappa_X^{1-a}}{a-1}+\frac{\kappa_Y^{1-a}}{a-1}\bigg)+\max\{\kappa_X^{2a+4},\kappa_Y^{2a+4}\}\cdot\mathcal{O}_p \big\{\mathcal{A}^2(n,p,q)\big\},
\end{eqnarray*}
where $\mathcal{A}(n,p,q)=C_{\bm{\Gamma}}\cdot\big(\text{Error}_{\mu}+\text{Error}_{H}\big)\cdot\max\big\{\log(p),\log(q)\big\}\cdot\big\{\sqrt{s_X}+\sqrt{s_Y}\big\}$.
Therefore,
\begin{eqnarray*}
&&\|\wtilde{\bm{A}}^X_1- \hat{\bm{A}}^X_1\|_p\\
&\lesssim& C\sqrt{\frac{\kappa_X^{1-a}}{a-1}+\frac{\kappa_Y^{1-a}}{a-1}}
  +\sqrt{\max\{\kappa_X^{2a+4},\kappa_Y^{2a+4}\}}\cdot\mathcal{O}_p \big\{\mathcal{A}(n,p,q)\big\}.
\end{eqnarray*}
Consequently, taking $\kappa_X,\kappa_Y\asymp$ $\mathcal{A}(n,p,q)^{-2/(3a+3)}$ gives
\begin{eqnarray*}
\|\wtilde{\bm{A}}^X_1- \hat{\bm{A}}^X_1\|_p
=\mathcal{O}_p\big(\mathcal{A}(n,p,q)^{(a-1)/(3a+3)}\big),
\end{eqnarray*}
as $n,p,q\rightarrow\infty$. 
For $r>1$, the proof can be obtained similarly.

\end{proof}

\

\section{Technical Lemmas}

\begin{Le}\label{A4}
Assume that $H_X(t_1,t_2)$ and $H_Y(t_1,t_2)$ are estimated using \eqref{min_opt_basis_X} and \eqref{min_opt_basis_Y}, respectively, with
\[
\min\Big\{\left(n N_X^2\right)^{-2l/(2l+2)},\, N_X^{-2l}\Big\} \;\lesssim\; \nu_H^X \;\lesssim\;
\max\Big\{\left(n N_X^2\right)^{-2l/(2l+2)},\, n^{-1}\Big\},
\]
and
\[
\min\Big\{\left(n N_Y^2\right)^{-2l/(2l+2)},\, N_Y^{-2l}\Big\} \;\lesssim\; \nu_H^Y \;\lesssim\;
\max\Big\{\left(n N_Y^2\right)^{-2l/(2l+2)},\, n^{-1}\Big\}.
\]
Suppose Assumptions~\ref{A1}--\ref{A6} hold. Then, as $n,p,q \to \infty$, the estimators $\hat{H}_X(t_1,t_2)$ and $\hat{H}_Y(t_1,t_2)$ satisfy
\begin{align}\label{rate_HH}
  \max\!\Big\{\mathbb{E}\big\|\hat{H}_X - H_X\big\|^2,\ \mathbb{E}\big\|\hat{H}_Y - H_Y\big\|^2\Big\} 
  &= \mathcal{O}\!\left(\text{Error}^2_{\mu} + \text{Error}^2_{H}\right),
\end{align}
where
\[
\text{Error}_{\mu}
= C_{\mu}\cdot\sqrt{\log(p+q)}\cdot 
\Big\{(n N_X)^{-l/(2l+1)} + (n N_Y)^{-l/(2l+1)} + n^{-1/2}\Big\}
\]
and
\[
\text{Error}_{H}
= (n N_X^2)^{-l/(2l+2)} + (n N_Y^2)^{-l/(2l+2)} + n^{-1/2}.
\]
\end{Le}

\begin{proof}
Under Assumptions \ref{A1} -- \ref{A6}, Lemma 4.2 in \citet{xiao2020asymptotic} indicates that
\begin{eqnarray}
  \mathbb{E} \big|\hat{\mu}_j^X(t) - \mu_j^X(t)|^4 &\leq& C_1 \alpha_{n,X}^4,\label{rate_X} \\
  \mathbb{E} |\hat{\mu}_m^Y(t) - \mu_m^Y(t)|^4 &\leq& C_2 \alpha_{n,Y}^4,\label{rate_Y}
\end{eqnarray}
where $\alpha_{n,X}$ and $\alpha_{n,Y}$ are $o(1)$ as $n\rightarrow\infty$, denoting the convergence rates of the mean function estimators for the two datasets, and $C_1$ and $C_2$ are constants that are independent of $j$, $m$, and $t$ under Assumption~\ref{mean_assup}. 
See Equation~(4.2) in \citet{xiao2020asymptotic} for the definitions of $\alpha_{n,X}$ and $\alpha_{n,Y}$. 
If the tuning parameters for the mean functions are taken as their optimal orders, then $\alpha_{n,X}$ and $\alpha_{n,Y}$ have the optimal orders $(nN_X)^{-l/(2l+1)} + n^{-1/2}$ and $(nN_Y)^{-l/(2l+1)} + n^{-1/2}$, respectively \citet{xiao2020asymptotic}. Let $\alpha_n=(nN_X)^{-l/(2l+1)}+ (nN_Y)^{-l/(2l+1)} + n^{-1/2}$.

In the following, we only prove the case for the first dataset $X$. 
Recall that
\begin{eqnarray*}
   U_i^X(T_{ig_1},T_{ig_2}):=\frac{1}{pq} \sum_{j=1}^p\sum_{m=1}^q\frac{1}{N_i^Y}\sum_{h=1}^{N_i^Y}G_{i,jm}^{XY}(T_{ig_1},T_{ih})\cdot G_{i,jm}^{XY}(T_{ig_2},T_{ih})
\end{eqnarray*}
with
\begin{gather*}
    G_{i,jm}^{XY}(T_{ig},T_{ih}):=\left\{x_{ijg}-\hat{\mu}_{j}^X(T_{ig})\right\}\cdot \left\{y_{imh}-\hat{\mu}^Y_{m}(T_{ih})\right\}.
\end{gather*}
Denote $R_{jm}^{XY}(t_1,t_2):=[\bm{R}_{XY}(t_1,t_2)]_{j,m}$.

First note that
\begin{eqnarray*}
&&G_{i,jm}^{XY}(T_{ig},T_{ih})\\
&=&  \left\{x_{ijg}-{\mu}_{j}^X(T_{ig})+{\mu}_{j}^X(T_{ig})-\hat{\mu}_{j}^X(T_{ig})\right\}\cdot \left\{y_{imh}-{\mu}^Y_{m}(T_{ih})+{\mu}^Y_{m}(T_{ih})-\hat{\mu}^Y_{m}(T_{ih})\right\}  \\
&=&R_{jm}^{XY}(T_{ig},T_{ih})+ \left\{x_{ijg}-{\mu}_{j}^X(T_{ig})\right\}\cdot \left\{y_{imh}-{\mu}^Y_{m}(T_{ih})\right\} - R_{jm}^{XY}(T_{ig},T_{ih}) \\
&&+
\left\{{\mu}_{j}^X(T_{ig})-\hat{\mu}_{j}^X(T_{ig})\right\}\cdot \left\{y_{imh}-{\mu}^Y_{m}(T_{ih})\right\} + \left\{x_{ijg}-{\mu}_{j}^X(T_{ig})\right\}\cdot \left\{{\mu}^Y_{m}(T_{ih})-\hat{\mu}^Y_{m}(T_{ih})\right\} \\
&&+\left\{{\mu}_{j}^X(T_{ig})-\hat{\mu}_{j}^X(T_{ig})\right\}\cdot \left\{{\mu}^Y_{m}(T_{ih})-\hat{\mu}^Y_{m}(T_{ih})\right\}.
\end{eqnarray*}
By \eqref{rate_X} and \eqref{rate_Y}, the terms 
$\bigl\{\mu_{j}^X(T_{ig}) - \hat{\mu}_{j}^X(T_{ig})\bigr\}$ and 
$\bigl\{\mu_{m}^Y(T_{ih}) - \hat{\mu}_{m}^Y(T_{ih})\bigr\}$ 
can be uniformly controlled by the rate $\alpha_n$ for any $j$, $m$.
By the definition of $C_{\mu}$, 
$|\mu_{j}^X(T_{ig}) - \hat{\mu}_{j}^X(T_{ig})|$ or $|\mu_{m}^Y(T_{ih}) - \hat{\mu}_{m}^Y(T_{ih})|$
is a sub-Gaussian random variable with its sub-Gaussian norm bounded by 
$
C_{\mu}\cdot C \cdot \alpha_n
$; see Proposition 2.5.2 in \citet{vershynin2018high}. Then 
\begin{eqnarray*}
  \max_{1\leq j\leq p,1\leq m\leq q}\{|\mu_{j}^X(T_{ig}) - \hat{\mu}_{j}^X(T_{ig})|,|\mu_{m}^Y(T_{ih}) - \hat{\mu}_{m}^Y(T_{ih})|\}=\mathcal{O}_p(C_{\mu}\cdot\sqrt{\log(p+q)}\cdot \alpha_n),
\end{eqnarray*}
by Exercise 2.5.10 in \citet{vershynin2018high}.
As such, we write
\begin{eqnarray*}
&&G_{i,jm}^{XY}(T_{ig},T_{ih})\\
&=&R_{jm}^{XY}(T_{ig},T_{ih})+ \bigg\{\left\{x_{ijg}-{\mu}_{j}^X(T_{ig})\right\}\cdot \left\{y_{imh}-{\mu}^Y_{m}(T_{ih})\right\} - R_{jm}^{XY}(T_{ig},T_{ih})\bigg\} +\mathcal{O}_p(\text{Error}_{\mu})\\
&=&R_{jm}^{XY}(T_{ig},T_{ih})+ \bigg\{\left(\varepsilon^X_{ij}(T_{ig})+\tau^X_{ijg}\right)\cdot \left(\varepsilon^Y_{im}(T_{ih})+\tau^Y_{imh}\right) - R_{jm}^{XY}(T_{ig},T_{ih})\bigg\} +\mathcal{O}_p(\text{Error}_{\mu})\\
&:=&R_{jm}^{XY}(T_{ig},T_{ih})+A_{jmh}(T_{ig},T_{ih})+\mathcal{O}_p(\text{Error}_{\mu}).
\end{eqnarray*}
where $\text{Error}_{\mu}=C_{\mu}\cdot\sqrt{\log(p+q)}\cdot \bigg\{\left(n N_X\right)^{-l/(2l+1)}+ \left(n N_Y\right)^{-l/(2l+1)} + n^{-1/2}\bigg\}$.
Accordingly,
\begin{eqnarray}\label{U_eqaution}
&&U_i^X(T_{ig_1},T_{ig_2})\nonumber\\
&=& H_X(T_{ig_1},T_{ig_2})\nonumber\\
&&+ \frac{1}{pq} \sum_{j=1}^p\sum_{m=1}^q\frac{1}{N_i^Y}\sum_{h=1}^{N_i^Y}R_{jm}^{XY}(T_{ig_1},T_{ih})\cdot R_{jm}^{XY}(T_{ig_2},T_{ih})-H_X(T_{ig_1},T_{ig_2})\nonumber\\
&& +  \frac{1}{pq} \sum_{j=1}^p\sum_{m=1}^q\frac{1}{N_i^Y}\sum_{h=1}^{N_i^Y}R_{jm}^{XY}(T_{ig_1},T_{ih})\cdot A_{jmh}(T_{ig_2},T_{ih})\nonumber\\
&& +  \frac{1}{pq} \sum_{j=1}^p\sum_{m=1}^q\frac{1}{N_i^Y}\sum_{h=1}^{N_i^Y}R_{jm}^{XY}(T_{ig_2},T_{ih})\cdot A_{jmh}(T_{ig_1},T_{ih})\nonumber\\
&& + \frac{1}{pq} \sum_{j=1}^p\sum_{m=1}^q\frac{1}{N_i^Y}\sum_{h=1}^{N_i^Y}A_{jmh}(T_{ig_1},T_{ih})\cdot A_{jmh}(T_{ig_2},T_{ih})+ \mathcal{O}_p(\alpha_n)\nonumber\\
&:=&H_X(T_{ig_1},T_{ig_2}) + B(T_{ig_1},T_{ig_2})+D+\mathcal{O}_p(\alpha_n). 
\end{eqnarray}
where
\begin{eqnarray*}
&&B(T_{ig_1},T_{ig_2}) \\
&=&  
\frac{1}{pq} \sum_{j=1}^p\sum_{m=1}^q\frac{1}{N_i^Y}\sum_{h=1}^{N_i^Y}R_{jm}^{XY}(T_{ig_1},T_{ih})\cdot R_{jm}^{XY}(T_{ig_2},T_{ih})-H_X(T_{ig_1},T_{ig_2})\\
&& +  \frac{1}{pq} \sum_{j=1}^p\sum_{m=1}^q\frac{1}{N_i^Y}\sum_{h=1}^{N_i^Y}R_{jm}^{XY}(T_{ig_1},T_{ih})\cdot A_{jmh}(T_{ig_2},T_{ih})\\
&& +  \frac{1}{pq} \sum_{j=1}^p\sum_{m=1}^q\frac{1}{N_i^Y}\sum_{h=1}^{N_i^Y}R_{jm}^{XY}(T_{ig_2},T_{ih})\cdot A_{jmh}(T_{ig_1},T_{ih})\\
&& + \frac{1}{pq} \sum_{j=1}^p\sum_{m=1}^q\frac{1}{N_i^Y}\sum_{h=1}^{N_i^Y}A_{jmh}(T_{ig_1},T_{ih})\cdot A_{jmh}(T_{ig_2},T_{ih}) - D
\end{eqnarray*}
and 
$$
D=\frac{1}{pq} \sum_{j=1}^p\sum_{m=1}^q\frac{1}{N_i^Y}\sum_{h=1}^{N_i^Y}\mathbb{E}A_{jmh}(T_{ig_1},T_{ih})\cdot A_{jmh}(T_{ig_2},T_{ih}).
$$
Note that $\mathbb{E} B(T_{ig_1},T_{ig_2})=0$ and $\mathbb{E}B^2(T_{ig_1},T_{ig_2})\leq C$. In the following, we focus on the term $D$.

Note that when $j \notin \mathcal{S}_X$ or
$m \notin \mathcal{S}_Y$,
$R_{jm}^{XY}(T_{ig},T_{ih})=0$.
Moreover, by Assumption~\ref{A1},
the functional signals vanish outside their
respective active sets.
Let
$Z_{imh}^Y
=\varepsilon_{im}^Y(T_{ih})+\tau_{imh}^Y$.
Then
\begin{align*}
D
=&\frac{1}{pq}
\sum_{j\in\mathcal{S}_X}
\sum_{m=1}^q
\mathbb{E}
\bigl[
A_{jmh}(T_{ig_1},T_{ih})
A_{jmh}(T_{ig_2},T_{ih})
\bigr]\\
&+\frac{1}{pq}
\sum_{j\notin\mathcal{S}_X}
\sum_{m\in\mathcal{S}_Y}
\mathbb{E}
\bigl[
\tau_{ijg_1}^X\tau_{ijg_2}^X
(Z_{imh}^Y)^2
\bigr]\\
&+\frac{1}{pq}
\sum_{j\notin\mathcal{S}_X}
\sum_{m\notin\mathcal{S}_Y}
\mathbb{E}
\bigl[
\tau_{ijg_1}^X\tau_{ijg_2}^X
(\tau_{imh}^Y)^2
\bigr].
\end{align*}
By the moment conditions in Assumptions~\ref{A1}
and~\ref{A3}, together with the observation-time
assumptions, the expectations in the first two
terms are uniformly bounded.
Therefore,
\begin{align*}
|D|
&\lesssim
\frac{s_X}{p}+\frac{s_Y}{q}\\
&\quad+
\frac{1}{pq}
\sum_{j\notin\mathcal{S}_X}
\left|
\operatorname{Cov}
\bigl(\tau_{ijg_1}^X,\tau_{ijg_2}^X\bigr)
\right|
\sum_{m\notin\mathcal{S}_Y}
\mathbb{E}(\tau_{imh}^Y)^2\\
&\lesssim
\frac{s_X}{p}
+\frac{s_Y}{q}
+\frac{1}{\sqrt{p}}.
\end{align*}
Under Assumptions~\ref{A1}, \ref{A3},
and~\ref{A6}, provided that
$s_X\lesssim\sqrt{p}$,
$s_Y\lesssim\sqrt{q}$,
and $p,q\gtrsim n$,
we obtain $|D|\lesssim n^{-1/2}$ and
\begin{eqnarray}\label{error_eq}
U_i^X(T_{ig_1},T_{ig_2})=   H_X(T_{ig_1},T_{ig_2}) + \text{Error}_{i,g_1,g_2}+\mathcal{O}(n^{-1/2})+\mathcal{O}_p(\text{Error}_{\mu}), 
\end{eqnarray}
where $\mathbb{E}\text{Error}_{i,g_1,g_2}=0$ and $\mathbb{E}\text{Error}^2_{i,g_1,g_2}<C$.

Recall that
\begingroup
\setlength{\arraycolsep}{2pt}
\begin{eqnarray*}
 \hat{H}_X:=\argmin_{f\in \mathcal{B}_{L_H,4} \otimes \mathcal{B}_{L_H,4}}  \frac{1}{n}\sum_{i=1}^n\frac{1}{(N_i^X-1)N_i^X}\sum_{1\leq g_1<g_2\leq N_i^X}\big\{U_i^X(T_{ig_1},T_{ig_2})-f(T_{ig_1},T_{ig_2})\big\}^2+\nu_H^X \mathcal{L}(f).
\end{eqnarray*}
\endgroup
Equation~\eqref{error_eq} indicates that the above optimization is a classical problem of covariance function estimation \citep{cai2010nonparametric, hsing2015theoretical,xiao2020asymptotic}. 
We then obtain \eqref{rate_HH} by Corollary 4.4 in \citet{xiao2020asymptotic}.

\end{proof}

\

\begin{Le}\label{Le: Iterative_bound}
Let $\bm{\hat{b}}\in \mathbb{S}_{q\cdot\kappa_Y}$ and
\begin{eqnarray*}
    \hat{\bm{a}}:=\argmin_{\bm{a}\in \mathbb{R}^{p\cdot\kappa_X}} \big\|\hat{\bm{\Gamma}}-\bm{a}\hat{\bm{b}}^\top\big\|_F^2+\varrho_X\sum_{j=1}^p\sqrt{\sum_{k=1}^{\kappa_X}(a_{j}^{(k)})^2}.
\end{eqnarray*}
Assume $\varrho_X$ is chosen such that $\sup_{1\leq j\leq p} \|(\bm{\Gamma}\bm{\hat{b}}-\hat{\bm{\Gamma}}\bm{\hat{b}})_j\|_E\leq C\varrho_X$ with $C\in [0, 1/2]$. Then
\begin{eqnarray*}
\operatorname{dist}(\hat{\bm{a}},\bm{a})   \leq C\cdot
\frac{\varrho_X}{\eta_1}\cdot\sqrt{s_X}+\frac{\eta_2}{\eta_1}\cdot\operatorname{dist}(\hat{\bm{b}},\bm{b})+\operatorname{dist}^2(\bm{b},\bm{\hat{b}}),
\end{eqnarray*}
where $\bm{a}=\bm{\Gamma}\bm{b}$ with $\bm{b}$ being the first right singular vector of $\bm\Gamma$.
\end{Le}

\begin{proof}
Without loss of generality, $\langle \bm{\hat{b}},\bm{b}\rangle \geq 0$.
Define $\bar{\bm{a}}:=\bm{\Gamma}\hat{\bm{b}}$.
First note that
\begin{eqnarray*}
\big\|\hat{\bm{\Gamma}}-\bm{\hat{a}}\hat{\bm{b}}^\top\big\|_F^2+\varrho_X\sum_{j=1}^p\|(\bm{\hat{a}})_j\|_E  \leq \big\|\hat{\bm{\Gamma}}-\bar{\bm{a}}\hat{\bm{b}}^\top\big\|_F^2+\varrho_X\sum_{j=1}^p\|(\bar{\bm{a}})_j\|_E.  
\end{eqnarray*}
Rearrange the above inequality, then
\begin{eqnarray*}
    \|\bm{\hat{a}}-\bar{\bm{a}}\|_E^2+\varrho_X\sum_{j=1}^p\|(\bm{\hat{a}})_j\|_E\leq 2\langle \hat{\bm{\Gamma}}\bm{\hat{b}}-\bm{\Gamma}\hat{\bm{b}},\bm{\hat{a}}-\bar{\bm{a}}\rangle+\varrho_X\sum_{j=1}^p\|(\bar{\bm{a}})_j\|_E.
\end{eqnarray*}
Since for $j \notin \mathcal{S}_X$ and $m \notin \mathcal{S}_Y$, we have $[\bm{R}_{XY}(t_1,t_2)]_{j,m} = 0$. Consequently, $[\bm{\Lambda}^{XY}_{k_1,k_2}]_{j,m} = 0$ for all $k_1$ and $k_2$, which implies that $(\bm{a}_r)_j = 0$ for $j \notin \mathcal{S}_X$. This leads to $\big(\bar{\bm a})_j=\big(\bm{\Gamma}\hat{\bm{b}})_j=(\sum_{r=1}^R\eta_r\bm{a}_r\bm{b}_r^\top\hat{\bm{b}}\big)_j=\bm{0}$ for $j\notin \mathcal{S}_X$.
Therefore,
\begin{eqnarray*}
&& \langle {\bm{\hat\Gamma}}\bm{\hat{b}}-\bm{\Gamma}\bm{\hat{b}},\bm{\hat{a}}-\bar{\bm{a}}\rangle\\
&\leq &  \sup_{1\leq j\leq p} \|({\bm{\hat\Gamma}}\bm{\hat{b}}-\bm{\Gamma}\bm{\hat{b}})_j\|_E \cdot \sum_{j=1}^p\|(\bm{\hat{a}}-\bar{\bm{a}})_j\|_E\\
&\leq &\sup_{1\leq j\leq p} \|(\bm{\Gamma}\bm{\hat{b}}-\hat{\bm{\Gamma}}\bm{\hat{b}})_j\|_E \cdot \bigg(\sum_{j\in \mathcal{S}_X}\|(\bm{\hat{a}}-\bar{\bm{a}})_j\|_E+\sum_{j\notin \mathcal{S}_X}\|(\bm{\hat{a}}-\bar{\bm{a}})_j\|_E\bigg)\\
&= & \sup_{1\leq j\leq p} \|(\bm{\Gamma}\bm{\hat{b}}-\hat{\bm{\Gamma}}\bm{\hat{b}})_j\|_E \cdot \bigg(\sum_{j\in \mathcal{S}_X}\|(\bm{\hat{a}}-\bar{\bm{a}})_j\|_E+\sum_{j\notin \mathcal{S}_X}\|\bm{\hat{a}}\|_E\bigg).
\end{eqnarray*}
Combining the above results and the fact $\sup_{1\leq j\leq p} \|(\bm{\Gamma}\bm{\hat{b}}-\hat{\bm{\Gamma}}\bm{\hat{b}})_j\|_E\leq C\varrho_X$ with $C\in [0, 1/2]$, we obtain
\begin{eqnarray*}
 &&   \|\bm{\hat{a}}-\bar{\bm{a}}\|^2_E\\
    &\leq& \varrho_X\sum_{j\in \mathcal{S}_X}\|(\bar{\bm{a}})_j\|_E-\varrho_X\sum_{j\in \mathcal{S}_X}\|(\bm{\hat{a}})_j\|_E+ 2\sup_{1\leq j\leq p} \|(\bm{\Gamma}\bm{\hat{b}}-\hat{\bm{\Gamma}}\bm{\hat{b}})_j\|_E\cdot \sum_{j\in \mathcal{S}_X}\|(\bm{\hat{a}}-\bar{\bm{a}})_j\|_E \\
    && +\big(2\sup_{1\leq j\leq p} \|(\bm{\Gamma}\bm{\hat{b}}-\hat{\bm{\Gamma}}\bm{\hat{b}})_j\|_E-\varrho_X\big)\cdot\sum_{j\notin \mathcal{S}_X}\|(\bm{\hat{a}})_j\|_E\\
    &\leq&\bigg(\varrho_X + 2\sup_{1\leq j\leq p} \|(\bm{\Gamma}\bm{\hat{b}}-\hat{\bm{\Gamma}}\bm{\hat{b}})_j\|_E\bigg)\cdot \sum_{j\in \mathcal{S}_X}\|(\bm{\hat{a}}-\bar{\bm{a}})_j\|_E \\
    &\leq&\bigg(\varrho_X + 2\sup_{1\leq j\leq p} \|(\bm{\Gamma}\bm{\hat{b}}-\hat{\bm{\Gamma}}\bm{\hat{b}})_j\|_E\bigg)\cdot\sqrt{s_X}\|\bm{\hat{a}}-\bar{\bm{a}}\|_E.
\end{eqnarray*}
It follows that
\begin{eqnarray*}
 \|\bm{\hat{a}}-\bar{\bm{a}}\|_E&\leq&\bigg(\varrho_X + 2\sup_{1\leq j\leq p} \|(\bm{\Gamma}\bm{\hat{b}}-\hat{\bm{\Gamma}}\bm{\hat{b}})_j\|_E\bigg)\cdot\sqrt{s_X}\\
 &\leq & C\cdot \varrho_X\cdot\sqrt{s_X}.
\end{eqnarray*}
Using Lemma \ref{Le: projection}, we finally obtain 
\begin{eqnarray*}
    \operatorname{dist}(\hat{\bm{a}},\bm{a})   \leq \frac{\|\hat{\bm{a}}-\bm{a}\|_E}{\eta_1}\leq \frac{\|\hat{\bm{a}}-\bar{\bm{a}}\|_E}{\eta_1}+\frac{\|\bm{\Gamma}\hat{\bm{b}}-\bm{\Gamma}\bm{b}\|_E}{\eta_1}\leq C\cdot \frac{\varrho_X}{\eta_1}\cdot\sqrt{s_X}+\frac{\|\bm{\Gamma}\hat{\bm{b}}-\bm{\Gamma}\bm{b}\|_E}{\eta_1},
\end{eqnarray*}
where
\begin{eqnarray*}
\frac{\|\bm{\Gamma}\hat{\bm{b}}-\bm{\Gamma}\bm{b}\|_E}{\eta_1}&=&\sqrt{\sum_{r=2}^R\eta_r^2/\eta_1^2\cdot \langle \hat{\bm{b}},\bm{b}_r\rangle^2+(1-\langle\hat{\bm{b}},\bm{b}\rangle)^2}\\
&\leq& \frac{\eta_2}{\eta_1}\cdot\operatorname{dist}(\hat{\bm{b}},\bm{b})+\big| 1-\sqrt{1-\operatorname{dist}^2(\bm{b},\bm{\hat{b}})}\big|\\
&\leq& \frac{\eta_2}{\eta_1}\cdot\operatorname{dist}(\hat{\bm{b}},\bm{b})+\frac{\operatorname{dist}^2(\bm{b},\bm{\hat{b}})}{1+\sqrt{1-\operatorname{dist}^2(\bm{b},\bm{\hat{b}})}}\\
&\leq& \frac{\eta_2}{\eta_1}\cdot\operatorname{dist}(\hat{\bm{b}},\bm{b})+\operatorname{dist}^2(\bm{b},\bm{\hat{b}}).
\end{eqnarray*}
We then complete the proof.
\end{proof}

\

\begin{Le}\label{Le: Iterative_bound_2}
Let $\bm{\hat{a}}\in \mathbb{S}_{p\cdot\kappa_X}$
\begin{eqnarray*}
     \hat{\bm{b}}:=\argmin_{\bm{b}\in \mathbb{R}^{q\cdot\kappa_Y}} \big\|\hat{\bm{\Gamma}}-\bm{\hat{a}}{\bm{b}}^\top\big\|_F^2+\varrho_Y\sum_{m=1}^q\sqrt{\sum_{k=1}^{\kappa_Y}(b_{m}^{(k)})^2}.
\end{eqnarray*}
Assume that $\varrho_Y$ is chosen such that $\sup_{1\leq m\leq q} \|\bm(\bm{\Gamma}^\top\bm{\hat{a}}-\bm{\hat{\Gamma}}^\top\bm{\hat{a}})_m\|_E\leq C\varrho_Y$ with $C\in [0,1/2]$. Then
\begin{eqnarray*}
\operatorname{dist}(\hat{\bm{b}},\bm{b}) \leq C\cdot \frac{\varrho_Y}{\eta_1}\cdot\sqrt{s_Y}+ \operatorname{dist}^2(\bm{a},\bm{\hat{a}})+\frac{\eta_2}{\eta_1} \cdot \operatorname{dist}(\bm{a},\bm{\hat{a}}),
\end{eqnarray*}
where $\bm{a}$ and $\bm{b}$ are the first singular vectors of $\bm\Gamma$.
\end{Le}

\begin{proof}
Define $\bm{\bar{b}}$ such that $(\bm{\bar{b}})_m = (\bm{\Gamma}^\top \bm{\hat{a}})_m$ for $m \in \mathcal{S}_Y$ and $(\bm{\bar{b}})_m=\bm{0}$ when $m \notin \mathcal{S}_Y$.
Using Lemma \ref{Le: projection}, we obtain for any $d\in \mathbb{R}$,

\begin{eqnarray*}
    \operatorname{dist}(\hat{\bm{b}},\bm{b})  \leq \frac{\|d\hat{\bm{b}}-\eta_1\bm{b}\|_E}{\eta_1}
    &\leq & \frac{\|d\hat{\bm{b}}-\bm{\bar{b}}\|_E}{\eta_1}+\frac{\|\bm{\bar{b}}-\eta_1\bm{b}\|_E}{\eta_1}\\
     &\leq& \frac{\|d\hat{\bm{b}}-\bm{\bar{b}}\|_E}{\eta_1}+\frac{\|\bm{\Gamma}^\top\bm{\hat{a}}-\eta_1\bm{b}\|_E}{\eta_1}\\
   &\leq& \frac{\|d\hat{\bm{b}}-\bm{\bar{b}}\|_E}{\eta_1}+\frac{\|\eta_1\bm{b}\bm{a}^\top\bm{\hat{a}}-\eta_1\bm{b}+\sum_{r>1}\eta_r\bm{b}_r\bm{a}_r^\top\bm{\hat{a}}\|_E}{\eta_1}\\
   &=& \frac{\|d\hat{\bm{b}}-\bm{\bar{b}}\|_E}{\eta_1}+\frac{\sqrt{\eta_1^2|\bm{a}^\top\bm{\hat{a}}-1|^2+\sum_{r>1}(\eta_r\bm{a}_r^\top\bm{\hat{a}})^2}}{\eta_1}\\
   &\leq& \frac{\|d\hat{\bm{b}}-\bm{\bar{b}}\|_E}{\eta_1}+
   |\bm{a}^\top\bm{\hat{a}}-1|+\frac{\eta_2}{\eta_1}\cdot\sqrt{\sum_{r>1}(\bm{a}_r^\top\bm{\hat{a}})^2}, 
\end{eqnarray*} 
where $\eta_r$, $\bm{b}_r$, and $\bm{a}_r$ are the $r$th singular value and singular vectors of $\bm{\Gamma}$, $r>1$. Without loss of generality, we assume that $\bm{a}^\top\bm{\hat{a}}\geq 0$. Then
\begin{eqnarray*}
   |\bm{a}^\top\bm{\hat{a}}-1|=\big| 1-\sqrt{1-\operatorname{dist}^2(\bm{a},\bm{\hat{a}})}\big|&=&\frac{\operatorname{dist}^2(\bm{a},\bm{\hat{a}})}{1+\sqrt{1-\operatorname{dist}^2(\bm{a},\bm{\hat{a}})}}\leq \operatorname{dist}^2(\bm{a},\bm{\hat{a}}). 
\end{eqnarray*}
In addition, 
\begin{eqnarray*}
{\sum_{r>1}(\bm{a}_r^\top\bm{\hat{a}})^2} \leq   \operatorname{dist}^2(\bm{a},\bm{\hat{a}}).
\end{eqnarray*}
Combine with the above results,
\begin{eqnarray}\label{equ_lemma_3}
\operatorname{dist}(\hat{\bm{b}},\bm{b}) \leq \frac{\|d\hat{\bm{b}}-\bm{\bar{b}}\|_E}{\eta_1}+ \operatorname{dist}^2(\bm{a},\bm{\hat{a}})+\frac{\eta_2}{\eta_1} \cdot \operatorname{dist}(\bm{a},\bm{\hat{a}}),
\end{eqnarray}
for all $d\in \mathbb{R}$.

In the following, we examine the error bound between $d\hat{\bm{b}}$ and $\bm{\bar{b}}$ when $d=1$.
First note that
\begin{eqnarray*}
\big\|\hat{\bm{\Gamma}}-\bm{\hat{a}}({\bm{\hat{b}}})^\top\big\|_F^2+\varrho_Y\sum_{m=1}^q\|({\bm{\hat{b}}})_m\|_E \leq \big\|\hat{\bm{\Gamma}}-\bm{\hat{a}}{\bm{\bar{b}}}^\top\big\|_F^2+\varrho_Y\sum_{m=1}^q\|({\bm{\bar{b}}})_m\|_E.  
\end{eqnarray*}
Rearranging the above inequality, we have
\begin{eqnarray*}
&& \|\bm{\hat{b}}-\bm{\bar{b}}\|^2_E+\varrho_Y\sum_{m=1}^q\|({\bm{\hat{b}}})_m\|_E\\
    &\leq& 2\langle \hat{\bm{\Gamma}}^\top\bm{\hat{a}}-\bar{\bm{b}},\bm{\hat{b}}-\bm{\bar{b}}\rangle+\varrho_Y\sum_{m=1}^q\|({\bm{\bar{b}}})_m\|_E\\
    &\leq&  2\sum_{m=1}^q\|(\hat{\bm{\Gamma}}^\top\bm{\hat{a}}-\bm{\bar{b}})_m\|_E\cdot\|({\bm{\hat{b}}}-\bm{\bar{b}})_m\|_E+\varrho_Y\sum_{m\in \mathcal{S}_Y}\|({\bm{\bar{b}}})_m\|_E\\
&\leq &C{\varrho_Y} \cdot \sum_{m\in \mathcal{S}_Y}\|({\bm{\hat{b}}}-\bm{\bar{b}})_m\|_E+ \sum_{m\notin \mathcal{S}_Y}\|(\hat{\bm{\Gamma}}^\top\bm{\hat{a}})_m\|_E\cdot\|({\bm{\hat{b}}})_m\|_E + \varrho_Y\sum_{m\in \mathcal{S}_Y}\|({\bm{\bar{b}}})_m\|_E.   
\end{eqnarray*}
Notice that
\begin{eqnarray*}
&&\sum_{m\notin \mathcal{S}_Y}\|(\hat{\bm{\Gamma}}^\top\bm{\hat{a}})_m\|_E\cdot\|({\bm{\hat{b}}})_m\|_E\\
&\leq&
\sum_{m\notin \mathcal{S}_Y}\|(\hat{\bm{\Gamma}}^\top\bm{\hat{a}}-{\bm{\Gamma}}^\top\bm{\hat{a}})_m\|_E\cdot\|({\bm{\hat{b}}})_m\|_E+
\sum_{m\notin \mathcal{S}_Y}\|({\bm{\Gamma}}^\top\bm{\hat{a}})_m\|_E\cdot\|({\bm{\hat{b}}})_m\|_E\\
&\leq&
C\varrho_Y\cdot
\sum_{m\notin \mathcal{S}_Y}\|({\bm{\hat{b}}})_m\|_E+
\sum_{m\notin \mathcal{S}_Y}\|({\bm{\Gamma}}^\top\bm{\hat{a}})_m\|_E\cdot\|({\bm{\hat{b}}})_m\|_E.
\end{eqnarray*}
Combining the above two inequalities, we obtain
\begin{eqnarray*}
\|\bm{\hat{b}}-\bm{\bar{b}}\|^2_E
\leq \frac{3}{2}\cdot {\varrho_Y} \cdot \sum_{m\in \mathcal{S}_Y}\|({\bm{\hat{b}}}-\bm{\bar{b}})_m\|_E  +\sum_{m\notin \mathcal{S}_Y}\|({\bm{\Gamma}}^\top\bm{\hat{a}})_m\|_E\cdot\|({\bm{\hat{b}}})_m\|_E.   
\end{eqnarray*}
Notice that
\begin{eqnarray*}
({\bm{\Gamma}}^\top\bm{\hat{a}})_m=\bigg(\sum_{r=1}^R\eta_r\bm{b}_r\bm{a}_r^\top\bm{\hat{a}}\bigg)_m= \sum_{r=1}^R\eta_r(\bm{b}_r)_m\bm{a}_r^\top\bm{\hat{a}}.
\end{eqnarray*}
Since for $j \notin \mathcal{S}_X$ and $m \notin \mathcal{S}_Y$, we have $[\bm{R}_{XY}(t_1,t_2)]_{j,m} = 0$. Consequently, $[\bm{\Lambda}^{XY}_{k_1,k_2}]_{j,m} = 0$ for all $k_1$ and $k_2$. This implies that $(\bm{b}_r)_m = 0$ for $m \notin \mathcal{S}_Y$.
As a result,
\begin{eqnarray*}
\|\bm{\hat{b}}-\bm{\bar{b}}\|_E^2
\leq \frac{3}{2}\cdot {\varrho_Y} \cdot \sum_{m\in \mathcal{S}_Y}\|({\bm{\hat{b}}}-\bm{\bar{b}})_m\|_E\lesssim \varrho_Y\sqrt{s_Y} \|\bm{\hat{b}}-\bm{\bar{b}}\|_E,  
\end{eqnarray*}
which in turn leads to
\begin{eqnarray*}
    \|\bm{\hat{b}}-\bm{\bar{b}}\|_E\lesssim  \varrho_Y\sqrt{s_Y}. 
\end{eqnarray*}
Combining the above result with \eqref{equ_lemma_3}, we then complete our proof.
\end{proof}

\

\newpage

\begin{Le}\label{Le: bound_gamma}
Recall that $\bm{\Lambda}^{XY}_{k_1,k_2}=\int_0^1\int_0^1 \bm{R}_{XY}(t_1,t_2)\wtilde\psi_{k_1}^X(t_1)\wtilde\psi_{k_2}^Y(t_2)\ \mathrm{d}t_1\mathrm{d}t_2\in \mathbb{R}^{p\times q}$, and we denote the $(j,m)$th element of $\bm{\Lambda}^{XY}_{k_1,k_2}$ as 
\[
{\Lambda}^{XY}_{jm,k_1,k_2}= \int_0^1\int_0^1 {R}_{jm,XY}(t_1,t_2)\wtilde\psi_{k_1}^X(t_1)\wtilde\psi_{k_2}^Y(t_2)\ \mathrm{d}t_1\mathrm{d}t_2,
\]
for $j\leq p$ and $m\leq q$. Similarly, denote
\[
\hat{{\Lambda}}^{XY}_{jm,k_1,k_2}:=\frac{1}{n}\sum_{i=1}^n\frac{1}{N_i^XN_i^Y}\sum_{g=1}^{N_i^X}\sum_{h=1}^{N_i^Y}{G}_{i,jm}^{XY}(T_{ig},T_{ih}) \hat{\psi}_{k_1}^X(T_{ig})\hat{\psi}_{k_2}^Y(T_{ih}).
\]
where 
$$
G_{i,jm}^{XY}(T_{ig},T_{ih}):=\big\{\varepsilon^X_{ij}(T_{ig})+\tau^X_{ijg}\big\}\cdot\big\{\varepsilon^Y_{im}(T_{ih})+\tau^Y_{imh}\big\}.
$$    
Suppose Assumptions \ref{A1} -- \ref{A6} hold, $\max_{k\leq\kappa_X}\mathbb{E}\big\|\wtilde\psi_{k}^X-\hat\psi_{k}^X\big\|^2=o(1)$ and $\max_{k\leq\kappa_Y}\mathbb{E}\big\|\wtilde\psi_{k}^Y-\hat\psi_{k}^Y\big\|^2=o(1)$ as $n\rightarrow\infty$. Then
\begin{eqnarray*}
&&\max_{1\leq j\leq p,1\leq  m\leq q,1\leq k_1\leq \kappa_X,1\leq k_2\leq \kappa_Y}\mathbb{E}|{\Lambda}^{XY}_{jm,k_1,k_2}- \hat{{\Lambda}}^{XY}_{jm,k_1,k_2}|^2\\
&=&\mathcal{O}\bigg(n^{-1}+\max_{k\leq\kappa_X}\mathbb{E}\big\|\wtilde\psi_{k}^X-\hat\psi_{k}^X\big\|^2+\max_{k\leq\kappa_Y}\mathbb{E}\big\|\wtilde\psi_{k}^Y-\hat\psi_{k}^Y\big\|^2\bigg)
\end{eqnarray*}
as $n,p,q\rightarrow\infty$.
\end{Le}

\begin{proof}
Without loss of generality, we consider the setting where the observation times
$T_{ig}$ and $T_{ih}$ are independently and uniformly
distributed on $\mathcal{T}=[0,1]$.
To account for the dependence between the estimated
eigenfunctions and the observations in their
estimation, we assume that
$\max_{k_1\leq\kappa_X}
\left[
\mathbb{E}
\left\{
\sup_{t\in\mathcal{T}}
\left|
\wtilde{\psi}_{k_1}^X(t)
-\hat{\psi}_{k_1}^X(t)
\right|^4
\right\}
\right]^{1/2}
\lesssim
\max_{k_1\leq\kappa_X}
\mathbb{E}
\left\|
\wtilde{\psi}_{k_1}^X
-\hat{\psi}_{k_1}^X
\right\|^2$
and
$\max_{k_2\leq\kappa_Y}$
$
\left[
\mathbb{E}\allowbreak
\left\{
\sup_{t\in\mathcal{T}}
\left|
\wtilde{\psi}_{k_2}^Y(t)
-\hat{\psi}_{k_2}^Y(t)
\right|^4
\right\}
\right]^{1/2}\lesssim
\max_{k_2\leq\kappa_Y}
\mathbb{E}
\left\|
\wtilde{\psi}_{k_2}^Y
-\hat{\psi}_{k_2}^Y
\right\|^2.$
We also assume that the empirical cross-products
have uniformly bounded fourth moments:
\[
\max_{\substack{
1\leq i\leq n,\;1\leq j\leq p,\;1\leq m\leq q\\
1\leq g\leq N_i^X,\;1\leq h\leq N_i^Y}}
\mathbb{E}
\left|
G_{i,jm}^{XY}(T_{ig},T_{ih})
\right|^4
\leq C.
\]

Notice that
\begin{align*}
&\left|
\Lambda^{XY}_{jm,k_1,k_2}
-\hat{\Lambda}^{XY}_{jm,k_1,k_2}
\right|\\
&\leq
\left|
\iint_0^1
R_{jm}^{XY}(t_1,t_2)
\wtilde{\psi}_{k_1}^X(t_1)
\wtilde{\psi}_{k_2}^Y(t_2)
\,\mathrm{d}t_1\mathrm{d}t_2
\right.\\
&\qquad\left.
-\frac{1}{n}\sum_{i=1}^n
\frac{1}{N_i^XN_i^Y}
\sum_{g=1}^{N_i^X}
\sum_{h=1}^{N_i^Y}
R_{jm}^{XY}(T_{ig},T_{ih})
\wtilde{\psi}_{k_1}^X(T_{ig})
\wtilde{\psi}_{k_2}^Y(T_{ih})
\right|\\
&\quad+
\frac{1}{n}\sum_{i=1}^n
\frac{1}{N_i^XN_i^Y}
\sum_{g=1}^{N_i^X}
\sum_{h=1}^{N_i^Y}
\left|R_{jm}^{XY}(T_{ig},T_{ih})\right|
\left|\wtilde{\psi}_{k_1}^X(T_{ig})\right|
\left|
\wtilde{\psi}_{k_2}^Y(T_{ih})
-\hat{\psi}_{k_2}^Y(T_{ih})
\right|\\
&\quad+
\frac{1}{n}\sum_{i=1}^n
\frac{1}{N_i^XN_i^Y}
\sum_{g=1}^{N_i^X}
\sum_{h=1}^{N_i^Y}
\left|R_{jm}^{XY}(T_{ig},T_{ih})\right|
\left|
\wtilde{\psi}_{k_1}^X(T_{ig})
-\hat{\psi}_{k_1}^X(T_{ig})
\right|
\left|\hat{\psi}_{k_2}^Y(T_{ih})\right|\\
&\quad+
\left|
\frac{1}{n}\sum_{i=1}^n
\frac{1}{N_i^XN_i^Y}
\sum_{g=1}^{N_i^X}
\sum_{h=1}^{N_i^Y}
\left\{
R_{jm}^{XY}(T_{ig},T_{ih})
-G_{i,jm}^{XY}(T_{ig},T_{ih})
\right\}
\wtilde{\psi}_{k_1}^X(T_{ig})
\wtilde{\psi}_{k_2}^Y(T_{ih})
\right|\\
&\quad+
\frac{1}{n}\sum_{i=1}^n
\frac{1}{N_i^XN_i^Y}
\sum_{g=1}^{N_i^X}
\sum_{h=1}^{N_i^Y}
\left|
R_{jm}^{XY}(T_{ig},T_{ih})
-G_{i,jm}^{XY}(T_{ig},T_{ih})
\right|\\
&\qquad\qquad\times
\left|
\hat{\psi}_{k_1}^X(T_{ig})
\hat{\psi}_{k_2}^Y(T_{ih})
-
\wtilde{\psi}_{k_1}^X(T_{ig})
\wtilde{\psi}_{k_2}^Y(T_{ih})
\right|\\
&:= (A)+(B)+(C)+(D)+(E).
\end{align*}

We first note that
\begin{align*}
&\iint_0^1
\left|R_{jm}^{XY}(t_1,t_2)\right|^2
\,\mathrm{d}t_1\mathrm{d}t_2\\
&\leq
\iint_0^1
\mathbb{E}
\left|
\varepsilon_{ij}^X(t_1)
\varepsilon_{im}^Y(t_2)
\right|^2
\,\mathrm{d}t_1\mathrm{d}t_2\\
&=
\mathbb{E}
\left(
\|\varepsilon_{ij}^X\|^2
\|\varepsilon_{im}^Y\|^2
\right)\\
&\leq
\sqrt{
\mathbb{E}\|\varepsilon_{ij}^X\|^4
\cdot
\mathbb{E}\|\varepsilon_{im}^Y\|^4
}
\leq C,
\end{align*}
by Assumption~\ref{A1}.

Moreover, since
\[
\mathbb{E}
\left\{
G_{i,jm}^{XY}(T_{ig},T_{ih})
\mid T_{ig},T_{ih}
\right\}
=
R_{jm}^{XY}(T_{ig},T_{ih}),
\]
Jensen's inequality and the fourth-moment
condition imply
\[
\mathbb{E}
\left|
R_{jm}^{XY}(T_{ig},T_{ih})
\right|^4
\leq C.
\]

For the first term, by the uniform distribution
of observation times and independence across
subjects, we have
\begin{align*}
\mathbb{E}(A)^2
&=
\operatorname{Var}
\bigg\{
\frac{1}{n}\sum_{i=1}^n
\frac{1}{N_i^XN_i^Y}
\sum_{g=1}^{N_i^X}
\sum_{h=1}^{N_i^Y}
R_{jm}^{XY}(T_{ig},T_{ih})
\wtilde{\psi}_{k_1}^X(T_{ig})
\wtilde{\psi}_{k_2}^Y(T_{ih})
\bigg\}\\
&=
\frac{1}{n^2}\sum_{i=1}^n
\operatorname{Var}
\bigg\{
\frac{1}{N_i^XN_i^Y}
\sum_{g=1}^{N_i^X}
\sum_{h=1}^{N_i^Y}
R_{jm}^{XY}(T_{ig},T_{ih})
\wtilde{\psi}_{k_1}^X(T_{ig})
\wtilde{\psi}_{k_2}^Y(T_{ih})
\bigg\}\\
&\leq
\frac{1}{n^2}\sum_{i=1}^n
\mathbb{E}
\bigg\{
\frac{1}{N_i^XN_i^Y}
\sum_{g=1}^{N_i^X}
\sum_{h=1}^{N_i^Y}
R_{jm}^{XY}(T_{ig},T_{ih})
\wtilde{\psi}_{k_1}^X(T_{ig})
\wtilde{\psi}_{k_2}^Y(T_{ih})
\bigg\}^2\\
&\lesssim
\frac{1}{n^2}\sum_{i=1}^n
\iint_0^1
\left|R_{jm}^{XY}(t_1,t_2)\right|^2
\,\mathrm{d}t_1\mathrm{d}t_2\\
&\lesssim \frac{1}{n},
\end{align*}
where the last two inequalities follow from
Jensen's inequality and the uniform boundedness
of the eigenfunctions.

Therefore,
\[
\mathbb{E}(A)^2=\mathcal{O}(n^{-1}).
\]

For the second term, by the Cauchy--Schwarz
inequality and the uniform boundedness of
$\wtilde{\psi}_{k_1}^X$, we have
\begin{align*}
(B)^2
&\lesssim
\bigg\{
\frac{1}{n}\sum_{i=1}^n
\frac{1}{N_i^XN_i^Y}
\sum_{g=1}^{N_i^X}
\sum_{h=1}^{N_i^Y}
\left|
R_{jm}^{XY}(T_{ig},T_{ih})
\right|^2
\bigg\}\\
&\qquad\times
\bigg\{
\frac{1}{n}\sum_{i=1}^n
\frac{1}{N_i^XN_i^Y}
\sum_{g=1}^{N_i^X}
\sum_{h=1}^{N_i^Y}
\left|
\wtilde{\psi}_{k_2}^Y(T_{ih})
-\hat{\psi}_{k_2}^Y(T_{ih})
\right|^2
\bigg\}\\
&\leq
\bigg\{
\frac{1}{n}\sum_{i=1}^n
\frac{1}{N_i^XN_i^Y}
\sum_{g=1}^{N_i^X}
\sum_{h=1}^{N_i^Y}
\left|
R_{jm}^{XY}(T_{ig},T_{ih})
\right|^2
\bigg\}\cdot
\sup_{t\in\mathcal{T}}
\left|
\wtilde{\psi}_{k_2}^Y(t)
-\hat{\psi}_{k_2}^Y(t)
\right|^2.
\end{align*}

Taking expectations and applying the
Cauchy--Schwarz inequality again yields
\begin{align*}
\mathbb{E}(B)^2
&\lesssim
\bigg[
\mathbb{E}
\bigg\{
\frac{1}{n}\sum_{i=1}^n
\frac{1}{N_i^XN_i^Y}
\sum_{g=1}^{N_i^X}
\sum_{h=1}^{N_i^Y}
\left|
R_{jm}^{XY}(T_{ig},T_{ih})
\right|^2
\bigg\}^2
\bigg]^{1/2}\\
&\qquad\times
\bigg[
\mathbb{E}
\bigg\{
\sup_{t\in\mathcal{T}}
\left|
\wtilde{\psi}_{k_2}^Y(t)
-\hat{\psi}_{k_2}^Y(t)
\right|^4
\bigg\}
\bigg]^{1/2}\\
&\lesssim
\mathbb{E}
\left\|
\wtilde{\psi}_{k_2}^Y
-\hat{\psi}_{k_2}^Y
\right\|^2.
\end{align*}

Here, the last inequality follows from the
uniform fourth-moment bound and the assumed
uniform convergence condition for the estimated
eigenfunctions.

Therefore,
\[
\mathbb{E}(B)^2
=
\mathcal{O}
\left(
\max_{k_2\leq\kappa_Y}
\mathbb{E}
\left\|
\wtilde{\psi}_{k_2}^Y
-\hat{\psi}_{k_2}^Y
\right\|^2
\right).
\]

Similarly,
\[
\mathbb{E}(C)^2
=
\mathcal{O}
\left(
\max_{k_1\leq\kappa_X}
\mathbb{E}
\left\|
\wtilde{\psi}_{k_1}^X
-\hat{\psi}_{k_1}^X
\right\|^2
\right).
\]

For the fourth term, note that
\[
\mathbb{E}
\left\{
R_{jm}^{XY}(T_{ig},T_{ih})
-
G_{i,jm}^{XY}(T_{ig},T_{ih})
\mid T_{ig},T_{ih}
\right\}
=0.
\]

Using the independence across subjects
and the uniform boundedness of
$\wtilde{\psi}_{k_1}^X$ and
$\wtilde{\psi}_{k_2}^Y$, we obtain
\begin{align*}
\mathbb{E}(D)^2
&=
\operatorname{Var}
\bigg\{
\frac{1}{n}\sum_{i=1}^n
\frac{1}{N_i^XN_i^Y}
\sum_{g=1}^{N_i^X}
\sum_{h=1}^{N_i^Y}
\big\{
R_{jm}^{XY}(T_{ig},T_{ih})
-
G_{i,jm}^{XY}(T_{ig},T_{ih})
\big\}\\
&\qquad\qquad\times
\wtilde{\psi}_{k_1}^X(T_{ig})
\wtilde{\psi}_{k_2}^Y(T_{ih})
\bigg\}\\
&\leq
\frac{1}{n^2}\sum_{i=1}^n
\mathbb{E}
\bigg\{
\frac{1}{N_i^XN_i^Y}
\sum_{g=1}^{N_i^X}
\sum_{h=1}^{N_i^Y}
\big\{
R_{jm}^{XY}(T_{ig},T_{ih})
-
G_{i,jm}^{XY}(T_{ig},T_{ih})
\big\}\\
&\qquad\qquad\times
\wtilde{\psi}_{k_1}^X(T_{ig})
\wtilde{\psi}_{k_2}^Y(T_{ih})
\bigg\}^2\\
&\lesssim
\frac{1}{n^2}\sum_{i=1}^n
\frac{1}{N_i^XN_i^Y}
\sum_{g=1}^{N_i^X}
\sum_{h=1}^{N_i^Y}
\mathbb{E}
\left|
R_{jm}^{XY}(T_{ig},T_{ih})
-
G_{i,jm}^{XY}(T_{ig},T_{ih})
\right|^2\\
&\lesssim
\frac{1}{n}.
\end{align*}

Therefore,
\[
\mathbb{E}(D)^2=\mathcal{O}(n^{-1}).
\]

Finally, for the fifth term, the uniform
boundedness of the true and estimated
eigenfunctions implies
\begin{align*}
&\left|
\hat{\psi}_{k_1}^X(T_{ig})
\hat{\psi}_{k_2}^Y(T_{ih})
-
\wtilde{\psi}_{k_1}^X(T_{ig})
\wtilde{\psi}_{k_2}^Y(T_{ih})
\right|\\
&\quad\lesssim
\left|
\hat{\psi}_{k_1}^X(T_{ig})
-
\wtilde{\psi}_{k_1}^X(T_{ig})
\right|
+
\left|
\hat{\psi}_{k_2}^Y(T_{ih})
-
\wtilde{\psi}_{k_2}^Y(T_{ih})
\right|.
\end{align*}

Consequently,
\begin{align*}
(E)^2
&\lesssim
\bigg\{
\sup_{t\in\mathcal{T}}
\left|
\hat{\psi}_{k_1}^X(t)
-
\wtilde{\psi}_{k_1}^X(t)
\right|^2
+
\sup_{t\in\mathcal{T}}
\left|
\hat{\psi}_{k_2}^Y(t)
-
\wtilde{\psi}_{k_2}^Y(t)
\right|^2
\bigg\}\\
&\qquad\times
\bigg\{
\frac{1}{n}\sum_{i=1}^n
\frac{1}{N_i^XN_i^Y}
\sum_{g=1}^{N_i^X}
\sum_{h=1}^{N_i^Y}
\left|
R_{jm}^{XY}(T_{ig},T_{ih})
-
G_{i,jm}^{XY}(T_{ig},T_{ih})
\right|
\bigg\}^2.
\end{align*}

By the Cauchy--Schwarz inequality,
\begin{align*}
\mathbb{E}(E)^2
&\lesssim
\bigg[
\mathbb{E}
\bigg\{
\sup_{t\in\mathcal{T}}
\left|
\hat{\psi}_{k_1}^X(t)
-
\wtilde{\psi}_{k_1}^X(t)
\right|^4
\bigg\}
\bigg]^{1/2}+
\bigg[
\mathbb{E}
\bigg\{
\sup_{t\in\mathcal{T}}
\left|
\hat{\psi}_{k_2}^Y(t)
-
\wtilde{\psi}_{k_2}^Y(t)
\right|^4
\bigg\}
\bigg]^{1/2}\\
&\lesssim
\max_{k_1\leq\kappa_X}
\mathbb{E}
\left\|
\wtilde{\psi}_{k_1}^X
-
\hat{\psi}_{k_1}^X
\right\|^2
+
\max_{k_2\leq\kappa_Y}
\mathbb{E}
\left\|
\wtilde{\psi}_{k_2}^Y
-
\hat{\psi}_{k_2}^Y
\right\|^2,
\end{align*}
where the first inequality uses the uniformly
bounded fourth moments of the empirical
cross-products, and the second follows from
the assumed uniform convergence conditions.

Combining the convergence orders of
$(A)$, $(B)$, $(C)$, $(D)$, and $(E)$,
we obtain
\begin{align*}
\mathbb{E}
\left|
\Lambda_{jm,k_1,k_2}^{XY}
-
\hat{\Lambda}_{jm,k_1,k_2}^{XY}
\right|^2=
\mathcal{O}
\bigg(
n^{-1}
+
\max_{k_1\leq\kappa_X}
\mathbb{E}
\left\|
\wtilde{\psi}_{k_1}^X
-
\hat{\psi}_{k_1}^X
\right\|^2
+
\max_{k_2\leq\kappa_Y}
\mathbb{E}
\left\|
\wtilde{\psi}_{k_2}^Y
-
\hat{\psi}_{k_2}^Y
\right\|^2
\bigg).
\end{align*}

Since the moment and convergence bounds
are uniform over $j,m,k_1,k_2$,
the claimed result follows.
\end{proof}

\

\begin{Le}\label{Le: projection}
       For any $d\in \mathbb{R}$, we have
       \begin{eqnarray*}
    \operatorname{dist}(\bm{u},\bm{v}) &\leq & \frac{\|\bm{u} - d \bm{v}\|_E}{\|\bm{u}\|_E}, \quad \forall \bm{u},\bm{v} \in \mathbb{R}^n.
\end{eqnarray*}
    \end{Le}
    \begin{proof}
        \begin{equation*}
            \begin{split}
                & \operatorname{dist}(\bm{u}, \bm{v}) = \sqrt{1 - \frac{\langle \bm{u}, \bm{v}\rangle^2}{\|\bm{u}\|_E^2 \|\bm{v}\|_E^2}} \leq \frac{\|\bm{u}-d \bm{v}\|_E}{\|\bm{u}\|_E} \\
                \Leftarrow \quad & 1 - \frac{\langle \bm{u}, \bm{v}\rangle^2}{\|\bm{u}\|_E^2 \|\bm{v}\|_E^2} \leq \frac{\|\bm{u}\|_E^2 - 2 d \langle \bm{u}, \bm{v}\rangle + d^2 \|\bm{v}\|_E^2}{\|\bm{u}\|_E^2} \\
                \Leftarrow \quad & 0 \leq \langle \bm{u}, \bm{v}\rangle^2 - 2 d \|\bm{v}\|_E^2 \langle \bm{u}, \bm{v}\rangle + d^2 \|\bm{v}\|_E^4 \\
                \Leftarrow \quad & 0 \leq \left( \langle \bm{u}, \bm{v}\rangle - d \|\bm{v}\|_E^2 \right)^2.
            \end{split}
        \end{equation*}
    \end{proof}

\

\section{Optimal Orders for the Tuning}\label{op_sec}

Recall that for $r \leq R$, $\hat{\bm{A}}^X_r$ and $\hat{\bm{A}}^Y_r$ are obtained from \eqref{est_functional_loading}, 
with $\hat{\bm{a}}_{r}$ and $\hat{\bm{b}}_{r}$ estimated via \eqref{opt_SVD}, 
where the tuning parameters are denoted by $\varrho^X_r$ and $\varrho^Y_r$.
The optimal orders of $\varrho^X_r$ and $\varrho^Y_r$ are
\[
C_{\bm{\Gamma}}\cdot
\Big(\sqrt{q}\big(\text{Error}_{\mu}+\text{Error}_{H}\big)\max\{\kappa_X^{a+2},\kappa_Y^{a+2}\}\Big)\cdot \log p
\]
and 
\[
C_{\bm{\Gamma}}\cdot
\Big(\sqrt{p}\big(\text{Error}_{\mu}+\text{Error}_{H}\big)\max\{\kappa_X^{a+2},\kappa_Y^{a+2}\}\Big)\cdot \log q,
\]
respectively.
In addition, the truncation levels $\kappa_X$ and $\kappa_Y$ take the optimal order
\begin{eqnarray*}
    \kappa_X,\kappa_Y\asymp 
    \left\{C_{\bm{\Gamma}}\cdot\big(\text{Error}_{\mu}+\text{Error}_{H}\big)\cdot\max\big\{\log(p),\log(q)\big\}\cdot\big\{\sqrt{s_X}+\sqrt{s_Y}\big\}\right\}^{-2/(3a+3)}.
\end{eqnarray*}
For $\nu_H^X$ and $\nu_H^Y$ in covariance kernel estimation, the optimal orders of $\nu_H^X$ and $\nu_H^Y$ are
\[
\min\Big\{\left(n N_X^2\right)^{-2l/(2l+2)},\, N_X^{-2l}\Big\} \;\lesssim\; \nu_H^X \;\lesssim\;
\max\Big\{\left(n N_X^2\right)^{-2l/(2l+2)},\, n^{-1}\Big\},
\]
and
\[
\min\Big\{\left(n N_Y^2\right)^{-2l/(2l+2)},\, N_Y^{-2l}\Big\} \;\lesssim\; \nu_H^Y \;\lesssim\;
\max\Big\{\left(n N_Y^2\right)^{-2l/(2l+2)},\, n^{-1}\Big\};
\]
For tuning parameters $\nu_{\mu,j}^X$ and $\nu_{\mu,j}^Y$ for mean function estimation, their optimal orders can be found in \citet{xiao2020asymptotic}.

\end{document}